\documentstyle[12pt,epsf]{article}

\newcounter{saveeqn}
\newcounter{App} 

\newcommand{\gapprox}{\raisebox{-.2ex}{$\stackrel{\textstyle>}
{\raisebox{-.6ex}[0ex][0ex]{$\sim$}}$}}
\newcommand{\lapprox}{\raisebox{-.2ex}{$\stackrel{\textstyle<}
{\raisebox{-.6ex}[0ex][0ex]{$\sim$}}$}}

\begin{document}


\begin{flushright}
CEBAF-TH-95-18 \\
Revised
\end{flushright}
\vspace{2cm}
\begin{center}
{\Large\bf Transition Form Factor
$\gamma\gamma^*\to\pi^0$
 and QCD sum rules}
\end{center}
\begin{center}
{A.V.RADYUSHKIN\footnote{Also Laboratory of Theoretical Physics,
JINR, Dubna, Russian Federation}
\\ {\em Physics Department, Old Dominion University, Norfolk, VA 23529,USA} \\
{\em and}\\ {\em Continuous Electron Beam Accelerator Facility,} \\
 {\em Newport News, VA 23606, USA}\\
R.RUSKOV \footnote{On leave of absence from Institute of Nuclear
Research and Nuclear Energy,
Bulgarian Academy of Sciences, 1784, Sofia, Bulgaria}\\
{\em Laboratory of Theoretical Physics, JINR, Dubna, Russian Federation}\\ }
\end{center}

\vspace{2cm}
\begin{abstract}

The transition  $\gamma^*(q_1)\gamma^*(q_2) \to \pi^\circ(p)$
is studied within the QCD sum rule framework.
As a first step,  we analyze the
kinematic situation when both  photon virtualities
are spacelike and large.
We  construct a QCD sum rule for
$F_{\gamma^*\gamma^*\pi^\circ}(q_1^2,q_2^2)$
and show that, in the asymptotic limit $ |q_1^2|, |q_2^2| \to \infty$,
it reproduces the leading-order pQCD result.
Then we  study the limit $|q_1^2| \to 0$, in which one of the photons
is (almost) real.   We develop a factorization procedure for the
infrared singularities $\ln(q_1^2), 1/q_1^2, 1/q_1^4, etc.,$ emerging in this limit.
The infrared-sensitive contributions are absorbed in this approach
by  bilocal correlators, which can be also interpreted
as the distribution amplitudes for  (almost) real photon.
Under explicitly formulated assumptions concerning
the form of these amplitudes,  we obtain a QCD sum rule
for $F_{\gamma^*\gamma^*\pi^\circ}(q_1^2=0,q_2^2=-Q^2)$
and study its $Q^2$-dependence.
In contrast to pQCD,  we make no assumptions
about the shape of the pion distribution amplitude $\varphi_{\pi}(x)$.
Our results  agree with the Brodsky-Lepage
proposal that the $Q^2$-dependence of this form factor
is given by an  interpolation
between its  $Q^2=0$ value fixed by the axial anomaly
and  $1/Q^2$ pQCD behaviour  for large $Q^2$,
provided that one interpolates to a value
close to that dictated by the asymptotic
form $\varphi_{\pi}^{as} (x)=6 f_{\pi} x (1-x)$ of
the pion distribution amplitude.
We interpret this as  an evidence that $\varphi_{\pi}(x)$
is rather close to the asymptotic form.

\end{abstract}

\newpage

\section{Introduction}

\setcounter{equation} 0

The studies of the transition form factor
for the process $\gamma^* \gamma^* \to \pi^0$
when two virtual photons $\gamma^*$  produce  a neutral pion
is apparently the cleanest case for testing
QCD predictions for elastic processes.
In contrast to  the  pion electromagnetic  form factor,
the perturbative QCD  subprocess
$\gamma^*(q_1) + \gamma^*(q_2) \to \bar q(\bar xp) + q (xp) $
appears at  the zeroth
order in the QCD coupling constant $\alpha_s$,
and  the asymptotically leading term has no suppression.
The  relevant diagram resembles
the handbag diagram for  the forward virtual
Compton amplitude  used in the studies of  deep inelastic
scattering.
This  gives  good reasons to  expect that
perturbative QCD for this process may
work at  accessible values of spacelike
photon virtualities $q_1^2 \equiv -q^2,
q_2^2 \equiv -Q^2$.
In the lowest order,  pQCD  predicts that \cite{bl80}
\begin{equation}
F_{\gamma^* \gamma^*  \pi^0 }^{LO}(q^2, Q^2) = \frac{4\pi}{3}
\int_0^1 {{\varphi_{\pi}(x)}\over{xQ^2+\bar x q^2}} \, dx  ,
\end{equation}
where $\varphi_{\pi}(x)$ is the pion distribution amplitude
and $x, \bar x \equiv 1-x $ are the fractions
of the pion light-cone momentum carried by the quarks.
In the region where  both photon virtualities are
large: $ q^2 \sim  Q^2 \gapprox \, 1 \, GeV^2$,
the pQCD predicts  the overall $1/Q^2$ fall-off
of the form factor, which differs from  the
naive vector meson dominance expectation
$F_{\gamma^* \gamma^*  \pi^0 }(q^2, Q^2) \sim 1/q^2 Q^2 \sim 1/Q^4$.
Thus, establishing  the $1/Q^2$ power law in this region
is a crucial    test of pQCD for this process.
The study of $F_{\gamma^* \gamma^*  \pi^0 }(q^2, Q^2)$
over  a wide range of the ratio  $q^2/Q^2$ of
two large photon virtualities
can  then provide  a  nontrivial information
about the shape of $\varphi_{\pi}(x)$.

However, experimentally most favourable
situation  is when one of the photons
is real $q^2=0$ or almost real.
For experiments on $e^+ e^-$-machines having one small virtuality
strongly increases the cross section.
It was also proposed to use
real photons to study the $\pi^0$ production
in $\gamma e$ collisions \cite{gin1}.
On fixed-target accelerators,
like CEBAF, one can attempt to study
the  $F_{\gamma^* \gamma^*  \pi^0 }(q^2=0, Q^2)$
form factor  through $\gamma^* \pi^*  \to \gamma$
processes \cite{afanas} with  a photon in the final state.
In the  $q^2 \to 0$ limit (with $Q^2$ large),  the
nonperturbative information about the pion is accumulated
in the same integral $I$ of  $\varphi_{\pi}(x)/x$ that appears in
the asymptotic pQCD expression for the one-gluon exchange
contribution for the pion electromagnetic form factor.
Hence, information extracted from
the studies of $F_{\gamma^* \gamma^*  \pi^0 }(q^2=0, Q^2)$
can be used to settle the bounds on the
pQCD hard contribution to the pion EM form factor.

Since the zeroth moment of the pion distribution function
is  normalized by the matrix element
of the axial current  ($i.e.,$ by the pion decay
constant $f_{\pi}$),
 the  value of $I$ is sensitive to the
shape of the
pion distribution amplitude $\varphi_{\pi}(x)$.
Two most popular   choices are the asymptotic form
$\varphi_{\pi}^{as}(x) = 6 f_{\pi} x(1-x)$ \cite{tmf78}-\cite{pl80}
and   the CZ model
$\varphi_{\pi}^{CZ}(x) = 30 f_{\pi} x(1-x)(1-2x)^2$ \cite{cz82}.
Using  $\varphi_{\pi}^{CZ}(x)$  increases the integral by
  an extra factor of 5/3 compared to the value based
on $\varphi_{\pi}^{as}(x)$. This observation can be used for
an experimental discrimination between these two models.
In fact, both the CELLO data \cite{CELLO}
and preliminary high-$Q^2$ CLEO data \cite{CLEO}  seem to
favour  the  leading-order
pQCD prediction with the normalization
corresponding to
a rather narrow distribution amplitude
close to $\varphi_{\pi}^{as}(x)$.
To perform  a detailed
comparison   of the (future) data
with  theoretical  predictions,
one should take into account the
pQCD radiative corrections.
These include the  one-loop  contributions  to the hard
scattering amplitude \cite{braaten,kmr},
which,  decreasing  the leading-order result by
about  $20 \%$, still leave a sizable
gap between the  predictions based on two
models mentioned above.
One should also  add
the terms  generated by  two-loop
evolution of the pion distribution amplitude
 \cite{ditrad,mikhrad,sarmadi}.
Originally, these  corrections were found to be tiny \cite{kmr}.
A recent progress \cite{muller1} in understanding the structure
of the two-loop evolution suggests  that the size of these
corrections  is  somewhat larger.
However, the numerical analysis
of the two-loop evolution  presented
in ref.\cite{muller2}  does not indicate
appreciable changes for  the integral
over the distribution amplitude.
Hence,  there are good chances that the controversial
subject of the shape of  $\varphi_{\pi}(x)$
may  soon be settled experimentally.

Within the   pQCD  approach, the pion distribution amplitude
$\varphi_{\pi}(x)$ is a phenomenological model
function whose shape  should be taken
either from experiment (this was not possible so far)
or calculated in some nonperturbative approach,
$e.g.,$ using QCD sum rules.
However,  applications of the QCD sum rules to
nonlocal hadronic characteristics (functions),
like distribution amplitudes
$\varphi(x)$ are much more involved  than those for the simpler
classic cases \cite{svz}  of  hadronic masses
and decay widths.
The trickiest  problem  is that the  underlying
operator product expansion
(OPE) for the relevant correlators
has a slow convergence because
some terms  are parametrically enhanced
by powers of $N$ for the $x^N$ moment
of the distribution amplitude \cite{cz82}.
For this reason,  one needs
a very  detailed information
about the nonperturbative QCD vacuum
to get a reliable sum rule.
Such information is not available yet,
and results for  $\varphi_{\pi}(x)$ have a strong
model dependence\cite{MR}. In the present paper,
instead of starting with the
pQCD approach and taking $\varphi_{\pi}(x)$ from QCD sum rules,
we  calculate  $F_{\gamma^* \gamma^*  \pi^0 }(q^2=0, Q^2)$
directly  from a QCD sum rule for the three-point function.
A serious problem for such an attempt  is that one of the photons
has a small virtuality and the relevant three-point amplitude
is sensitive to  nonperturbative  long-distance QCD dynamics.
For this reason, as an  intermediate  step, we
construct a QCD sum rule for a simpler kinematical
situation when both photon virtualities are large.
To take the $q^2 \to 0$ limit, we perform additional
factorization using the methods developed in
refs.\cite{BalDY}-\cite{BBK}.

The paper is organized as follows. In Section 2
we introduce our basic object:
the correlator of two vector and one axial current,
discuss its  behaviour in different
kinematical situations and interrelation between
QCD sum rules and pQCD approaches.
The situation  when both photon virtualities are
large is considered in Section 3.  We construct there
the QCD sum rule valid in this ``large-$q^2$'' kinematics and  analyze its
structure and some limiting cases.
In Section 4, we outline  specific problems one faces
trying to use  the operator product expansion
in the limit when one of the photon virtualities is small
(small-$q^2$ kinematics). The  basic features
of mass singularities, which appear in this limit,
are illustrated using  a  scalar model  as a toy example.
 The methods developed there are  then used in
 Section 5  to construct a factorization
procedure of long- and short-distance contributions  in
the realistic QCD situation.
We show that the long-distance contributions  in this case
have the structure of  bilocal correlators\footnote{The bilocal
correlators were originally introduced
in a similar context by  Balitsky \cite{bilocal}.}
which
are considered  in Section 6.
In particular, we discuss there the continuum and $\rho$-meson
contributions into the bilocal correlators.
In Section 7, we  study the  contact terms \cite{BalDY} which appear
in some bilocals. In Section 8, we collect together
the contributions calculated in preceding sections
and write down the QCD sum rule for the $F_{\gamma^* \gamma^*  \pi^0 }(q^2, Q^2)$
form factor valid in small-$q^2$ kinematics.
We analyze  the  $q^2 \to 0$ limit and present our results
for  $F_{\gamma^* \gamma^*  \pi^0 }(q^2=0, Q^2)$.
In the concluding section we summarize our findings.
Some technical details of our calculations can be found in the Appendix.

\section{ Form factor $\gamma^* \gamma^* \to \pi^0$  and three-point function}

\setcounter{equation} 0

\subsection{Definitions}

The form factor $F_{\gamma^*\gamma^* \to \pi^o} \left(q_1^2,q_2^2\right)$
of the $\gamma^*\gamma^* \to \pi^o$ transition
 is determined by the matrix element
\begin{equation}
4 \pi \int
\langle {\pi},\stackrel{\rightarrow}{p}
|T\left\{J_{\mu}(X)\,J_{\nu}(0)\right\}| 0 \rangle e^{-iq_1 X } d^4 X
 =  \sqrt{2} i {\epsilon}_{{\mu}{\nu}{q_1}{q_2}}\,
 F_{\gamma^*\gamma^* \to \pi^o}\left(q_1^2,q_2^2\right)
\label{eq:form}
\end{equation}
where  $J_{\mu}$\ is the electromagnetic current of the light quarks
(divided by the electron charge):
\begin{equation}
J_{\mu}=\left(\frac2{3}\bar{u}\gamma_{\mu}u-
\frac1{3}\bar{d}\gamma_{\mu}d\right)\, ,
\label{eq:tokem}
\end{equation}
 $|  {\pi},\stackrel{\rightarrow}{p} \rangle$  is a one-pion state
with  the 4-momentum $p$  and  we use throughout
the convention $\epsilon_{\mu \nu \alpha \beta} q_2^{\beta} \equiv
{\epsilon}_{\mu \nu \alpha q_2 }$, $\epsilon_{\mu \nu \alpha  \beta}
q_1^{\alpha} q_2^{\beta} \equiv  \epsilon_{\mu \nu q_1 q_2 }$, $etc.$

To study the form factor, we should construct first a formalism
in which the pion would  emerge as a QCD bound state in the
$\bar q q$ system. A possible  way is to start with a
  three-point correlation function
\begin{equation}
{\cal{F}}_{\alpha\mu\nu}(q_1,q_2)= 2
\pi i
\int
\langle 0 |T\left\{j_{\alpha}^5(Y)\,J_{\mu}(X)\,J_{\nu}(0)\right\}| 0 \rangle
e^{-iq_{1}X}\,e^{ipY}  d^4X\,d^4Y \,  ,
\label{eq:corr}
\end{equation}
 (cf. \cite{NeRa83})
where $p = q_1 + q_2$.
In addition to the two EM currents present in
eq.(\ref{eq:form}),
the correlator (\ref{eq:corr})
contains also the axial current $j_{\alpha}^5$
\begin{equation}
j_{\alpha}^5 =
\left(\bar{u}\gamma_{5}\gamma_{\alpha}u\,-\,
      \bar{d}\gamma_{5}\gamma_{\alpha}d\right) .
\label{eq:tokax}
\end{equation}
The latter has the necessary property that its
projection onto the neutral pion state is non-zero.
In fact,  this projection
is proportional to   the $\pi^- \to \mu \nu$ decay constant
$f_{\pi} \approx 130.7\, MeV$:
\begin{equation}
\langle 0 |j_{\alpha}^5(0)| {\pi^0},\stackrel{\rightarrow}{p}\rangle =
 -i\,\sqrt{2} \, f_{\pi} p_{\alpha}.
\label{eq:fpi}
\end{equation}

The three-point correlator
(\ref{eq:corr})  has a richer Lorentz structure than  the original amplitude
(\ref{eq:form}),
and not all  the tensor structures it contains are relevant to our study
of $F_{\gamma^*\gamma^* \to \pi^o}(q_1^2,q_2^2)$. Incorporating the  Lorentz invariance
properties of the  three-point function  and Bose symmetry for
the virtual photons, one can write the amplitude ${\cal{F}}_{\alpha\mu\nu}$
as
\begin{eqnarray*}
{\cal{F}}_{\alpha\mu\nu}(q_1,q_2) =
 p_{\alpha}{\epsilon}_{{\mu}{\nu}{q_1}{q_2}}
 {\cal{F}}_1\left(p^2,q_1^2,q_2^2\right)
+r_{\alpha}{\epsilon}_{{\mu}{\nu}{q_1}{q_2}}
 {\cal{A}}_1\left(p^2,q_1^2,q_2^2\right)\\
+ \left[{\epsilon}_{{\alpha}{\mu}{q_1}{q_2}}q_{{1}{\nu}}-
{\epsilon}_{{\alpha}{\nu}{q_1}{q_2}}q_{{2}{\mu}}\right]
 {\cal{F}}_2\left(p^2,q_1^2,q_2^2\right)
+\left[{\epsilon}_{{\alpha}{\mu}{q_1}{q_2}}q_{{2}{\nu}}-
{\epsilon}_{{\alpha}{\nu}{q_1}{q_2}}q_{{1}{\mu}}\right]
 {\cal{F}}_3\left(p^2,q_1^2,q_2^2\right)\\
+\left[{\epsilon}_{{\alpha}{\mu}{q_1}{q_2}}q_{{1}{\nu}}+
{\epsilon}_{{\alpha}{\nu}{q_1}{q_2}}q_{{2}{\mu}}\right]
 {\cal{A}}_2\left(p^2,q_1^2,q_2^2\right)
+\left[{\epsilon}_{{\alpha}{\mu}{q_1}{q_2}}q_{{2}{\nu}}+
{\epsilon}_{{\alpha}{\nu}{q_1}{q_2}}q_{{1}{\mu}}\right]
 {\cal{A}}_3\left(p^2,q_1^2,q_2^2\right)\\
+ {\epsilon}_{{\alpha}{\mu}{\nu}{r}}
\left[{p^2\over{2}}{\cal{F}}_4 +
{r^2\over{2}}{\cal{F}}_5 + (pr){\cal{A}}_6\right]
+ {\epsilon}_{{\alpha}{\mu}{\nu}{p}}\left[{p^2\over{2}}{\cal{A}}_4 +
{r^2\over{2}}{\cal{A}}_5 + (pr){\cal{F}}_6\right],
\end{eqnarray*}
where $p=q_1+q_2,$  $ r = q_1-q_2$.
The invariant amplitudes ${\cal{F}},{\cal{A}}$  have the following symmetry
properties:
${\cal{F}}_i\left(p^2,q_1^2,q_2^2\right)=
{\cal{F}}_i\left(p^2,q_2^2,q_1^2\right), \,
{\cal{A}}_i\left(p^2,q_1^2,q_2^2\right)=
- {\cal{A}}_i\left(p^2,q_2^2,q_1^2\right)$.

Utilizing   the fact that, in the four-dimensional space-time,
 there is no antisymmetric  tensor of rank 5, $i.e.,$
\begin{equation}
{\epsilon}_{{\alpha}{\mu}{\nu}{\gamma}}\,g_{{\delta}{\varepsilon}}+
{\epsilon}_{{\mu}{\nu}{\gamma}{\delta}}\,g_{{\alpha}{\varepsilon}}+
{\epsilon}_{{\nu}{\gamma}{\delta}{\alpha}}\,g_{{\mu}{\varepsilon}}+
{\epsilon}_{{\gamma}{\delta}{\alpha}{\mu}}\,g_{{\nu}{\varepsilon}}+
{\epsilon}_{{\delta}{\alpha}{\mu}{\nu}}\,g_{{\gamma}{\varepsilon}}=0 \  ,
\nonumber
\end{equation}
and using the conditions
 $q_1^{\mu}{\cal{F}}_{\alpha\mu\nu}=q_2^{\nu}{\cal{F}}_{\alpha\mu\nu}=0$
imposed by the EM current conservation,
we get finally the expansion in terms of four invariant amplitudes:
\begin{eqnarray}
{\cal{F}}_{\alpha\mu\nu}(q_1,q_2) & = &
{\epsilon}_{{\mu}{\nu}{q_1}{q_2}}\left[p_{\alpha}F
+r_{\alpha}A\right] \nonumber \\ &+&
 \left[q_{{2}{\nu}}{\epsilon}_{{\alpha}{\mu}{q_1}{q_2}}-
q_{{1}{\mu}}{\epsilon}_{{\alpha}{\nu}{q_1}{q_2}}\right]\widetilde  F
+\left[q_{{2}{\nu}}{\epsilon}_{{\alpha}{\mu}{q_1}{q_2}}+
q_{{1}{\mu}}{\epsilon}_{{\alpha}{\nu}{q_1}{q_2}}\right]
\widetilde  A
\nonumber \\
&+&{\epsilon}_{{\alpha}{\mu}{\nu}{r}}
\left[{{p^2+r^2}\over{4}}\widetilde  F - {(pr)\over2}\widetilde  A\right]
+{\epsilon}_{{\alpha}{\mu}{\nu}{p}}\left[ -
{(pr)\over{2}}\widetilde  F + {{p^2+r^2}\over{4}}\widetilde  A\right] \  .
\label{eq:tensor}
\end{eqnarray}

According to  eqs. (\ref{eq:form}), (\ref{eq:fpi}),
the three-point amplitude ${\cal{F}}_{\alpha\mu\nu}(q_1,q_2)$
has a  pole for $p^2=m_{\pi}^2$:
\begin{equation}
{\cal{F}}_{\alpha\mu\nu}(q_1,q_2) =
\frac{f_{\pi}}{p^2-m_{\pi}^2}
p_{\alpha}{\epsilon}_{{\mu}{\nu}{q_1}{q_2}}
F_{\gamma^*\gamma^* \to \pi^o}\left(q_1^2,q_2^2\right) + \ldots
\label{eq:pole}
\end{equation}
Thus, the tensor
structure of  the pion contribution
is  $p_{\alpha}{\epsilon}_{{\mu}{\nu}{q_1}{q_2}}$.  In  eq.(\ref{eq:tensor}),
it  corresponds  to the invariant amplitude $F\left(q_1^2,q_2^2,p^2\right)$.
In other words,
$F\left(q_1^2,q_2^2,p^2\right)$ has the pole  $1/(p^2-m_{\pi}^2)$, and the
form factor $F_{\gamma^*\gamma^* \to \pi^o}\left(q_1^2,q_2^2\right)$
can be extracted from the  residue of that  pole.

 However, the problem is that  the bound state poles
 are present only in full amplitudes,
formally corresponding to the total sums over all orders of
perturbation theory (in QCD one should not forget to add
also nonperturbative contributions).
Terms corresponding to  any finite order
do not have  such poles.
Fortunately, it is not always necessary to perform
an  explicit all-order summation (impossible in QCD)
to   extract information about a particular bound state.

\subsection{Implications of the axial anomaly}

Using the axial anomaly  \cite{ABJ} relation for massless quarks
\begin{equation}
\partial_{\alpha} j_{\alpha}^5 =
\frac{e^2}{16 \pi^2}
\epsilon_{\mu \nu \rho \sigma } F^{\mu \nu} F^{\rho \sigma }\, ,
\end{equation}
 we obtain the constraint
\begin{equation}
 p^2 F - (q_1^2 + q_2^2) \widetilde  F + (q_1^2 -q_2^2) (A + \widetilde  A) =
 \frac{1}{\pi}.
\label{eq:axconstr}
\end{equation}
For real photons ($q_1^2 = q_2^2=0$), this reduces to
 $p^2 F = 1/\pi$, provided that $\widetilde
F$ and $(A + \widetilde  A)$
do not have $1/q_1^2$ or  $1/q_2^2$ singularities,
which is true since  there are no massless
$\bar q q$  bound states in the vector channel.
Hence,  $F \sim 1/p^2$ for real photons, $i.e.,$  $F $  really
has a pole corresponding to a massless pion \cite{dz,consist}.
Furthermore, the anomaly relation (\ref{eq:axconstr}) fixes the   value
of the $F_{\gamma^*\gamma^* \to \pi^o}(q_1^2,q_2^2)$   \,
form factor when both
 virtualities of the photons are zero :
\begin{equation}
F_{\gamma^*\gamma^* \to \pi^o}(0,0) = {{1}\over{\pi f_{\pi}}}.
\label{eq:norma}
\end{equation}
It should be emphasized  that the anomaly relation
does not require that $F_{\gamma^*\gamma^* \to \pi^o}(0,Q^2)$
has this value for all $Q^2$: eq.(\ref{eq:axconstr})
can be satisfied even if $F_{\gamma^*\gamma^* \to \pi^o}(q^2,Q^2)$
has  a nontrivial dependence on photon virtualities.

Thus, the axial anomaly allows one to calculate exactly
one  particular  combination of invariant amplitudes
(\ref{eq:axconstr}) for  arbitrary values of the virtualities
$q_1^2,q_2^2,p^2$.  In QCD, this is a rather exceptional
situation. Normally, a reliable QCD calculation is possible
only in the region of large virtualities where one can
incorporate  the asymptotic freedom property of
the theory.

\begin{figure}[ht]
\mbox{
   \epsfxsize=12cm
 \epsfysize=5cm
 \hspace{0.5cm}
 \epsffile{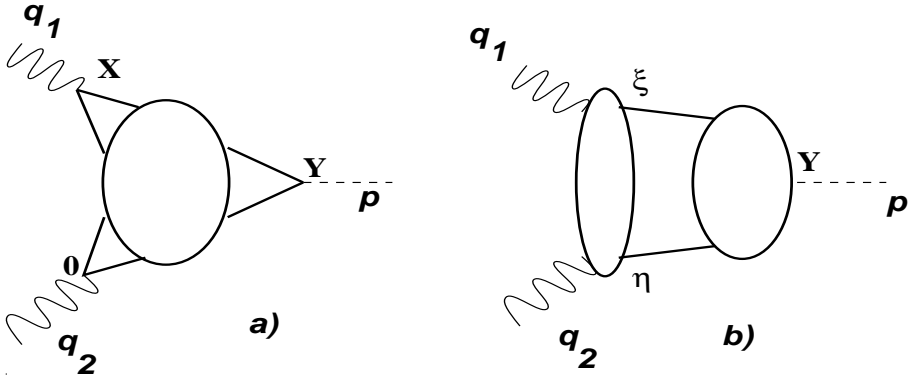}  }
  \vspace{0.7cm}
{\caption{\label{first}
$a)$ Three-point correlation function. $b)$ Structure of factorization
in the  limit \mbox{$Q^2,q^2 \gg |p^2|$}.
 }}
\end{figure}

\subsection{Factorizable contributions and perturbative QCD}

If both of the
 photon virtualities are large,
the leading  term of the $1/q^2$-expansion of any diagram
contributing to the amplitude  $F$
can be written in the factorized form (cf. \cite{20,20a}):
\begin{equation}
F\left(p^2,q_1^2,q_2^2\right) \sim
\int  C(\xi, \eta, q_1, q_2)  \,
\Pi(\xi, \eta, p) d^4 \xi d^4 \eta \,  ,
\end{equation}
where $C(\xi, \eta, q_1, q_2)$ is the short-distance
coefficient function and $\Pi(\xi, \eta, p)$ is the long-distance
factor given by a particular term of the PT expansion for the
correlator of the axial current with a composite
operator ${\cal O}(\xi , \eta) \sim \bar q (\xi) \ldots
 q(\eta)$:
\begin{equation}
\Pi(\xi, \eta, p) \sim  \int  \langle 0| T ({\cal O}(\xi , \eta)
 j(Y)) |0  \rangle \, e^{ipY} d^4 Y  .
\end{equation}
The correlator $\Pi(\xi, \eta, p)$
is formally given by a sum over all orders of PT,
and it  has a pole for $p^2 = m_{\pi}^2$:
\begin{equation}
\Pi(\xi, \eta; p) = \frac{f_{\pi}}{p^2 - m_{\pi}^2}
\langle 0| {\cal O}(\xi , \eta) | {\pi^0},\stackrel{\rightarrow}{p} \rangle \ .
\label{eq:corrpole}
\end{equation}
Comparing eqs. (\ref{eq:corrpole}) and (\ref{eq:pole}), we conclude
that the transition form factor is given  by
\begin{equation}
F_{\gamma^*\gamma^* \to \pi^o}(q_1^2,q_2^2) = \int C(\xi, \eta, q_1, q_2)
\langle 0| {\cal O}(\xi , \eta) | {\pi^0},\stackrel{\rightarrow}{p} \rangle \
d^4 \xi d^4 \eta \  .
\end{equation}
Thus,  the possibility
to factorize  short- and long-distance contributions
allows one to get an explicit expression for the form factor of
a bound state, the pion, without
explicitly obtaining the pion pole from
a  summation  or a bound-state formalism.

\begin{figure}[t]
\mbox{
   \epsfxsize=12cm
 \epsfysize=5cm
 \hspace{0.5cm}
 \epsffile{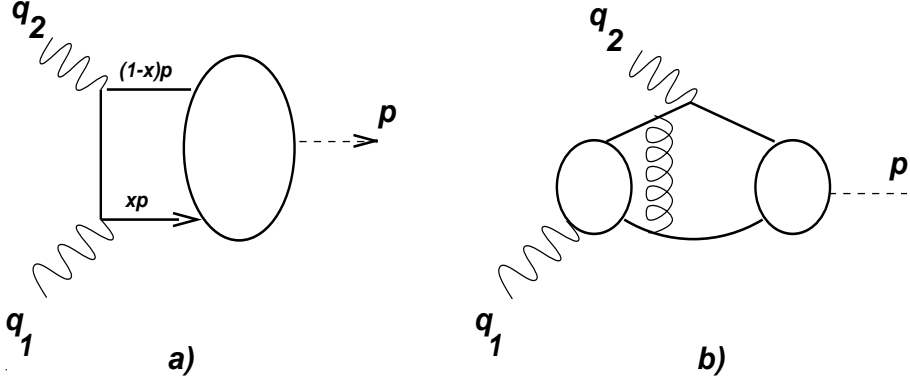}  }
  \vspace{0.5cm}
{\caption{\label{second}
$a)$ Lowest-order pQCD term. $b)$ Hard gluon exchange diagram.
 }}
\end{figure}

Now, introducing the distribution
amplitude $\varphi_{\pi}(x)$
\begin{equation}
\langle 0| {\cal O}(\xi , \eta) | {\pi^0},\stackrel{\rightarrow}{p} \rangle \
= \int_0^1  e^{i x (\xi p) + i \bar x (\eta p)} \varphi_{\pi} (x) dx  \  ,
\end{equation}
where $\bar x \equiv 1-x$, we obtain the hard scattering formula
\begin{equation}
  F_{\gamma^*\gamma^* \to \pi^o}(q_1^2,q_2^2) =
  \int_0^1 T(q_1,q_2; xp, \bar x p) \,
\varphi_{\pi}(x) \, dx  \  ,
\end{equation}
with $T(q_1,q_2; xp, \bar x p)$ being the amplitude for the subprocess
$\gamma^* \gamma^* \to \bar q q $.

\subsection{Perturbative QCD predictions}

Calculating the subprocess amplitude in the lowest order one  gets
the result \cite{bl80}
\begin{equation}
F^{LO}(q^2,Q^2) =  \frac{4 \pi}{3}
\int_0^1 \frac{\varphi_{\pi}(x)}{x Q^2+\bar x q^2} \, dx,
\end{equation}
which was already  discussed in the Introduction.
For definiteness, from now on we will adhere to the
convention that $q^2$ is the smaller of the two virtualities:
$q^2 \leq Q^2$.
The leading-order  pQCD formula  has a smooth limit when $q^2 \to 0$,
predicting that the asymptotic behaviour of the
$\gamma \gamma^* \to \pi^0$ form factor is \cite{bl80}:
\begin{equation}
F_{\gamma \gamma^*  \pi^0 }^{LO}(Q^2) = \frac{4\pi}{3}
\int_0^1 {{\varphi_{\pi}(x)}\over{xQ^2}} dx  +O(1/Q^4).
\label{eq:ggpipqcd}
\end{equation}
 The $x^n$-moments of the pion distribution amplitude  $\varphi_{\pi}(x)$  are
given by  the matrix elements of the twist-2 composite operators
$\bar q \gamma_5 \{ \gamma^{\nu}
D^{\nu_1} \ldots D^{\nu_n} \}q$. Furthermore, the  contributions
to $F_{\gamma \gamma^*  \pi^0 }^{pQCD}(Q^2)$  from the
higher twist two-body
operators $\bar q \gamma_5 \{
D^{\nu_1} \ldots D^{\nu_n} \} q$ and
 $\bar q \gamma_5  \sigma^{\mu \nu}
D^{\nu_1} \ldots D^{\nu_n} q$, which are potentially
dangerous due to large magnitude
$O(m_{\pi}^2/(m_u+m_d))$ of their matrix elements \cite{pl80,ynd}
 are proportional to the light quark masses,
and their net input is suppressed by
  $O(m_{\pi}^2/Q^2)$ factor for large $Q^2$.
Hence, one should not expect  unusually
large $(1/Q^2)^N$ corrections  to the leading twist term (\ref{eq:ggpipqcd}).

In  the first order in $\alpha_s(Q^2)$,
one can imagine a subprocess in which
the real photon first dissociates  into a quark-antiquark pair
(the relevant amplitude can be called the
photon distribution amplitude $\varphi_{\gamma}(y)$).
The $\bar q q$-pair
then interacts with the virtual photon
to produce the final state quark-antiquark pair
converting eventually  into the pion.
This contribution  is analogous to the pQCD hard scattering
term for the pion form factor.
However, in our case, such a subprocess  is formally
realized only at the next-to-leading twist level:
\begin{equation}
F_{\gamma \gamma^*  \pi^0 }^{NLO}(Q^2) \sim \alpha_s(Q^2)
\int_0^1 { { {\varphi_{\gamma}(y)} {\varphi^P_{\pi}(x)} }
\over{ y^2 x Q^4} } dx  +O(1/Q^6).
\label{eq:ggpipqcd2}
\end{equation}
A more careful inspection shows, however,
that if  $\varphi_{\gamma}(y) \sim y$
for small $y$, the integral over $y$ diverges.
This means that this contribution is not factorizable
in the usual sense and should be considered
within the QCD sum rule context (see discussion below).
We will see that  this term is rather small,
mainly because it is suppressed both by  $O(\alpha_s/\pi)$
and  $O(1/Q^2)$ factors compared
to the leading-order  term (\ref{eq:ggpipqcd}).

One-loop perturbative QCD corrections to eq.(\ref{eq:ggpipqcd})
are known \cite{AuCh81}-\cite{kmr} and they are
under control. For these reasons, the process $\gamma \gamma^* \to \pi^0$
provides  the cleanest test of
pQCD   for exclusive processes.

\subsection{Wave-function dependence
and Brodsky-Lepage interpolation formula}

To fix the absolute magnitude of the leading-order
pQCD prediction (\ref{eq:ggpipqcd}),
one should know the value of  the integral
\begin{equation}
I = \frac{1}{f_{\pi}}\int_0^1 {{\varphi_{\pi}(x)}\over{x}} dx \
\label{eq:I}  ,
\end{equation}
which  depends on the shape of the pion distribution
amplitude $\varphi_{\pi}(x)$.
In particular,  using  the
asymptotic form\cite{tmf78}-\cite{pl80}
\begin{equation}
\varphi_{\pi}^{as}(x) = 6 f_{\pi} x(1-x) \,  ,
\label{eq:phias}
\end{equation}
one obtains $I^{as}=3$ resulting in  the
absolute  prediction for the asymptotic
behaviour of the $F_{\gamma \gamma^*  \pi^0 }(Q^2)$
form factor \cite{bl80}
\begin{equation}
F_{\gamma \gamma^*  \pi^0 }^{as}(Q^2) =
\frac{4\pi  f_{\pi}}{Q^2} + O(1/Q^4).
\label{eq:fggpias}
\end{equation}
On the other hand, if one uses the
CZ-amplitude, the value of the
integral $I$ increases:  $I^{CZ}=5$,
and   experimental
data can, in principle,
  discriminate between these two possibilities.
Of course,  the asymptotic $1/Q^2$-dependence cannot
be a true behaviour
of $F_{\gamma \gamma^*  \pi^0 }(Q^2)$  for all $Q^2$-values:
it should be somehow modified in the  low-$Q^2$
region to comply with the  bound on
$F_{\gamma \gamma^*  \pi^0 }(Q^2)$  in the $Q^2=0$ limit
imposed by the anomaly relation (\ref{eq:norma}).
Brodsky and Lepage \cite{blin}
 proposed the interpolation formula
\begin{equation}
F_{\gamma \gamma^*  \pi^0 }^{BL(int)}(Q^2) =
{ {1} \over {\pi f_{\pi}
\left (1+{{Q^2}\over{4 \pi^2 f_{\pi}^2}} \right )}}
\label{eq:blin}
\end{equation}
which reproduces both the $Q^2 =0 $ value (\ref{eq:norma})
and the high-$Q^2$ asymptotics (\ref{eq:fggpias}) dictated by
the asymptotic form of the distribution amplitude (\ref{eq:phias}).
According to refs.\cite{CELLO,CLEO},
this formula   agrees   with  experimental
data. It also agrees with the results  obtained in several quark
model calculations  \cite{hiroshi}-\cite{frank}.
On the other hand, the curve based on the formula
\begin{equation}
F_{\gamma \gamma^*  \pi^0 }^{CZ(int)}(Q^2) =
{ {1} \over {\pi f_{\pi}
\left (1+{{3Q^2}\over{20 \pi^2 f_{\pi}^2}} \right )}},
\label{eq:czblin}
\end{equation}
interpolating between
the $Q^2=0$ value and the CZ-normalized high-$Q^2$ prediction,
is far from existing data points and quark model results.

\subsection{Pion wave function and QCD sum rules}

From the theoretical side, there are also doubts \cite{MR,BrFil} that
QCD sum rules really require that  the pion distribution function
has the  CZ shape.
In particular, the QCD sum rule calculation of $\varphi_\pi(x)$
at the middle point $x=1/2$ performed in ref. \cite{BrFil}
produced the value
$\varphi_\pi(1/2) \approx 1.2 \, f_{\pi}$,
to be compared with $\varphi_\pi^{as}(1/2) = 1.5 \, f_{\pi}$
and  $\varphi_\pi^{CZ}(1/2) =0$.
In ref. \cite{MR}, it was pointed out that keeping in  the
OPE the lowest condensates only does not provide information necessary
for a reliable determination of the pion distribution amplitude.
This is especially clear if one writes   the CZ sum rule
directly for $\varphi_\pi(x)$
\begin{eqnarray}
f_\pi\varphi_\pi(x)&=&\frac{3M^2}{2\pi^2}(1-e^{-s_0/M^2})x(1-x)
  +\frac{\alpha_s\langle GG\rangle}{24\pi M^2}[\delta(x)+\delta(1-x)]
\nonumber \\
		& & \hspace{10mm} +\frac{8}{81}\frac{\pi\alpha_s\langle\bar
		   qq\rangle^2}{M^4}
\{11[\delta(x)+\delta(1-x)]+2[\delta^\prime(x)+\delta^\prime(1-x)]\}.
\label{eq:wfsr}
\end{eqnarray}
As emphasized in \cite{MR}, it is the $\delta$-function terms
here which are crucial in generating  a humpy form for  $\varphi_\pi^{CZ}(x)$.
Adding higher condensates, $e.g.,$ $\langle \bar q D^2 q\rangle$,
one would get even higher derivatives of
$\delta(x)$ and $\delta(1-x)$.
All the subseries of such singular terms
can be treated as  an expansion
of some finite-width functions related to nonlocal condensates.
The sum rule  based on a model  for the nonlocal condensates
consistent with the earlier   estimates
for  $\langle \bar q D^2 q\rangle$ \cite{belioffe}
reduces  the values of the lowest nontrivial moments of
  $\varphi_\pi(x)$ bringing them very close to
those for the asymptotic form
(\ref{eq:phias}).  As a result, the model distribution amplitude
constructed in \cite{MR}  gives $I \sim 3$.
One can also try  to  construct a QCD sum rule directly
for the integral $I$.
However, it is clear that  such a sum rule cannot be derived
by a simple substitution of the original CZ sum rule into the integral
(\ref{eq:I}),
because of singularities generated by the $\delta(x)/x$ and
 $\delta^\prime(x)/x$  terms.
The singularities disappear if one uses  the nonlocal condensates
in the sum rule for $\varphi_\pi(x)$,
and the resulting value for $I$ is close to 3 \cite{ggmikh}.
A more radical way is to consider the sum rule for the original
$\gamma \gamma^* \to \pi^0$ amplitude as a whole,
 without approximating it by the
first term of the pQCD expansion. As we will see,  the  singularities
mentioned  above will appear then as the
 power $(1/q^2)^n$  infrared singularities
in the relevant operator product expansion  (OPE).
However, these singularities are caused by  a formal
extension  of the OPE formulas from the  large $q^2$ region
where they are valid,  into the small-$q^2$ region,
where the OPE  breaks down.
In fact,
the OPE should be modified in the small-$q^2$ region.
Such a modification is equivalent to the regularization of the
infrared singularities.
Our goal in the present paper is to construct  a QCD sum rule
for the transition form factor valid in the region of small $q^2$.

\subsection{Unfactorizable contributions and QCD sum rules }

If the photon virtualities are not very large,
then it is normally impossible to factorize
the $p^2$-dependence of the diagrams contributing to the
three-point amplitude
 $F\left(p^2,q_1^2,q_2^2\right)$ into  separate factors
one of which will produce the pion pole.
We know, of course,
that the full amplitude $F\left(p^2,q_1^2,q_2^2\right)$
must have the pion pole, but it is unclear how much a particular
finite-order diagram contributes to such a pole.

To display the pole structure of $F\left(p^2,q_1^2,q_2^2\right)$,
it is convenient to use the dispersion relation
for the three-point  amplitude:
 \begin{equation}
F\left(p^2,q_1^2,q_2^2\right)={1\over{\pi}}\int_0^{\infty}
\frac{{\rho}\left(s,q_1^2,q_2^2\right)}{s-p^2}\,ds
+ ``subtractions" .
\label{eq:disp1}
\end{equation}

The  pion  contribution to the spectral density is proportional to the
$F_{\gamma^*\gamma^* \to \pi^o}$ \ form factor:
\begin{equation}
{\rho}\left(s,q_1^2,q_2^2\right)=
\pi f_{\pi}\delta(s-m_\pi^2)
F_{\gamma^*\gamma^* \to \pi^o}\left(q_1^2,q_2^2\right)
+  ``higher \ states",
\end{equation}
but any approximation for
$F\left(p^2,q_1^2,q_2^2\right)$  obtained from a PT
expansion, gives information not only about the pion, but also about
the higher states. Hence, there are two  problems. First,
we should arrange a reliable PT expansion  for
$F\left(p^2,q_1^2,q_2^2\right)$. The only way is to take
$p^2$ spacelike and large enough to produce a reasonably converging OPE.
 The second problem is that when  $p^2$ is large,
the pion contribution does not overwhelmingly
dominate  the dispersion integral and, to
disentangle it, one should
subtract the contribution due to  higher states.
Note, that though  the contribution of a  state located
at $s= m^2$ is suppressed  by the  $-p^2/(m^2-p^2)$-factor
compared to that of a state at $s= 0$,
the  relative suppression of higher states
 disappears as $-p^2$ tends to infinity.

The higher states  include $A_1$ and higher  pseudovector resonances
which presumably become broader and broader, with their
sum rapidly approaching the   pQCD  spectral density.
So, the simplest model is to approximate all the higher states,
including the $A_1$, by the spectral density
${\rho}^{PT}(s,q_1^2,q_2^2)$ calculated in
perturbation theory:
\begin{equation}
{\rho}^{mod}\left(s,q_1^2,q_2^2\right) =
\pi f_{\pi}\delta(s)F_{\gamma^*\gamma^* \to \pi^o}\left(q_1^2,q_2^2\right)
+\theta(s-s_o){\rho}^{PT}(s,q_1^2,q_2^2),
\label{eq:rhoph}
\end{equation}
where the parameter $s_o$ is the effective threshold for the higher states.

The smaller $p^2$, the
bigger relative contribution of the lowest state. Hence,
the strategy is to take the smallest possible $p^2$
within the region where the $1/p^2$ expansion is still legitimate.
In fact, it is more convenient to use  the faster decreasing
exponential weight $\exp[-s/M^2]$ instead of $1/(s-p^2)$.
This is achieved by applying to $F(q_1^2,q_2^2,p^2)$
the SVZ-Borel transformation \cite{svz}:
\begin{eqnarray}
\hat B(p^2\rightarrow M^2)F(q_1^2,q_2^2,p^2)
\equiv \Phi (q_1^2,q_2^2,M^2)
=\frac{1}{\pi M^2} \int_0^\infty  e^{-s/M^2} {\rho}(s,q_1^2,q_2^2) \, ds .
\label{eq:fph}
\end{eqnarray}
Another merit of the
SVZ-Borel transformation is that using it one
gets a  factorially improved OPE power series:
$1/(-p^2)^N \to (1/M^2)^N / (N-1)!$.

In a sense, the QCD sum rules can be treated as
a method of extracting  information about the lowest state
from the behaviour of    $F(q_1^2,q_2^2,p^2)$
in  the  large-$p^2$ region.
To construct   a QCD sum rule, one  should  calculate
the SVZ transform   $\Phi (q_1^2,q_2^2,M^2)$
as a  power  expansion in $1/M^2$  for large $M^2$.
To this end, one should calculate first the
three-point function $T(p^2,q^2,Q^2)$
as a  power  expansion in $1/p^2$  for large $p^2$.
However,  a particular   form of the
$1/p^2$ expansion depends
on the  interrelation between the values of the photon virtualities
$q^2 $
and  $Q^2$.

The simplest    case  is when both
virtualities are sufficiently large and similar in  magnitude:
$Q^2 \sim q^2 \sim - p^2 > \mu^2$,  where $\mu$ is a typical
hadronic scale  $\mu^2 \sim 1 \, GeV^2$.
This case  will be referred to as  the ``large-$q^2$'' kinematics.
 Then all
 power-behaved $(1/M^2)^n$ contributions
correspond to the situation
when all the  currents $J_{\mu}(X)$, $J_{\nu}(0)$ and $j_{\alpha}^5(Y)$
are close to each other, $i.e.,$   all the
  intervals $X^2, Y^2,(X-Y)^2$\, are small.

A more complicated case is when  $q^2$ is small  $q^2 \ll  \mu ^2$,
while $Q^2$ is still large:  $Q^2 > \mu ^2 $ .
In this case,  to be referred to as ``small-$q^2$'' kinematics,
one should also take into account
the configuration
when  the electromagnetic current $J_{\mu}(X)$ related
to  the $q$-photon is  far away from  two other currents, $i.e.,$
when there is a possibility of
long-distance propagation in the $q$-channel.
In the limit $q^2 \to 0$, such a propagation is
responsible for the mass singularities ($ 1/q^2 , \, \ln(q^2)$, $etc.$)
in the Feynman diagrams contributing to the
three-point amplitude.

\section{QCD sum rules for the
$F_{\gamma^* \gamma^* \pi^0}(q^2,Q^2)$
form factor in large-$q^2$ kinematics}

\setcounter{equation} 0

Let us consider first  the simpler case, when both
$Q^2$ and  $q^2$  are large.  In this situation,
it is sometimes convenient  to introduce  another set of variables:
average  virtuality
$\widetilde Q^2 = (Q^2 + q^2)/2$ and the
 asymmetry parameter $\omega = (Q^2 - q^2)/(Q^2 + q^2)$.

\subsection{Lowest-order perturbative term}

The starting point   of the operator expansion is  the
perturbative triangle graph (Fig.\ref{eq:fig1}$a$).
Its contribution contains  almost all types of the
tensor structures  present in
eq. (\ref{eq:tensor}). Extracting the $F$ part and using the Feynman
parameterization we get:
\begin{equation}
F^{PT}(q^2,Q^2,p^2)=\frac{2}{\pi}\int_0^1 \
\frac{x_1x_2}{[q^2x_1x_3+Q^2x_2x_3-p^2x_1x_2]} \
\delta(1-\sum_{i=1}^3 x_i)\  dx_1dx_2dx_3 \ \ .
\label{eq:F1}
\end{equation}
In this representation, it  is straightforward  to apply
the SVZ-Borel  transformation to get
\begin{equation}
\Phi^{PT}(q^2,Q^2,M^2)=\frac{2}{\pi M^2}
\int_0^1
 \exp \left \{- \frac{q^2x_1x_3+Q^2x_2x_3}{x_1x_2M^2} \right \} \
\delta(1-\sum_{i=1}^3 x_i)\
dx_1dx_2dx_3\
\label{eq:Phipt}.
\end{equation}
Comparing this representation  with the definition of the
SVZ transform  (\ref{eq:fph}), we can immediately write down the  formula for the
  perturbative spectral density ${\rho}^{PT}(s,q^2,Q^2)$:
\begin{equation}
\rho^{PT}(s,q^2,Q^2)=2 \int_0^1
\delta \left ( s - \frac{q^2x_1x_3+Q^2x_2x_3}{x_1x_2} \right ) \
\delta(1-\sum_{i=1}^3 x_i)\
dx_1dx_2dx_3 .
\label{eq:rpts}
\end{equation}
Scaling  the integration variables:
$x_1+x_2 =  y$, $x_2 =  xy$, $x_1 =(1-x)y \equiv \bar x y$
and taking trivial integrals over $x_3$ and $y$, we get
\begin{equation}
\rho^{PT}(s,q^2,Q^2)=2\int_0^1 \frac{x\bar{x}(xQ^2+ \bar x q^2)^2}
{[s{x}\bar{x}+xQ^2+ \bar x q^2]^3} \,dx  \, ,
\label{eq:rhopt}
\end{equation}
or, in terms of $\widetilde Q^2$   and
  $\omega $:
\begin{equation}
\rho^{PT}(s,q^2,Q^2)=2 \int_0^1
\frac{x\bar{x} \widetilde Q^4(1+\omega(x-\bar{x}))^2}
{[s{x}\bar{x}+\widetilde Q^2(1+\omega(x-\bar{x}))]^3} \, dx.
\label{eq:rhoptw}
\end{equation}
It should be noted  that the variable $x$ in the integrals above
is the light-cone fraction of the total pion
momentum $p$ carried by one of the quarks.

A very simple result for $\rho^{PT}(s,q^2,Q^2)$  appears
when $q^2=0$:
\begin{equation}
\rho^{PT}(s,q^2=0,\, Q^2) = {{Q^2}\over{(s+Q^2)^2}}.
\label{eq:rhoq20}
\end{equation}
As $Q^2$ tends to  zero, the spectral density $\rho^{PT}(s,q^2=0,Q^2)$
becomes narrower and higher converting into   $\delta(s)$ in the
$Q^2 \to 0$ limit \cite{dz}:
\begin{equation}
\rho^{PT}(s,q^2=0,\, Q^2=0) = \delta(s).
\end{equation}
  Thus,  the perturbative triangle diagram
``tells us''  that, for massless quarks and $Q^2=0$, two photons can produce only
a single  massless pseudoscalar state, and
there are no other states in  the spectrum  of the final hadrons.
As $Q^2$ increases, the spectral function broadens,
$i.e.,$ higher states can also be produced.

\subsection{Gluon condensate corrections}

When the Borel parameter $M^2$ (or the probing virtuality  $-p^2$)
decreases, both perturbative (logarithmic or ${\cal O}(\alpha_s$) ) and
nonperturbative  (power or ${\cal O}(1/p^2)$) corrections come into play.
As  argued by SVZ  \cite{svz}, the power
corrections  proportional to quark and gluon condensates:
 ${\langle 0|\bar{\psi}\psi|0\rangle}^2$,
$\langle 0|G^a_{\mu\nu}G^a_{\mu\nu}|0 \rangle$,
$etc.$, are  much more important than
the higher order perturbative corrections.
In many cases, the latter can be  safely  neglected.

\begin{figure}[thb]
\mbox{
   \epsfxsize=16cm
 \epsfysize=8cm
 \epsffile{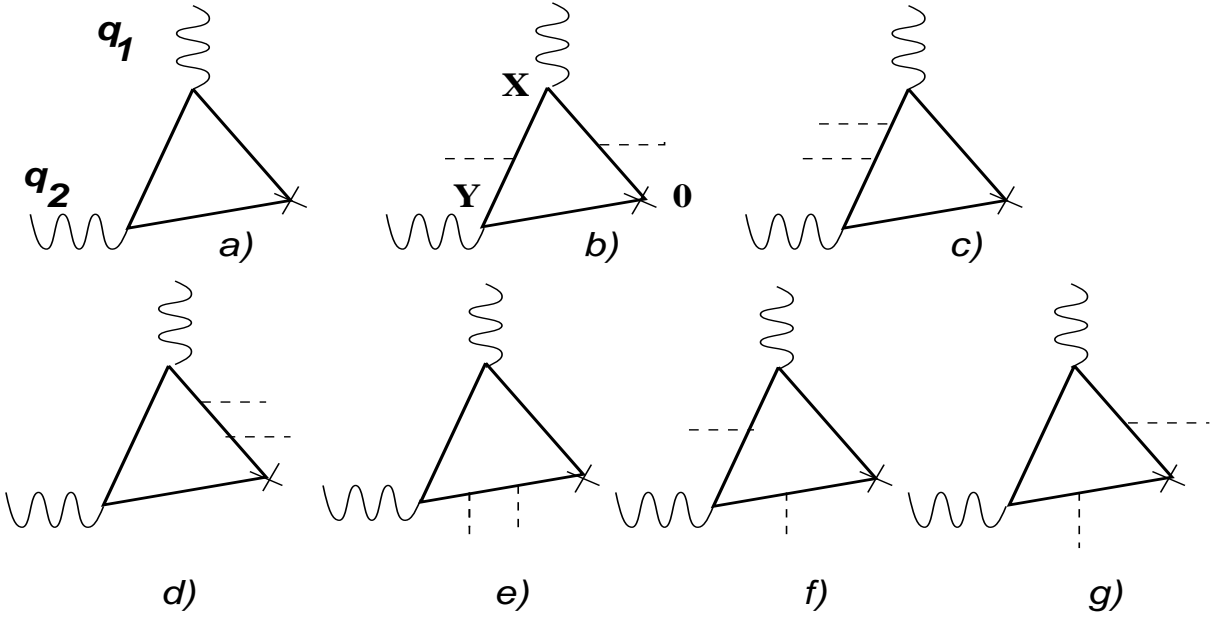}  }
  \vspace{1cm}
{\caption{\label{eq:fig1}
$a)$ Lowest-order perturbative contribution. $b)-g)$
Gluon condensate corrections.
 }}
\end{figure}

The lowest-order diagrams proportional to
 the gluon condensate
are shown in Fig.\ref{eq:fig1}$b$--$g$.  They
take into account the fact that propagating through
the QCD vacuum, quarks interact  with the  nonperturbative
gluon fluctuations which can be treated as a background field.
The most straightforward way to calculate these diagrams
 is to take quark propagators in the coordinate representation
using the Fock-Schwinger gauge $z_\mu A_\mu(z)=0$
for the background field $A_\mu(z)$:
\begin{eqnarray}
\lefteqn{
\hat S(X,Y)=\frac{\hat \Delta}{2\pi^2\Delta^4} -
\frac{1}{8\pi^2}\frac{\Delta_{\alpha}}{\Delta^2}
{\widetilde G}_{\alpha\beta}(0)\gamma_{\beta}\gamma_5 }
\nonumber\\ & &
+ \frac{i}{4\pi^2}\frac{\hat{\Delta}}{\Delta^4} Y_\rho X_\mu
G_{\rho\mu}(0) -
\frac{1}{192\pi^2}\frac{\hat{\Delta}}{\Delta^4}(X^2Y^2-(XY)^2)
G_{\beta\chi}(0)G_{\beta\chi}(0).
\label{eq:qprop}
\end{eqnarray}
Here $\Delta=X-Y$ and
${\widetilde G}_{\alpha\beta}=
\frac{1}{2}{\epsilon}_{{\alpha}{\beta}{\rho}{\sigma}}G_{\rho\sigma}$.
Using this expression for each quark propagator of the original
triangle diagram
and retaining the $O(GG)$ terms, one obtains all the  diagrams
of Fig.\ref{eq:fig1}$b$--$g$.
In particular, one can immediately see that contributions
 \ref{eq:fig1}$d$ and  \ref{eq:fig1}$e$ vanish due to our
choice of the coordinate origin at the axial current vertex
(such a choice, in which
the two photons are treated symmetrically, is more
convenient in this calculation than that
implied by our original  definition   (\ref{eq:corr})).
The remaining diagrams
 \ref{eq:fig1}$b,c,f,g$
are easily calculated in  the coordinate
representation. After performing the Fourier transformation
to the momentum space, we  extract the
tensor structure corresponding to the $F$ form factor
and  then apply the SVZ-Borel transformation  to the
relevant contribution $F^{\langle GG\rangle}(q^2,Q^2,p^2)$.
The final result  for the sum of the  $O(GG)$ diagrams reads:
\begin{eqnarray}
\Phi^{\langle GG\rangle}(q^2,Q^2,M^2) =
\frac{\pi^2}{9}
{\langle \frac{\alpha_s}{\pi}GG \rangle}
\left(\frac{1}{2M^4 Q^2} + \frac{1}{2M^4 q^2}
 - \frac1{M^2 Q^2 q^2}\right)   \nonumber \\
 =\frac{\pi^2}{9}
{\langle \frac{\alpha_s}{\pi}GG \rangle}
\left(\frac{1}{\widetilde Q^2M^4}-\frac{1}{\widetilde Q^4M^2}\right)
\frac{1}{1-\omega^2}. \  \  \
\label{eq:fgg}
\end{eqnarray}

\begin{figure}[t]
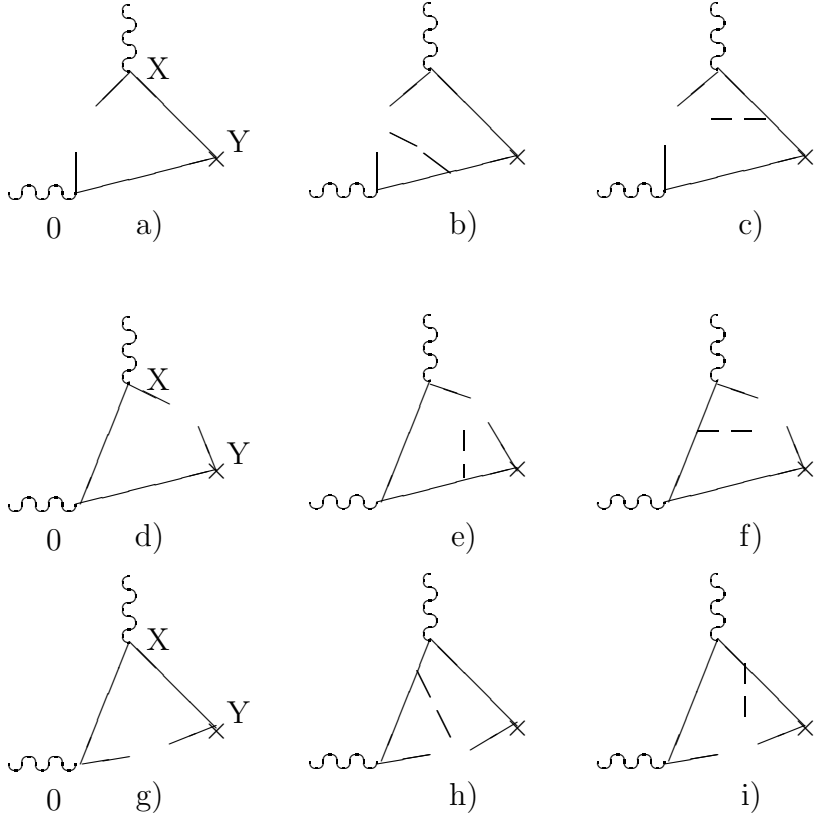

\input{./r61mod.pic}
\vspace{0.7cm}
\input{./r62mod.pic}
\input{./r63mod.pic}
\caption{Quark condensate corrections  involving ``soft'' gluons.}
\label{eq:fig3}
\end{figure}

\subsection{Quark condensate corrections}

For massless $u$-  and $d$-quarks, the  quark condensate contribution
starts with  terms proportional to
$\langle 0|\bar{q}\Gamma q \bar{q}\Gamma q|0\rangle$.
Using the usual  vacuum dominance hypothesis \cite{svz},
these can be reduced to ${\langle 0|\bar{q}q|0\rangle}^2$.
There are two types of diagrams producing
the ${\langle 0|\bar{q}q|0\rangle}^2$ contributions.

\begin{figure}[p]
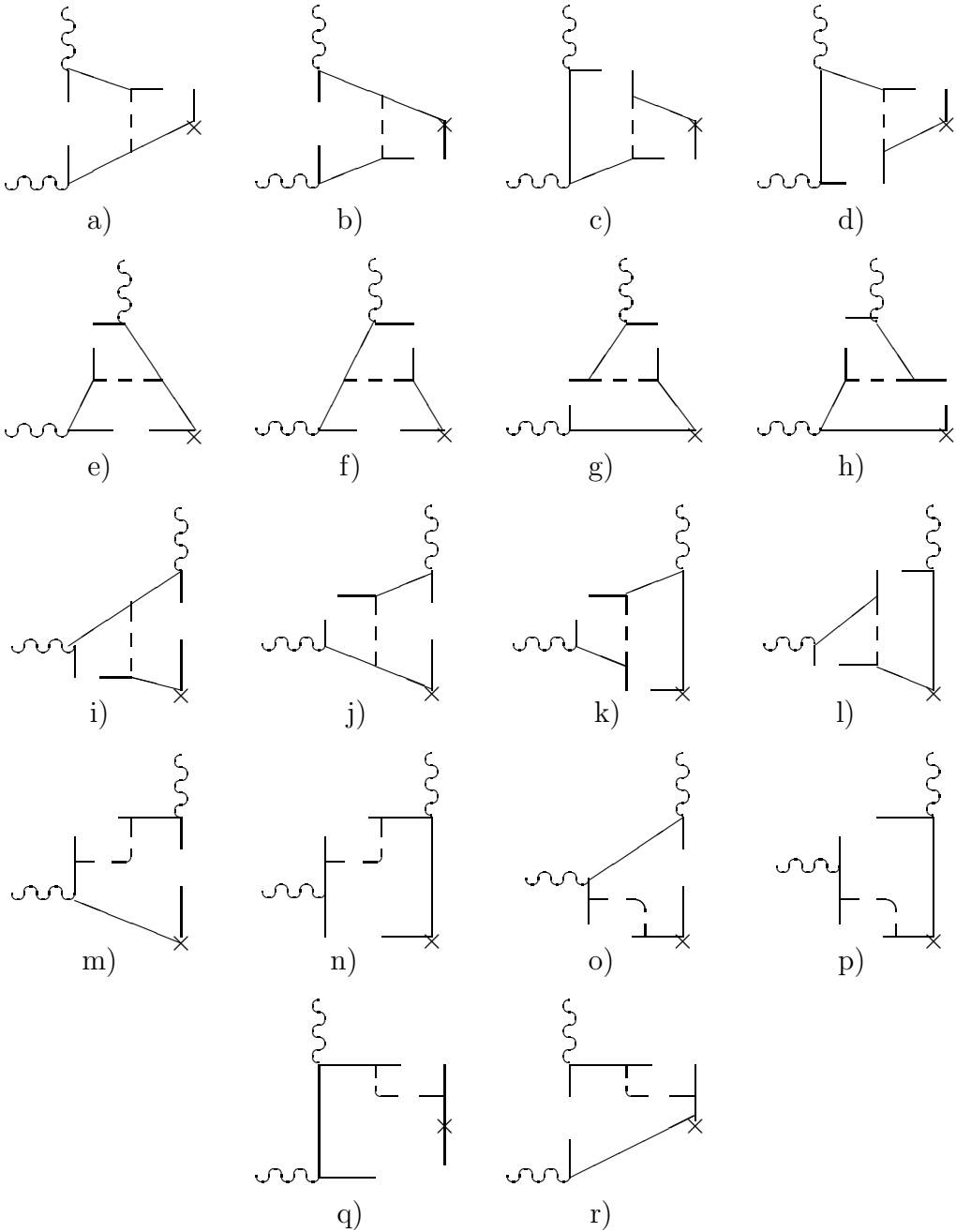

\input{./r71.pic}
\vspace{0.5cm}
\input{./r72.pic}
\vspace{0.5cm}
\input{./r73.pic}
\vspace{0.5cm}
\input{./r74.pic}
\vspace{0.5cm}
\input{./r75.pic}
\caption{Quark condensate corrections involving  ``hard'' gluons.}
\label{eq:fig4}
\end{figure}

First, there are  diagrams shown in Fig.\ref{eq:fig3}$a$--$i$\
(sometimes called the soft gluon
diagrams) corresponding to local operators of
$\bar q DDD q$, $\bar q GD q$ and $\bar q DG q$ type,
with the covariant derivatives $DDD$,
and the $GD$, $DG$ factors producing
an  extra $\bar q  q$ term by  equations of motion.

The  diagrams shown in Fig.\ref{eq:fig4}$a$--$r$  (and corresponding  to a
hard gluon exchange)   give
the ${\langle 0|\bar{q}q|0\rangle}^2$ structure directly.
However, only the diagrams \ref{eq:fig4}$a$--$d$ contribute
to the $F$ form factor.
 The total  $O({\langle \bar{q}q \rangle}^2)$
contribution to the  SVZ-borelized amplitude is
\begin{eqnarray}
\Phi^{<\bar{q}q>^2}=
\frac{64\pi^2 \alpha_s{\langle \bar{q}q \rangle}^2}{243 M^2}
\left( \frac1{M^4}
\left [ \frac{Q^2}{q^4}+ \frac9{2q^2}+\frac9{2Q^2}+\frac{q^2}{Q^4} \right ] +
\frac9{Q^2 q^4} +\frac9{Q^4 q^2} \right ) \nonumber \\
= \frac{64\pi^2 \alpha_s{\langle \bar{q}q \rangle}^2} {243 M^2}
\left(\frac{11-3\omega^2}{\widetilde Q^2M^4}+\frac{18}{\widetilde Q^6}\right)
\frac{1}{(1-\omega^2)^2}.
\label{eq:fqq}
\end{eqnarray}

\subsection{QCD sum rule}

Combining now eqs.(\ref{eq:rhoph}), (\ref{eq:fph}),
 (\ref{eq:rhopt}), (\ref{eq:fgg}) and (\ref{eq:fqq}), we obtain
the QCD sum rule for the $F$ form factor valid in the region
where both
virtualities of the photons are large:
\begin{eqnarray}
\pi f_{\pi} \mbox{$F_{\gamma^*\gamma^*\pi^\circ}$}(q^2,Q^2)=
 2\int_0^{s_o} ds \, e^{-s/{M^2}}
\int_0^1 \frac{x\bar{x}(xQ^2+ \bar x q^2)^2}
{[s{x}\bar{x}+xQ^2+ \bar x q^2]^3} \,dx  \,
\nonumber \\
+\frac{\pi^2}{9}
{\langle \frac{\alpha_s}{\pi}GG \rangle}
\left(\frac{1}{2M^2 Q^2} + \frac{1}{2M^2 q^2}
 - \frac1{Q^2 q^2}\right)
 \nonumber\\
+ \frac{64}{243}\pi^3\alpha_s{\langle \bar{q}q\rangle}^2
\left( \frac1{M^4}
\left [ \frac{Q^2}{q^4}+ \frac9{2q^2}+\frac9{2Q^2}+\frac{q^2}{Q^4} \right ] +
\frac9{Q^2 q^4} +\frac9{Q^4 q^2} \right )  ,
\label{eq:SRLarge}
\end{eqnarray}
or, in terms of  the variables $ \widetilde Q^2$ and $\omega$:
\begin{eqnarray}
F_{\gamma^*\gamma^* \to \pi^o}(q^2,Q^2)= \frac{1}{\pi f_{\pi}}
\left\{ 2\int_0^{s_o} ds \, e^{-s/{M^2}}  \int_0^1 dx \,
\frac{x\bar{x} \widetilde Q^4(1+\omega(x-\bar{x}))^2}
{[s{x}\bar{x}+\widetilde Q^2(1+\omega(x-\bar{x}))]^3}
\right.\label{eq:SR1}\\
\left. + \frac{\pi^2}{9}
\langle \frac{\alpha_s}{\pi}GG\rangle
\left(\frac{1}{\widetilde Q^2M^2}-
\frac{1}{\widetilde Q^4}\right)\frac{1}{1-\omega^2}
+\frac{64}{243}\pi^3\alpha_s{\langle \bar{q}q\rangle}^2
\left(\frac{11-3\omega^2}{\widetilde Q^2M^4}+\frac{18}{\widetilde Q^6}\right)
\frac{1}{(1-\omega^2)^2} \right\}
\nonumber
\end{eqnarray}
In particular, taking the exactly symmetric kinematics,
when $\omega = 0$ and  $q^2=Q^2 $,
we obtain  the sum rule
\begin{eqnarray}
F_{\gamma^*\gamma^* \to \pi^o}( Q^2, Q^2)= \frac{1}{\pi f_{\pi}}
\left\{ 2\int_0^{s_o} ds \, e^{-s/{M^2}}  \int_0^1 dx \,
\frac{x\bar{x} \ Q^4}
{[s{x}\bar{x}+ Q^2]^3} +
\right. \label{eq:SRsymm} \\
\left. +
\frac{\pi^2}{9}
\langle \frac{\alpha_s}{\pi}GG \rangle
\left(\frac{1}{ Q^2M^2}-\frac{1}{ Q^4}\right)
+ \frac{64}{243}\pi^3\alpha_s{\langle \bar{q}q \rangle}^2
\left(\frac{11}{ Q^2M^4}+\frac{18}{ Q^6}\right)
 \right\}
\nonumber
\end{eqnarray}
derived in ref.\cite{NeRa83}.

\subsection{$\widetilde Q^2 \to \infty $ limit
and transition to perturbative QCD}

In the  limit  $\widetilde Q^2 \to \infty$
with $\omega$  fixed, eq.(\ref{eq:SR1}) reproduces the sum rule
\begin{eqnarray}
F_{\gamma^*\gamma^* \to \pi^o}
( (1-\omega)\widetilde Q^2 ,(1+\omega)\widetilde Q^2  )
=\frac{1}{f_{\pi}\widetilde Q^2}
  \left \{ \frac{2M^2}{\pi}(1-e^{-s_0/M^2})
\int_0^1 \frac{x\bar{x}\,dx}{ (1+\omega(x-\bar{x}))}
 \right.   \nonumber \\
+ \left. \frac{\pi}{9M^2 (1-\omega^2)}
\langle \frac{\alpha_s}{\pi}GG\rangle
+  \frac{64\pi^2 }{243 M^4}
\frac{(11-3\omega^2)}{(1-\omega^2)^2}\alpha_s{\langle \bar{q}q\rangle}^2
 \right \}  \label{eq:SRlargeQ2}
\end{eqnarray}
obtained by Gorsky \cite{Gor}, who calculated
the leading $1/\widetilde Q^2$  contribution only.

To make a connection with the perturbative QCD approach, it is instructive
to rewrite eq.(\ref{eq:SRlargeQ2}) as
\begin{eqnarray}
F_{\gamma^*\gamma^* \to \pi^o}
( (1-\omega)\widetilde Q^2 ,(1+\omega)\widetilde Q^2  )
= \frac{4\pi}{3f_{\pi}\widetilde Q^2}
   \int_0^1 \frac{dx}{ (1+\omega(x-\bar{x}))} \,
\left \{ \frac{3M^2}{2\pi^2}(1-e^{-s_0/M^2}) x\bar{x} \right. \nonumber \\
+ \frac{1}{24M^2}
\langle \frac{\alpha_s}{\pi}GG\rangle [\delta(x) + \delta (\bar{x})] 
+ \left. \frac{8}{81M^4}\pi\alpha_s{\langle \bar{q}q\rangle}^2
 \biggl ( 11[\delta(x) + \delta (\bar{x})] +
2[\delta^{\prime}(x) + \delta ^{\prime}(\bar{x})]
\biggr ) \right \}  .\label{eq:SRlargeQ2wf} 
\end{eqnarray}

Note, that the expression in large curly brackets
coincides with the QCD sum rule (\ref{eq:wfsr}) for
$f_{\pi} \varphi_{\pi}(x)$.
Hence, in the large-$\widetilde Q^2$  limit,  the QCD sum rule
(\ref{eq:SRlargeQ2wf})
exactly reproduces  the perturbative QCD result
\begin{equation}
F_{\gamma^*\gamma^* \to \pi^o}
\left ( (1-\omega)\widetilde Q^2 ,(1+\omega)\widetilde Q^2  \right )
= \frac{4\pi}{3\widetilde Q^2}
   \int_0^1 \frac{\varphi_{\pi}(x)\, dx}{ (1+\omega(x-\bar{x}))}.
\label{eq:ffpert}
\end{equation}

\section{Operator product expansion in small-$q^2$ kinematics}

\setcounter{equation} 0

In the   nonsymmetric kinematics  $q^2 \ll Q^2 \sim 1{\mbox{ GeV}}^2$,
it is more convenient  to  use the ``old''
variables $q^2=-q_1^2$ and $Q^2=-q_2^2$
instead of $\omega$ and $\widetilde Q^2$.

\subsection{General features of the small-$q^2$ kinematics}

\setcounter{equation} 0

\begin{figure}[b]
\input{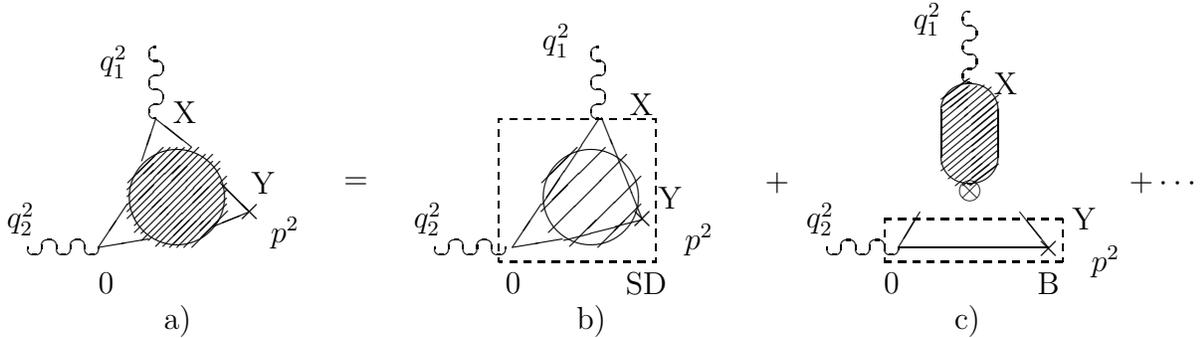}
\caption{Configurations responsible for the power-behaved contributions.}
\label{eq:fig5}
\end{figure}

Incorporating a  general analysis  of   the asymptotic
behaviour of  Feynman diagrams (see Appendix A and \cite{20,20a}),
one can show that, in the small-$q^2$ kinematics,
there are two different regions of
integration capable of producing a
contribution to the correlator (see Fig.\ref{eq:fig5}), which
behaves like an inverse  power of $p^2$.
 The first one is the  purely short-distance  region
corresponding to a situation
when all three currents are separated by short intervals,
 $i.e.,$ all the intervals $X^2,Y^2,(X-Y)^2$ are small.
The second region corresponds to another
short distance regime, in which   the vertex $X$
related  to the  small virtuality photon is
separated by a long  interval  from  two other currents
(this means that  $Y^2$ is small, but  $X^2$  and $(X-Y)^2$\ are large).
This is illustrated by  Fig.\ref{eq:fig5}, where Fig.\ref{eq:fig5}$a$
shows the full correlator ${\cal{F}}_{\alpha\mu\nu}(q_1,q_2)$,  whereas
Figs.\ref{eq:fig5}$b,c$   represent the two  possibilities
of getting the power-law  contributions.

In Fig.\ref{eq:fig5}$c$,  the long distance contribution (the dashed blob)
is given by a two-point (bilocal, cf. \cite{bilocal}) correlator $\Pi(q)$ of
the electromagnetic current
$J_{\mu}$ and some composite operator $\cal O$ of quark and gluon fields,
the latter represented  by  $\otimes$.
At low $q^2$, the correlator $\Pi(q)$ cannot be
calculated  in  perturbation theory.  The standard strategy  is to model
this nonperturbative object
by the ``first resonance plus continuum'' ansatz,
with the parameters of the spectrum determined
from a QCD sum rule for such a correlator.
In other words, one should  calculate $\Pi(q)$  at large $q^2$
using the OPE, then extract, in a standard way, the parameters
of the model spectrum in the $q$-channel and, finally, use
 the model spectral density   in a dispersion relation for
$\Pi(q)$ to obtain $\Pi(q)$  at small $q^2$.

\begin{figure}[t]
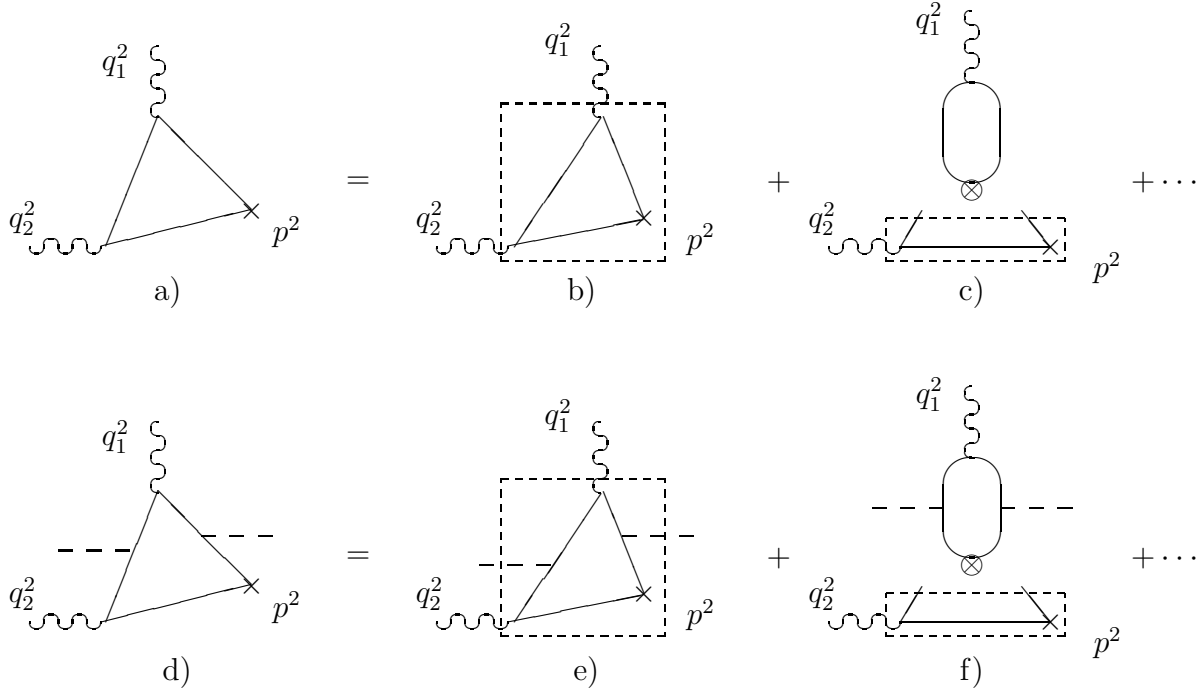

\input{./r21.pic}
\vspace{1.0cm}
\input{./r22.pic}
\caption{Separation of  $SD$- and $B$-contributions for
 perturbative diagrams and gluon condensate corrections.
 }
\label{eq:fig6}
\end{figure}

On the diagrammatic level, the factorization procedure
should be applied  for
each contributing diagram: the lowest-order triangle
graph (Fig.\ref{eq:fig6}$a,b,c$),
gluon-condensate graphs (Fig. \ref{eq:fig6}$d,e,f$), $etc.$
In this way, each diagram is represented as the sum of its
purely  short-distance ($SD$) and bilocal ($B$) parts,
the latter  obtained   through the relevant
contribution  into the operator product expansion  for
$J_{\nu}(Y)\,j_{\alpha}^5(0)$.
In fact, it is more convenient to define the $SD$-part of a diagram
as the difference between the original unfactorized expression
and its $B$-part determined $via$ the OPE.

\newpage
The total   SD-contribution  is given
by a sum of the SD-parts of all relevant diagrams:
\begin{figure}[h]
\input{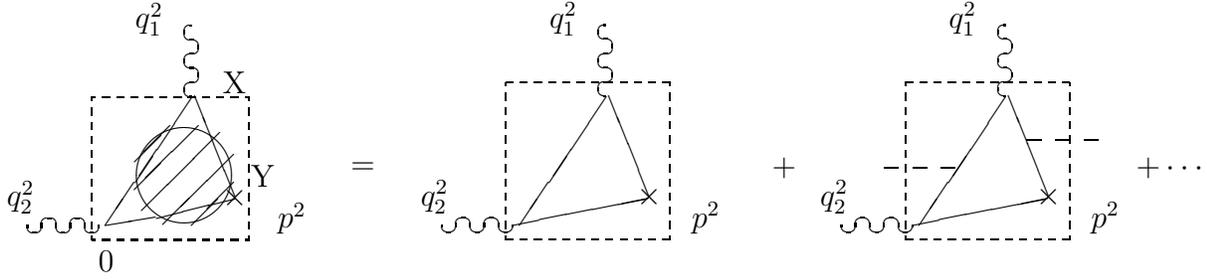}
\caption{$SD$-contribution.}
\label{eq:fig7}
\end{figure}

\begin{figure}[h]
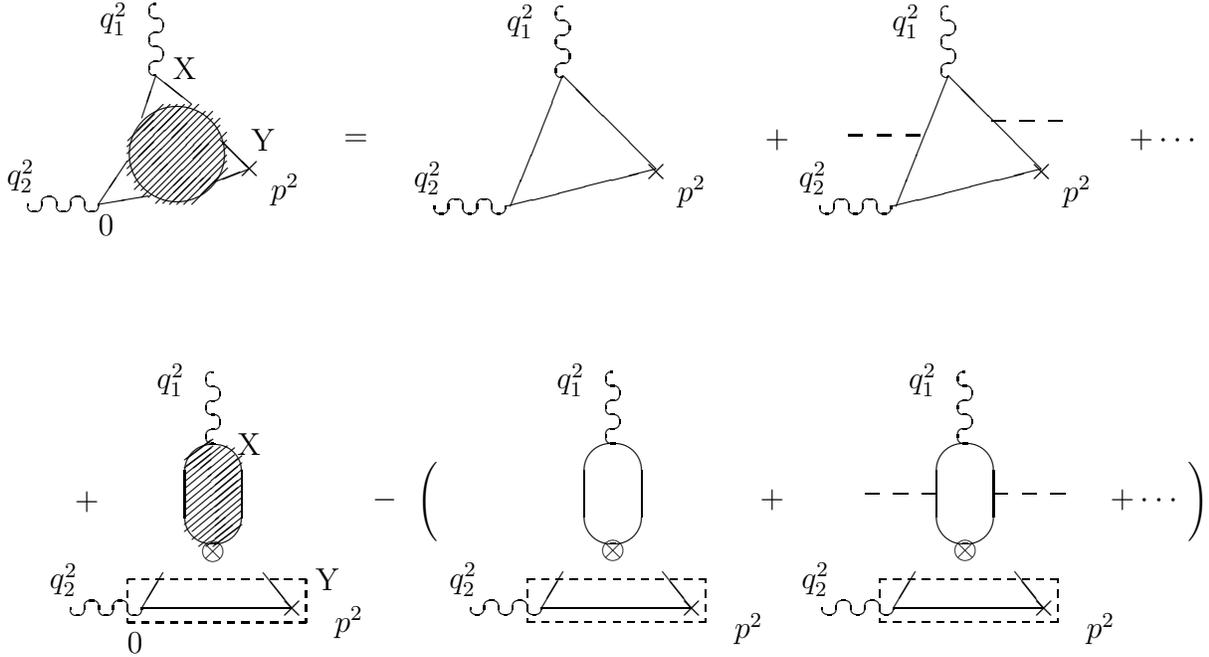

\input{./r41mod.pic}
\vspace{1.5cm}
\input{./r42mod.pic}
\caption{Structure of factorization in the small-$q^2$ kinematics.}
\label{eq:fig8}
\end{figure}

\noindent Now, substituting the $SD$-terms by the ``original minus $B$-terms'',
we  obtain  the  expansion   shown on  Fig.\ref{eq:fig8}.
The dots there stand for the rest of the condensate diagrams and higher-order
corrections.
The first row in Fig.\ref{eq:fig8}
corresponds to the usual OPE for the correlator (\ref{eq:corr}) in the large-$q^2$
kinematics $(\,q^2\sim Q^2\sim -p^2 \sim \mu^2)$. The second
row corresponds to additional terms which should be  taken into
account in  the case of small-$q^2$ kinematics
 $(\,q^2\ll Q^2\sim-p^2\sim\mu^2\,)$.
Note, that at large values of $q^2$, these
additional terms  should  die out to convert   the modified OPE
specific to the small-$q^2$ kinematics
into the standard OPE for the large-$q^2$  kinematics
(cf. \cite{NeRa84b,BeiNeRa88}).
Another statement is that, at low $q^2$, the ``subtraction'' terms
(those in the parentheses  in the second row of Fig.\ref{eq:fig8})
have  exactly the same singularities as the corresponding terms in  the
first row of Fig.\ref{eq:fig8} and, therefore,
the total  expression is regular in  this
kinematic limit.

\subsection{Logarithmic singularities }

According to eqs.(\ref{eq:SRLarge}),
the condensate terms contain $1/q^2$, $1/q^4,$ $etc.$ singularities in
the \mbox{$q^2\to 0$} limit. On the other hand, the perturbative expression,
though finite as  $q^2\to 0$,
contains contributions which
are  non-analytic at this point.
To  study the structure of the non-analytic terms,
it is convenient to use the following method.
First, let us introduce another set of integration variables
in  the expression for the Borel transform $\Phi^{PT}(q^2,Q^2,M^2)$
(\ref{eq:Phipt}):
$$x_1+x_3=\lambda ,  \  \ x_3=y \lambda, \ \ x_1=(1-y)
\lambda \equiv\bar{y}\lambda, \ \ x_2=1-\lambda.$$
This gives
\begin{equation}
\Phi^{PT}(q^2,Q^2,M^2)  =
\frac{2}{\pi M^2} \int_0^1 \lambda  d\lambda \, dy
e^{{-q^2 y \lambda}/{M^2\bar{\lambda}}}
e^{ {-y Q^2 }/{\bar{y} M^2} }.
\label{eq:Phipt2}
\end{equation}
Performing a  formal $q^2$-expansion
of the exponential in the integrand,
we will get  divergent integrals for all
the coefficients of the $(q^2)^n$ expansion,
except for the lowest term.
To get a more sensible result, we
use a continuous version of the series expansion
for the exponential, $i.e.,$ the
Mellin representation:
\begin{equation}
e^{-A}=\frac{1}{2\pi i} \int_{- \delta -i \infty}^{- \delta +
i \infty}\,A^J\,\Gamma(-J)\,dJ \  .\nonumber
\end{equation}

Now, the integral over $\lambda$ can be taken easily,
and the next step is to calculate the $J$-integral by taking
residues at $J=0,1, \ldots$. This gives:
\begin{eqnarray}
\Phi^{PT}(q^2,Q^2,M^2)&=&\frac{1}{\pi M^2}\int_0^1 dy
e^{-{Q^2y}/{M^2\bar{y}}}
\left\{\ 1+ \frac{q^2y}{M^2}e^{{q^2y}/{M^2}} +
 \left[2\frac{q^2y}{M^2}+
\frac{q^4y^2}{M^4}\right] e^{{q^2y}/{M^2}}
\ln{\left ( \frac{q^2y}{M^2}\right
)}\right.
\nonumber 
\\
& &\qquad \qquad \left.
-  \sum_{n=1}^{\infty}
{\left(\frac{q^2y}{M^2}\right)}^n
\frac{\psi(n)(n+1)}{(n-1)!}\right\}
\label{eq:ptq0} 
\end{eqnarray}
where $ \psi(z)\equiv{\Gamma^\prime(z)}/{\Gamma(z)}$ and
$\Gamma(z)$  is the Euler Gamma function.

The non-analytic terms $ q^2\ln{q^2}$ and  $ q^4\ln{q^2}$
are  typical examples of mass singularities \cite{23}.
They appear
when a long-distance propagation
of particles is possible. In our case, the
mass singularities  are related to the
possibility to create, for massless quarks, a $q\bar{q}$-pair by a
zero-virtuality photon.
To apply perturbative QCD methods, one  should  first factorize the
 contributions due to  short and long distance dynamics,
as outlined  above (see Figs.\ref{eq:fig6},\ref{eq:fig7}).

\subsection{Scalar model}

\begin{figure}[htb]
\input{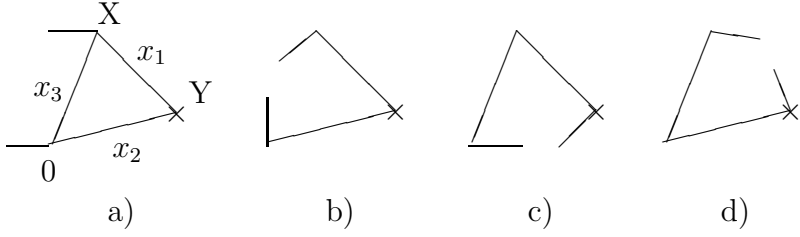}
\caption{Lowest terms of the OPE in a scalar toy model.}
\label{eq:fig9}
\end{figure}

To  clarify  the origin of the singularities and  outline
the method of their
factorization, let us consider first the analogue of
our form factor in a simple scalar theory
$g\phi^3_{(4)}$:
\begin{equation}
{\cal{F}}(q_1,q_2)=\int d^4X\,d^4Y\ e^{-iq_{1}X}\,e^{ipY}
\langle 0 |T\left\{j(X)\,j(Y)\,j(0)\right\}| 0 \rangle
\label{eq:a1},
\end{equation}
where $j(X)=:\!\phi(X)\phi(X)\!:$ \ .
The perturbative contribution and some of
the power corrections are shown  in Fig.\ref{eq:fig9}.

Calculating the perturbative contribution (Fig.\ref{eq:fig9}$a)$ in the
$\alpha$-representation or using the Feynman parameters,
we get
\begin{eqnarray}
\Phi_{(a)} & \equiv & \hat B(-p^2\rightarrow M^2){\cal{F}}_{(a)} =
\int_0^1 dx_1 dx_3 \theta(x_1+x_3<1) \frac{1}{x_2x_3}
e^{-\frac{q^2 x_1 x_3+Q^2 x_2 x_3}{x_1 x_2 M^2}} \nonumber\\
& &=\int_0^1 \frac{d\lambda dy}{\bar{\lambda}\bar y M^2}
e^{{-y q^2\lambda}/{M^2\bar{\lambda}}}
e^{ {-y Q^2}/{M^2\bar y} } \  .
\label{eq:a2}
\end{eqnarray}
Using the Mellin representation, as described above,
we obtain the expression
\begin{equation}
\Phi_{(a)}=\frac{1}{2\pi^2}\int_0^1 dy e^{-y{Q^2}/{\bar y M^2}}
\frac{1}{\bar y M^2}
\left\{ - e^{{y q^2}/{M^2}} \ln{\left (\frac{y q^2}{M^2} \right )} \right.
\left.+ \sum_{n=0}^{\infty} {\left(\frac{y q^2}{M^2}\right)}^n
\frac{\psi(n+1)}{n!}\right\}
\label{eq:a3},
\end{equation}
in which  the term containing the logarithmic singularity is
explicitly displayed.

The diagrams \ref{eq:fig9}$b,c,d$  corresponding to power corrections give
\begin{equation}
\Phi_{(b)}=\hat B(-p^2\rightarrow M^2)
\left(-\frac{8\langle\phi^2\rangle}{q^2Q^2}\right)=0,\quad
\Phi_{(c)}=-\frac{8\langle\phi^2\rangle}{Q^2M^2},\quad
\Phi_{(d)}=-\frac{8\langle\phi^2\rangle}{q^2M^2} \  .
\label{eq:a4}
\end{equation}
 Since
the diagrams \ref{eq:fig9}$a,d$  contain terms singular in $q^2$,
it is necessary
to perform an additional factorization of  short- and
long-distance contributions for these diagrams.

To get the bilocal contribution for the  amplitude (\ref{eq:a1}), let us
extract terms related  to the simplest
coefficient function proportional
to the propagator $S(Y)=i/{4\pi^2(Y^2-i0)}$
(see Fig.\ref{eq:fig5}$c$):
\begin{eqnarray}
\lefteqn{ {\cal{F}}^{B}= \int d^4Y\,e^{ipY}\
\frac{1}{\pi^2Y^2}\ \sum_{n=0}^{\infty}\frac{1}{n!}\
Y^{\mu_1}\ldots Y^{\mu_n} }\nonumber\\
& &\times \int d^4X\,e^{-iq_1X}\ \langle 0 |T\left\{j(X)\,
\phi(0)({\partial}_{\mu_1}\ldots{\partial}_{\mu_n})\phi(0)\right\}| 0 \rangle \,  .
\label{eq:a5}
\end{eqnarray}
Here, the long-distance
contribution is described  by the two-point correlator
\begin{eqnarray}
\Pi_{\mu_1\ldots\mu_n}(q_1) \equiv
\int d^4X\,e^{-iq_1X}\ \langle 0 |T\left\{j(X)\,\phi(0)\{
{\partial}_{\mu_1}\ldots{\partial}_{\mu_n}\}
\phi(0)\right\}| 0 \rangle
\nonumber \\
 = (-i)^n q_{1\mu_1}\ldots q_{1\mu_n}\ \Pi_n(q_1^2)  +  \ldots \ ,
\label{eq:two-point}
\end{eqnarray}
where  the dots denote the terms containing $g_{\mu_i \mu_j}$.

The large-$p^2$ behaviour of  ${\cal{F}}^{B}$ is governed
by the leading light-cone  singularity of the integrand.
To  display the $Y^2$-structure of the integrand
in eq.(\ref{eq:two-point}), we  reexpand the composite operator
over the  traceless operators possessing a  definite twist \cite{20a}:
\begin{eqnarray}
\lefteqn{Y^{\mu_1}\ldots Y^{\mu_n}\
(\phi{\partial}_{\mu_1}\ldots{\partial}_{\mu_n}\phi)= }\nonumber\\
& &=\sum_{l=0}^{[n/2]}\frac{n!(n-2l+1)}{l!(n-l+1)!}
\left(\frac{Y^2}{4}\right)^l\{Y^{\mu_1}\ldots Y^{\mu_{n-2l}}\}
(\phi({{\partial^2}})^l
\{{\partial}_{\mu_1}\ldots{\partial}_{\mu_{n-2l}}\}\phi) \ .
\label{eq:a6}
\end{eqnarray}
In (\ref{eq:a6}),  we  used the standard  notation
 $\{{\partial}_{\mu_1}\ldots{\partial}_{\mu_n}\}$
for a traceless group of indices $\mu_1\ldots\mu_n$:
 $g^{\mu_i\mu_j}{{O}}_{\{\ldots\mu_i\ldots\mu_j\ldots\}}=0$.
The leading $1/p^2$ contribution in (\ref{eq:a5}) in the
 $p^2\rightarrow \infty$ limit
comes from the lowest twist operators ($t=2$ in this case).
All other operators
are accompanied by the factors $\sim Y^2,\,(Y^2)^2$, $etc.$,
which cancel singularity of the propagator $\sim 1/Y^2$ and,
hence, give no contribution in the large-$p^2$ limit.

To get the B-regime  contribution for the perturbative triangle loop,
one should  substitute in eq. (\ref{eq:a5})  the perturbative
expression for  the twist-2 part of the two-point correlator
\begin{equation}
\Pi_n(q_1^2) \to  \Pi_n^{PT}(q^2)=\frac{1}{8\pi^2}\int_0^1
dy\,y^n\,\ln{\frac{q^2y\bar{y}}{\mu^2}}
\label{eq:a8}
\end{equation}
where $\mu^2$ is the UV cut-off parameter for the composite operator
$\phi\{{\partial}_{\mu_1}\ldots{\partial}_{\mu_n}\}\phi$,
and $q^2$ serves as a cut-off for  the low-momentum
region of integration.  As expected, in   the limit $q^2 \to 0$,
one obtains a mass singularity.

Substituting  (\ref{eq:a8}) into  (\ref{eq:a5}) and
taking into account that  $\{q_{1\mu_1}\ldots q_{1\mu_n}\}$ differs
from $q_{1\mu_1}\ldots q_{1\mu_n}$ only in
inessential terms $\sim Y^2$, it is straightforward to perform
summation over $n$ to get
\begin{equation}
 {\cal{F}}^{B}= \int_0^1 dy \, \ln{\frac{q^2y\bar{y}}{\mu^2}}
\int \frac{d^4Y}{\pi^2Y^2}\,e^{i(pY)-iy (q_1Y)} =
\int_0^1 dy \, \frac1{(p-y q_1)^2}
\ln{\frac{q^2y\bar{y}}{\mu^2}}.
\label{eq:summa}
\end{equation}
Since $(p-y q_1)^2=p^2-2y (pq_1)+y^2 q_1^2$ and $(p-q_1)^2=-Q^2$,
we can write $(p-y q_1)^2= \bar y  p^2 +y \bar q^2 - y Q^2$.
Applying the SVZ-Borel transformation, we get
\begin{equation}
\Phi^{B(PT)}= - \frac{1}{2\pi^2}\int_0^1 \frac{dy }{y M^2}
e^{-{y  Q^2}/{\bar{y} M^2}}
\left\{ e^{{y  q^2}/{M^2}} \ln{\left (\frac{y  \bar{y }q^2}{\mu^2} \right )}
\right\}
\label{eq:a9}.
\end{equation}
Comparing this result with the exact expression (\ref{eq:a3}),
one can observe that the non-analytic terms of two expressions coincide.
Hence, their difference, $i.e.,$ the
coefficient function of the SD regime (see Fig.\ref{eq:fig6}$b$,
Figs.\ref{eq:fig7},\ref{eq:fig8} )  does not have non-analytic terms
(mass singularities)   in the $q^2\to 0$  limit
(cf. \cite{tkachev},\cite{BeiNeRa88}).

For the ``$\langle\phi^2\rangle$-condensate'' correction given by the
diagram shown in Fig.\ref{eq:fig9}$d$,  the
contribution corresponding to  the lowest twist operators exactly reproduces the
singular term (\ref{eq:a4}) $\sim 1/q^2$.  Thus, this singularity  will
not appear in the modified OPE suitable for the  nonsymmetric
kinematics $q^2\ll Q^2$.

As we  stressed  above, the two-point correlator in (\ref{eq:a5}) is
responsible for the long-distance contribution $\sim 1/|q_1|$ and is not
directly calculable in  perturbation theory.
However, we can write it  through  a dispersion
relation
\begin{equation}
\Pi_n(q_1^2)=\frac{1}{\pi}\int_0^{\infty}  \frac{\delta\Pi_n(s)}{s-q_1^2}
ds
+ (\mbox{ subtractions }) ,
\label{eq:a10}
\end{equation}
where $\delta\Pi_n(s)\equiv (\Pi_n(s+i0)-\Pi_n(s-i0))/{2i}$
is the relevant spectral density.
As usual, we will model it by a
``first resonance  plus continuum''  ansatz.
A similar dispersion relation can be written  for the
perturbative correlator $\Pi_n^{PT} (q_1^2)$
(\ref{eq:a8}),  with the perturbative spectral density
$\delta\Pi_n^{PT} (s)$ substituting  the exact
one. It  is easy to realize  that the ambiguity  in the value of
the  $\mu^2$-parameter in eqs.
(\ref{eq:a8}),(\ref{eq:a9}) corresponds to the ambiguity  in the UV
subtraction procedure for the correlators.
However, the large-$s$ behavior of the exact and perturbative
spectral densities  is the same,  and it is  not necessary to specify
the subtraction procedure because the  correlators
appear in the modified  OPE only through the difference
$\Pi_n(q_1^2) - \Pi_n^{PT} (q_1^2)$
(see Fig.\ref{eq:fig8}), which is UV finite.

Writing explicitly the sum over the  ``hadronic '' states in
(\ref{eq:a10}),  we get
\begin{eqnarray}
i^{n+1} \Pi_n(q_1,Y)&\equiv& i^{n+1} \Pi_{\{\mu_1\ldots\mu_n\}}(q_1)\
Y^{\mu_1}\ldots Y^{\mu_n}= \nonumber\\
&=&\frac{f_{\phi}^2  (Yq_1)^n\langle y ^n\rangle}{m^2_{\phi}-q_1^2}+
\frac{1}{\pi}\int_{s_{\phi}}^{\infty} ds \frac{i\delta\Pi^{PT}_n(s)
(Yq_1)^n}{s-q_1^2}
+ (\mbox{ subtractions }) .
\label{eq:a11}
\end{eqnarray}
By analogy
with the definition of  the matrix element for the $\pi$-meson,
we define $\langle 0| j(0)|\phi,{p}\rangle = if_\phi$.
We can  introduce the twist-2 distribution amplitude $\phi(y )$
by treating the matrix elements of composite operators as its  moments:
 \begin{equation}
\langle \phi,{p}| \phi(0)(Y\partial)^n\phi(0)|0 \rangle=
i^n (Yp)^n (-i f_{\phi}) \int_0^1   y ^n \varphi(y ) dy .
\label{eq:a12}
\end{equation}
We also  used a convenient shorthand notation $\langle y ^n\rangle
\equiv \int_0^1 \varphi(y ) y^n dy $.
The contribution of the higher excited states (continuum) in eq.(\ref{eq:a12})
 is approximated, as usual,  by the perturbative spectral density
$\delta \Pi_n^{PT}(s) = -({1}/{8\pi}) \int_0^1   y^n  dy $
\label{eq:a13}
starting from the continuum threshold $s_{\phi}$.

The parameters of the model spectral density
in eq.(\ref{eq:a11}),  namely,  the mass $m_\phi$,
the residue of the first pole $f_\phi$, the
moments of the distribution amplitude $\langle y^n\rangle$ and
the threshold $s_{\phi}$ should  be  extracted  from the
auxiliary  sum rule for the moments of the scalar meson distribution amplitude.
In this case, the additional terms in the modified OPE will decrease
with increasing $q^2$. As a result,  the modified SR will
reproduce the original SR  for the three-point correlator
valid in the region where both $q^2$ and $Q^2$ are large
(see Fig.\ref{eq:fig9}). Treating  these parameters as known,  we
substitute (\ref{eq:a11}) in (\ref{eq:a5}) and define the ``bilocal''
contribution in the r.h.s. of the SR for the three-point correlator.
As a result,  all the additional terms
in the OPE (see the second row of Fig.\ref{eq:fig8})
can be written  in the following form:
\newpage
\begin{eqnarray}
\lefteqn{\Delta\Phi \equiv
\Phi^{bilocal}-\left(\Phi^{B}_{(a)}+
\Phi^{B}_{(d)}+\cdots\right) 
} \label{eq:a14}
\\
 & &= \int_0^1 \frac{1}{y M^2}\,
  e^{-{Q^2\bar{y }}/{M^2y }}\,e^{{q^2\bar{y}}/{M^2}}
\left[ \frac{4f_{\phi}^2\varphi(y )}{m_{\phi}^2+q^2}+
\frac{1}{2\pi^2}
\ln{\left(\frac{q^2}{s_{\phi}+q^2}\right)} +
8\frac{\langle\phi^2\rangle}{q^2}(\delta(y )+\delta(\bar{y }))\right] .
\nonumber 
\end{eqnarray}
Now we can write our final  expression for the r.h.s. of the modified sum rule:
\begin{eqnarray}
\lefteqn{\Phi(q^2,Q^2,M^2)=
\Phi_{(a)}+\Phi_{(b)}+\Phi_{(c)}+\Phi_{(d)}+\Delta\Phi
}\nonumber\\
& & = \int_0^1 \frac{1}{y M^2}\,e^{-{Q^2\bar{y }}/{M^2y }}\left\{
\frac{4f_{\phi}^2\varphi(y )}{m_{\phi}^2+q^2}\,e^{{q^2\bar{y }}/{M^2}}
- \frac{1}{2\pi^2} \ln{\frac{(s_{\phi}+q^2)\bar{y }}{M^2} }\,e^{{q^2\bar{y
}}/{M^2}}+\right.\nonumber\\
& & \left.+ \sum_{n=0}^{\infty} {\left(\frac{q^2\bar{y }}{M^2}\right)}^n
 \frac{\psi(n+1)}{n!}\right\} -\frac{8\langle\phi^2\rangle}{Q^2M^2} .
\label{eq:a15}
\end{eqnarray}
It is straightforward to observe that this  expression is well-defined
for $q^2=0$. Below, in our analysis of the QCD sum rule for the
$F_{\gamma^*\gamma^* \to \pi^o}(q^2,Q^2)$ form factor
we will follow a similar strategy.
 Of course, instead of the toy scalar meson, the $\rho$-meson will
play the role of the  lowest state in the bilocal contributions.

\section{Mass singularities in the QCD case}

\setcounter{equation} 0

As discussed above,  the B-regime  for the correlator (\ref{eq:corr})
 corresponds to a situation when
only the points  $Y$ and $0$ are separated by
 a  small interval (see Fig.\ref{eq:fig5}$b)$. In  simple cases,  the
short-distance
 coefficient function is given by a single quark propagator
or a product  of propagators. The long-distance contribution is represented
by a particular two-point correlator of the
electromagnetic current $J_{\mu}(X)$ and some composite operator  (see
\cite{BalY83,NeRa84b}). As we will see,
only the operators of two lowest twists appear in this expansion.

\subsection{Terms with single-propagator coefficient  function}

The simplest case is when
the coefficient function is given by a single  quark propagator
$S(Y)={\hat Y}/{2\pi^2(Y^2-i0)^2}$.
The relevant contribution
(Fig.\ref{eq:fig5}$b$)  can be written as
\begin{eqnarray}
\lefteqn{{\cal{F}}_{\alpha\mu\nu}^{B}=-\frac{2\pi}{3} \int d^4Y\,e^{ipY}\
\frac{Y^\beta}{2\pi^2Y^4}\
\sum_{n=0}^{\infty}\frac{1}{n!}\ Y^{\mu_1}\ldots Y^{\mu_n} }\nonumber\\
& &\times\left[\left\{\ -S_{\nu\beta\alpha\sigma}
\int d^4X\,e^{-iq_1X}\ \langle 0 |T\{J_{\mu}(X)\,
\bar{u}(0)({\stackrel{\leftarrow}{\partial}}_{\mu_1}\ldots
{\stackrel{\leftarrow}{\partial}}_{\mu_n})
\gamma_{\sigma}\gamma_5 u(0)\}| 0 \rangle \right.\right.\nonumber\\
& &\left.\qquad
+i{\epsilon}_{\nu\beta\alpha\sigma} \int d^4X\,e^{-iq_1X}\ \langle 0
|T\{J_{\mu}(X)\,
\bar{u}(0)({\stackrel{\leftarrow}{\partial}}_{\mu_1}
\ldots{\stackrel{\leftarrow}{\partial}}_{\mu_n})
\gamma_{\sigma}u(0)\}| 0 \rangle\ \right\}
\label{eq:sdII}\\
& &+\left\{\ S_{\nu\beta\alpha\sigma}
\int d^4X\,e^{-iq_1X}\ \langle 0 |T\{J_{\mu}(X)\,
\bar{u}(0)({\stackrel{\rightarrow}{\partial}}_{\mu_1}
\ldots{\stackrel{\rightarrow}{\partial}}_{\mu_n})
\gamma_{\sigma}\gamma_5 u(0)\}| 0 \rangle \right.\nonumber\\
& &\left.\left.\qquad
+i{\epsilon}_{\nu\beta\alpha\sigma} \int d^4X\,e^{-iq_1X}\ \langle 0
|T\{J_{\mu}(X)\,
\bar{u}(0)({\stackrel{\rightarrow}{\partial}}_{\mu_1}
\ldots{\stackrel{\rightarrow}{\partial}}_{\mu_n})
\gamma_{\sigma}u(0)\}| 0 \rangle\ \right\}\right]\nonumber ,
\end{eqnarray}
where\
$S_{\nu\beta\alpha\sigma}\equiv
(g_{\nu\beta}\,g_{\alpha\sigma}-g_{\nu\alpha}\,g_{\beta\sigma}
+g_{\nu\sigma}\,g_{\alpha\beta})$.

Let us consider  the
bilocal correlators  with the right-sided derivatives
\begin{eqnarray}
{ R}_n^5(q_1,Y)=\int d^4X\,e^{-iq_1X}\ \langle 0 |T\{J_{\mu}(X)\,
\bar{u}(0)(Y{\stackrel{\rightarrow}{\partial}})^n
\gamma_{\sigma}\gamma_5 u(0)\}| 0 \rangle   , \nonumber\\
{ R}_n(q_1,Y)=\int d^4X\,e^{-iq_1X}\ \langle 0 |T\{J_{\mu}(X)\,
\bar{u}(0)(Y{\stackrel{\rightarrow}{\partial}})^n
\gamma_{\sigma}u(0)\}| 0 \rangle .
\label{eq:defbilocr}
\end{eqnarray}
The bilocals  ${L}_n^5(q_1,Y)$ and ${ L}_n(q_1,Y)$  with the left-sided
derivatives
can be treated in the same way.

For any $n$, we can  expand the current with derivatives over a
set of traceless operators. More precisely, one should
 deal with traceless
combinations of the indices $\beta,\mu_1,\ldots,\mu_n$. Therefore the
essential
contribution to eq.(\ref{eq:sdII})  comes from two types of  operators, $viz.,$
the  lowest twist operators  which  correspond to the traceless
$\{\beta,\mu_1,\ldots,\mu_n\}$
combination and  the next-to-leading  twist operators which  contain
one contraction of $\sim g_{\beta\mu_i}$ or
$\sim g_{\mu_i\mu_j}$ type. The operators with higher twists
are accompanied by the  factors
$(Y^2)^2$, $(Y^2)^3$, $etc.$  which cancel  the singularity of the quark
propagator $1/Y^4$ and, hence, do not produce
mass singularities.

\subsection{Factorization of the perturbative term}

The factorization procedure for the perturbative loop
is illustrated  on Fig.\ref{eq:fig6}$a,b,c$.
The diagram
\ref{eq:fig6}$c$ corresponds to the
expression (\ref{eq:sdII}) with  the  two-point
correlators given by their  lowest-order perturbative form:
\begin{equation}
{\left\{\begin{array}{cc}
{R}_n^5(q_1,Y)&\!\! \\
{R}_n(q_1,Y)&\!\!
\end{array}\right\}}= 12{\left[\begin{array}{cc}
- i {\epsilon}_{\mu\alpha\sigma\beta}&\!\!  \\
S_{\mu\alpha\sigma\beta}& \!\!
\end{array}\right]}\int d^D\hat k\ \frac{(-iYk)^n[k_\alpha k_\beta - k_\alpha
q_{1\beta}]}{k^2(k-q_1)^2}
\label{eq:ptcorr}
\end{equation}
where $d^D\hat k\equiv d^D k/(2\pi)^D,\ D=4-2\varepsilon$; here
and in the following we use  dimensional regularization for
the UV divergencies and the $\overline{MS}$ subtraction scheme.

To extract the contribution of the lowest two twists
it is sufficient to keep  only the
terms up to $Y^2$ in the expansion  of the integral
(\ref{eq:ptcorr}) in powers of $Y^2$.  Indeed, the twist-2
contribution in (\ref{eq:sdII}) is obtained by taking
 formally $Y^2=0$ in the numerator of the integrand. Terms
proportional to $Y^2$ give  the contribution of the twist-4
operators. Using this expansion  (see Appendix B) and (\ref{eq:sdII}),
it is possible to perform summation over $n$ and to integrate
over $d^4Y$. After simple
but  lengthy  calculations   we find:
\begin{eqnarray}
\lefteqn{\Phi^{B(PT)}(q^2,Q^2,M^2)=\frac{2}{\pi}\int_0^1 dy
e^{-{Q^2y }/{M^2\bar{y }}} \frac{1}{2M^2} }\nonumber\\
& &\times\left\{ e^{{q^2y }/{M^2}} \ln{\frac{q^2y \bar{y }}{\mu^2}}
\left[2\frac{q^2y }{M^2}+\frac{q^4y ^2}{M^4}\right]
+2\frac{q^2y }{M^2}-\frac{3}{2}\frac{q^4y ^2}{M^4}\right\} .
\label{eq:ptq0sd}
\end{eqnarray}
As expected, the terms proportional
to $q^2\ln{q^2}$ (given by the twist-2 operators)
and $q^4\ln{q^2}$ (produced by the twist-4 operators), coincide with the
non-analytic terms in (\ref{eq:ptq0}).

\subsection{Factorization  for terms proportional to the gluonic condensate}

In a similar way, we can consider   the diagrams proportional to
the gluon condensate (see Fig.\ref{eq:fig1}). In this case,
it is convenient to perform factorization  diagram by diagram, because
different groups of diagrams correspond to  different coefficient functions
(CF) of the B regime.  Applying the same technique
that was  used for the perturbative contribution,
we get the following representation for
Fig.\ref{eq:fig1}$b,c$:
\begin{eqnarray}
\lefteqn{\Phi_{b}(q^2,Q^2,M^2)=-\frac{\pi}{18M^6}
\langle\frac{\alpha_s}{\pi}GG\rangle \int_0^1 dy
\frac{\bar{y}}{y^2} e^{-{Q^2\bar{y}}/{M^2y}}  }\nonumber\\
& &\times\left\{ \frac{M^2}{q^2\bar{y}} + e^{{q^2\bar{y}}/{M^2}}
\ln{\frac{q^2\bar{y}}{M^2}}
- \sum_{n=1}^{\infty} {\left(\frac{q^2\bar{y}}{M^2}\right)}^n
\frac{\psi(n+1)}{n!}\right\}
\label{eq:ggbq0}
\end{eqnarray}
\begin{eqnarray}
\lefteqn{\Phi_{c}(q^2,Q^2,M^2)=\frac{\pi}{18M^4}
\langle\frac{\alpha_s}{\pi}GG\rangle
\left\{\frac{1}{q^2}+\frac{1}{Q^2}\right\} }\nonumber\\
& &- \frac{\pi}{9M^6}
\langle\frac{\alpha_s}{\pi}GG\rangle \int_0^1 dy e^{-{Q^2\bar{y}}/{M^2y}}
\left\{ \frac{M^2}{q^2y} + \left(\frac{\bar{y}}{y}-\frac{1}{y^2}\right)
e^{{q^2\bar{y}}/{M^2}} \ln{\frac{q^2\bar{y}}{M^2}}\right. \nonumber\\
& &\left.-\left(\frac{\bar{y}}{y}-\frac{1}{y^2}\right)
\sum_{n=1}^{\infty} {\left(\frac{q^2\bar{y}}{M^2}\right)}^n
\frac{\psi(n+1)}{n!}\right\} \,  .
\label{eq:ggcq0}
\end{eqnarray}
The factorization procedure for these diagrams can be performed
just like it was done  in the
perturbative case. The only  difference is that now
we should take the $\langle{GG}\rangle$ part
of the correlator  (\ref{eq:defbilocr}) rather than its  perturbative part.
For Fig.\ref{eq:fig1}$b$, the bilocal contribution is
\begin{eqnarray}
\lefteqn{\Phi_{b}^{B}(q^2,Q^2,M^2)=-\frac{\pi}{18}
\langle\frac{\alpha_s}{\pi}GG\rangle \frac{1}{M^6} \int_0^1 dy
\frac{{y}}{\bar y^2} e^{-{Q^2{y}}/{\bar y M^2}}  }\nonumber\\
& &\times\left\{ \frac{M^2}{y q^2}e^{{y q^2}/{M^2}}
 + e^{{y q^2}/{M^2}} \ln{\frac{y\bar{y}q^2}{\mu^2}}
 + e^{{y q^2}/{M^2}}(2{y}-\frac{7}{6}) \right\}
\label{eq:ggbq0sd},
\end{eqnarray}
and for  Fig.\ref{eq:fig1}$c$ we obtain
\begin{eqnarray}
\lefteqn{\Phi_{c}^{B}(q^2,Q^2,M^2)=\frac{\pi}{18}
\langle\frac{\alpha_s}{\pi}GG\rangle \frac{1}{M^4} \frac{1}{q^2} }
\label{eq:ggcq0sd} \\
& &-\frac{\pi}{18} \langle\frac{\alpha_s}{\pi}GG\rangle \frac{2}{M^6}
\int_0^1 dy e^{-{y Q^2}/{\bar y M^2}}
\left\{ \frac{M^2}{\bar y q^2}e^{{yq^2}/{M^2}}
 + \left(\frac{{y}}{\bar y}-\frac{1}{\bar y^2}\right)
e^{{y q^2}/{M^2}} \ln{\frac{y\bar{y}q^2}{\mu^2}}\right\}. \nonumber
\end{eqnarray}
Note, that the terms proportional to $1/q^2$ are due to the traceless
combination of indices $\beta,\mu_1,\ldots,\mu_n$
in (\ref{eq:sdII}), whereas the
terms proportional to $\ln{q^2}$ correspond to the next-to-leading twist
operators.

Diagrams \ref{eq:fig1}$f,g$ can
be treated in a similar way,
their total contribution being
\begin{eqnarray}
\lefteqn{\Phi_{f+g}(q^2,Q^2,M^2)=-\frac{\pi}{18}
\langle\frac{\alpha_s}{\pi}GG\rangle \frac{1}{M^6} \int_0^1 dy e^{-{y
Q^2}/{\bar{y} M^2}}  }\label{eq:ggfgq0}
\\
& &\times\left\{\ \frac{(2\bar y^2+y)}{\bar y^2}\left(\ e^{{y q^2}/{M^2}}
\ln{\frac{y q^2}{M^2}}
- \sum_{n=1}^{\infty} {\left(\frac{y q^2}{M^2}\right)}^n \frac{\psi(n+1)}{n!}\
\right)
-\frac{(1-2y)}{\bar y^2}\frac{M^2}{q^2}\ \right\} \, .
\nonumber 
\end{eqnarray}
In this case, the short-distance part of the relevant bilocal
contribution is given by a product of two quark propagators
(see eq.(\ref{eq:CF2sd2}) below).
Extracting
the $\langle{GG}\rangle$ contribution, we get
\begin{eqnarray}
\lefteqn{\Phi_{f+g}^{B}(q^2,Q^2,M^2)=-\frac{\pi}{18}
\langle\frac{\alpha_s}{\pi}GG\rangle \frac{1}{M^6}
\int_0^1 dy e^{-{y Q^2}/{\bar y M^2}}  }\nonumber\\
& &\times\left\{ \frac{(2\bar y^2+y)}{\bar y^2} e^{{y q^2}/{M^2}}
\ln{\frac{y\bar y q^2}{\mu^2}}
-\frac{(1- 2y)}{\bar y^2}\frac{M^2}{q^2} e^{{y q^2}/{M^2}}
+e^{{y q^2}/{M^2}}(\ldots  ) \right\} ,
\label{eq:ggfgq0sd}
\end{eqnarray}
where $(\ldots)$ denote terms regular for $q^2=0$.

\subsection{Factorization  for quark condensate corrections}

The factorization procedure should be applied also  to the
diagrams with the  quark condensate. The relevant contributions
are  analogous to the  power corrections
of the toy scalar model.

Consider first the diagrams with a soft gluon (Fig.\ref{eq:fig3}).
The first three
diagrams (Figs.\ref{eq:fig3}$a,b,c$)  produce a $p^2$-independent
contribution  into  $F(p^2,q_1^2,q_2^2)$
which  vanishes after the SVZ-Borel transformation in $p^2$.
 The last three  diagrams
(Figs.\ref{eq:fig3}$g,h,i$)  are regular in the $q^2 \to 0 $ limit,
and there is no need  for an additional factorization.
Finally, the contributions of the remaining three  diagrams
(Figs.\ref{eq:fig3}$d,e,f$) has a manifestly factorized form
(see (\ref{eq:sdII}),(\ref{eq:CF2sd2})), $i.e.,$ they are completely
absorbed by the bilocal terms in the modified OPE.

For the diagrams with a hard gluon exchange
(Fig.\ref{eq:fig4}), the situation is very similar.
The first three diagrams  (Figs.\ref{eq:fig4}$a,b,c$ )
produce a manifestly factorized contributions into $F(p^2,q_1^2,q_2^2)$.
The basic difference between them is that the coefficient function
of the $B$-regime is given by a
product  of two quark propagators while those
related to the diagrams \ref{eq:fig4}$b,c$ is formed by a
product  of two quark propagators and a  gluon one.
The contribution of
Fig.\ref{eq:fig4}$d$ is  regular in the $q^2 \to 0$ limit, and there is
no need  for an additional factorization.

As mentioned earlier,
the remaining diagrams \ref{eq:fig4}$e$--$r$
do not contribute to   $F(p^2,q_1^2,q_2^2)$ in the large-$q^2$ regime.
However, a  more  careful  analysis (see Section 7)
of the bilocal objects corresponding to some of these diagrams shows that
 there may appear  nontrivial contributions into
$F(p^2,q_1^2,q_2^2)$ due to the so-called contact terms \cite{BalY83}.
Within the QCD sum rule method,
the contact terms, in particular, play a crucial
role in establishing the correct  normalization for
the electromagnetic form factors at zero momentum
transfer \cite{NeRa84b,BeiNeRa88}.

After application of the subtraction
procedure illustrated  on Fig.\ref{eq:fig8}, all the infrared
singularities of the original sum rule cancel with the corresponding singular
contributions from the diagrams in which the $B$-term is extracted explicitly.
In our case, it is sufficient to consider
the bilocal objects  corresponding to  operators
of the lowest two twists.   The  terms
remaining  after this subtraction are regular in the $q^2 \to 0$
limit and  contribute to the  modified sum rule.

\section{Bilocal correlators}

\setcounter{equation} 0

\subsection{Bilocals related to a single-propagator coefficient function}

Let us start with  the  case when
the coefficient function of the $B$-regime  is given by a
single  quark propagator.
Its  contribution into the three-point
amplitude ${\cal F}_{\alpha\mu\nu}(q_1,p)$ can be written as
\begin{equation}
{\cal F}^B_{\alpha\mu\nu}(q_1,p) \sim \int e^{ipY}
\frac{Y^{\beta} d^4 Y}{(Y^2-i0)^2}
 \int  e^{-iq_1 X}
\langle 0 | T \bigl( J_{\mu}(X) (\bar \psi(0) \gamma_{\nu} \gamma_{\beta}
\gamma_5
\gamma_{\alpha} \psi(Y) \bigr) | 0 \rangle d^4 X  \, .
\end{equation}
As emphasized  before, the   correlators (bilocals)
$\langle 0| T \{ J_{\mu }(X)
{\cal O}^{(i)}  (0) \}| 0 \rangle$ of the $J_{\mu }(X)$-current
with  composite operators
$ {\cal O}^{(i)}  (0)$
are sensitive to the  long-distance
$(\sim 1/|q_1|)$  dynamics and, for this reason, they are not directly
calculable in perturbation
theory. The standard way out is to  write them in a
dispersion form and assume the simplest ansatz
(``lowest resonance $+$ continuum'') for
the  model spectral density,   with the continuum starting
at some effective threshold $s_{\rho}$.
The parameters of the model spectrum  can be  extracted from an auxiliary
QCD sum rule.
The quantum numbers of the electromagnetic current $J_{\mu}(X)$,
which appears in all the bilocals, dictates that the lowest resonance
should be  represented by the $\rho^0$-meson \footnote{Strictly speaking, we
deal with an isotopic mixture of $\rho^0$- and $\omega$-contributions which
degenerate in the chiral limit $m_u=m_d=0$ to an effective state
with the mass of the $\rho$-meson (cf. \cite{svz}).}.
In particular,   for the bilocals
corresponding to the-single propagator coefficient function,
we can write
\begin{eqnarray}
\lefteqn{{ R}_n(q_1,Y)= \frac{1}{\pi}\int_0^{\infty}  \frac{\delta
R_n(s)}{s-q_1^2} ds
+ (\mbox{ subtractions }) }\nonumber\\
& &=\mbox{ ``$\rho^o$-meson contribution'' }+
\frac{1}{\pi}\int_{s_{\rho}}^{\infty} ds \frac{\delta{ R}^{PT}_n(s,Y)}{s-q_1^2}
+ (\mbox{ subtractions }),
\label{eq:SE2}
\end{eqnarray}
where $\delta{ R}_n(s,Y)$ is the relevant discontinuity:
$\delta{ R}_n(s,Y)\equiv ({R}_n(s+i0,Y)-{R}_n(s-i0,Y))/{2i}$.

Picking out  the $\rho^o$-meson term
\begin{equation}
\delta^{(+)}(p^2-m_{\rho}^2)\ \frac{d^4p}{(2\pi)^3}\sum_{\lambda=-1}^{1}
|\rho^o_{\lambda};\stackrel{\rightarrow}{p}\rangle\
\langle \rho^o_{\lambda};\stackrel{\rightarrow}{p}|\ \subset\ {\bf\hat{I}},
\label{eq:SE3}
\end{equation}
 in  the sum over  physical hadronic  states (with  $\lambda$ being  the
helicity of the $\rho^o$),
we extract the $\rho^0$ contribution.
As a result, we obtain a set of matrix elements
which can be  parameterized as
\begin{eqnarray}
\langle 0|\bar{\psi}(0)\gamma_{\sigma}\psi(Y) |
\rho^o_{\lambda=0};\stackrel{\rightarrow}{p}\rangle\
= i p_{\sigma}\ f_{\rho}^V\,\phi_{\rho}^V(Yp,\mu^2) + \cdots
\label{eq:SE4} \\
\langle 0|\bar{\psi}(0)\gamma_{\sigma}\psi(Y) |
\rho^o_{|\lambda|=1};\stackrel{\rightarrow}{p}\rangle\
= \varepsilon_{\sigma}^{\bot}\ f_{\rho}^V m_{\rho}\,\phi_{\rho_{\bot}}^V(Yp,\mu^2)
+ i a_{V1}\, p_{\sigma}\ f_{\rho}^V m_{\rho} (\varepsilon^{\bot}Y)\,
\phi_{\rho_{\bot}}^{V1}(Yp,\mu^2)
+ \cdots
\label{eq:SE5}\\
\langle 0|\bar{\psi}(0)\gamma_{\sigma}\gamma_5\psi(Y) |
\rho^o_{|\lambda|=1};\stackrel{\rightarrow}{p}\rangle\
= \epsilon_{\sigma\alpha\beta\delta}\,\varepsilon_{\alpha}^{\bot}\, p_{\beta}\, Y_{\delta}\ f_{\rho}^A\,
\phi_{\rho}^A(Yp,\mu^2) + \cdots \ .
\label{eq:SE6}
\end{eqnarray}
Here only the twist 2 terms are written explicitly, and
the dots stand for the higher twist
 contributions,
 $\varepsilon_{\sigma}$ is the polarization vector of the $\rho^o$-meson,
and the helicity components have an evident interpretation in terms of the
longitudinal and
transverse  polarizations:
 $\rho^o_{\lambda=0}\equiv\rho^o_L$ , $\rho^o_{|\lambda|=1}\equiv\rho^o_{\bot}$.

In a standard way, the functions  $\phi_{\rho}(Yp,\mu^2)$  can be
related to the usual wave functions $\varphi_{\rho}(y,\mu^2)$  describing
the light-cone momentum distribution  inside the $\rho$:
\begin{equation}
\phi_{\rho}^{(i)}(Yp,\mu^2)=\int_0^1 dy e^{-i(Yp)y}
\varphi_{\rho}^{(i)}(x,\mu^2) ,
\label{eq:SE7}
\end{equation}
with $\mu^2$ being the renormalization parameter for the relevant composite
operators.
The constant $f_{\rho}^V$ fixing the normalization of the simplest
wave function  is known from previous  QCD sum rule studies:
$f_{\rho}^V \simeq 200 \,  \mbox{MeV}$ \cite{svz,CZ84},
while  the constants $f_{\rho}^A$ and $a_{V1}$  in eqs. (\ref{eq:SE5}),
(\ref{eq:SE6})
can be  fixed by   equations of motion (cf. \cite{Gor1,BraF90}), which
form an infinite set of  relations connecting the moments of different  wave
functions
 (see Appendix C).

For our purposes, it is more convenient to  write down
the   matrix elements in a form suitable for an
arbitrary polarization of the $\rho^o$-meson:
\begin{eqnarray}
\langle 0|\bar{\psi}(0)\gamma_{\sigma}\psi(Y) |
\rho_{\lambda}^o;\stackrel{\rightarrow}{p}\rangle\
= \varepsilon_{\sigma}^{(\lambda)}\ f_{\rho}^V m_{\rho}\,
\left[\,\phi_{\rho_{\bot}}^V(Yp,\mu^2) +
C_{V4}\, Y^2 \phi_{\rho_{\bot}}^{V4}(Yp,\mu^2) + \cdots\ \right] \nonumber\\
+ i a_{V1}\, p_{\sigma}\ f_{\rho}^V m_{\rho} (\varepsilon^{(\lambda)} Y)\,
\left[\,\phi_{\rho_{\bot}}^{V1}(Yp,\mu^2) +
C_{[V1]4}\, Y^2 \phi_{\rho_{\bot}}^{[V1]4}(Yp,\mu^2) + \cdots\ \right]
\nonumber\\
+ f_{\rho}^V m_{\rho} C_{[V2]4}\,
\left(\,Y_{\sigma} (\varepsilon^{(\lambda)} Y)\,-\,\varepsilon_{\sigma}^{(\lambda)} {Y^2}/{4}\,\right)\,
\phi_{\rho_{\bot}}^{[V2]4}(Yp,\mu^2) +\cdots
\label{eq:SE9}
\end{eqnarray}
\begin{eqnarray}
\langle 0|\bar{\psi}(0)\gamma_{\sigma}\gamma_5\psi(Y) |
\rho_{\lambda}^o;\stackrel{\rightarrow}{p}\rangle\
= \epsilon_{\sigma\alpha\beta\rho}\,\varepsilon_{\alpha}^{(\lambda)}\, p_{\beta}\, Y_{\rho}\ f_{\rho}^A\,
\left[\,\phi_{\rho}^A(Yp,\mu^2)
+ C_{A4}\, Y^2 \phi_{\rho}^{A4}(Yp,\mu^2) + \cdots\ \right].
\label{eq:SE10}
\end{eqnarray}

Since the C-parity of the $\rho^o$-meson is negative,
its wave functions have the following properties (here and below
$\bar y \equiv 1-y$):
\begin{eqnarray}
\varphi_{\rho_{\bot}}^{V,V4,[V2]4,A,A4}(y) =
\varphi_{\rho_{\bot}}^{V,V4,[V2]4,A,A4}(\bar y)&,&
\varphi_{\rho_{\bot}}^{V1,[V1]4}(y)  = -\varphi_{\rho_{\bot}}^{V1,[V1]4}(\bar y),
\nonumber\\
\int_0^1 dy\ \varphi_{\rho_{\bot}}^{V,V4,[V2]4,A,A4}(y) = 1&,&
\int_0^1 dy\ y\,\varphi_{\rho_{\bot}}^{V1,[V1]4}(y) =1 .
\label{eq:norm}
\end{eqnarray}

In the relations above, we have  explicitly displayed
wave functions  up to twist 4.
Note that, for a longitudinally polarized $\rho^o$-meson,
$$\varepsilon_{\sigma}^{\lambda=0}\simeq i\, {p_{\sigma}}/{m_{\rho}}+
{\cal{O}}\left({m_{\rho}}/{p_z}\right)$$
as  $p_z \to \infty$,
and the twist-2 part in eq.(\ref{eq:SE9}) coincides with the well known
definition (\ref{eq:SE4}).

Let us consider first the twist-2 $\rho^o$-meson contribution in
eq. (\ref{eq:SE2}). Applying (\ref{eq:SE3}), (\ref{eq:SE9}) and (\ref{eq:SE10}),
we obtain:
\begin{eqnarray}
{\left(\sum_{n=0}^{\infty} \frac{1}{n!} R_n(q_1,Y)\right)}
_{\rho^o} =
\frac{(-i)(f_{\rho}^V m_{\rho})^2}{m_{\rho}^2 - q_1^2}
\left\{
 \,\varepsilon_{\mu}\varepsilon_{\sigma}^*
\int_0^1\, dy\, e^{-i(Y q_1)\bar{y}}\,
\varphi_{\rho_{\bot}}^V(y)
\right.
\nonumber \\
\left.
 - i a_{V1}\, q_{1\sigma} \varepsilon_{\mu}\, (\varepsilon^* Y)
\int_0^1\, dy\, e^{-i(Y q_1)\bar{y}}\, \varphi_{\rho_{\bot}}^{V1}(y) \,
\right\}
\label{eq:SE11}
\end{eqnarray}
\begin{eqnarray}
\left(\sum_{n=0}^{\infty} \frac{1}{n!}  R_n^5(q_1,Y)\right)
_{\rho^o} =
\frac{(-i)(f_{\rho}^A f_{\rho}^V m_{\rho})}{m_{\rho}^2 - q_1^2}
\,\varepsilon_{\mu}\varepsilon_{\delta}^*\,\epsilon_{\sigma\delta\beta\rho}\,q_{1\beta}Y_{\rho}
\int_0^1\, dy\ e^{-i(Y q_1)\bar{y}}\ \varphi_{\rho_{\bot}}^A(y) .
\label{eq:SE12}
\end{eqnarray}
Here we use the shorthand  notation:
\begin{equation}
\varepsilon_{\mu}\varepsilon_{\sigma}^*\equiv
\sum_{\lambda=0,\pm 1}\,\varepsilon_{\mu}^{\lambda}\varepsilon_{\sigma}^{\lambda^*}=
- g_{\mu\sigma} + \frac{q_{1\mu}q_{1\sigma}}{m_{\rho}^2} .
\label{eq:SE13}
\end{equation}

Substituting eqs. (\ref{eq:SE11}), (\ref{eq:SE12}) into  eq. (\ref{eq:sdII})
and extracting the
proper  tensor structure we get:
\begin{equation}
F^{B(\rho)} = \frac{4\pi}{3}
\frac{f_{\rho}^V m_{\rho}}{m_{\rho}^2-q_1^2} \int_0^1 dy \frac{1}{\tilde{p}^4}
\left[\,-a_{V1}\,f_{\rho}^V m_{\rho}\,\varphi_{\rho_{\bot}}^{V1}(y)
        - f_{\rho}^A\,\varphi_{\rho_{\bot}}^{A}(y)(1+2\bar{y})\ \right] ,
\label{eq:SE14}
\end{equation}
where $$\tilde{p\,}^2 \equiv (q_2+\bar y q_1)^2 =-q_1^2\,y\bar{y} + q_2^2\,y +
p^2\,\bar{y}$$
is the virtuality of the hard quark written in the ``parton'' form.
This formula includes an extra  factor of $2$
which appears when  one adds the
contribution of the correlators $ L_n,  L_n^5$  and uses the symmetry
properties of the wave functions (\ref{eq:norm}).
It should be noted that the wave functions
$\varphi_{\rho_{\bot}}^V$,  $\varphi_{\rho_{\bot}}^{V4}$ and
$\varphi_{\rho_{\bot}}^{[V2]4}$ do not contribute to the form
factor $F$ which is  considered here.

Deriving,  in a similar way,  the  twist-4 contribution
 and  applying the SVZ-Borel transformation  to  the  resulting amplitude, we
obtain:
\begin{eqnarray}
\Phi^{B(\rho)} &=& \frac{4\pi}{3}
\frac{f_{\rho}^V m_{\rho}}{m_{\rho}^2+q^2}\, \int_0^1 dy
\frac{1}{\bar{y}^2 M^4}\, e^{{-Q^2 y}/{M^2 \bar{y}}}\,e^{{q^2 y}/{M^2}}
\label{eq:SE15}\\
&\times & \left[\,-a_{V1}\,f_{\rho}^V m_{\rho}\left(\varphi_{\rho_{\bot}}^{V1}(y) -
\frac{4 C_{[V1]4}}{\bar{y} M^2}\,\varphi_{\rho_{\bot}}^{[V1]4}(y)\right)
     -  f_{\rho}^A (1+2\bar{y})\left(\varphi_{\rho_{\bot}}^{A}(y) -
\frac{4 C_{A4}}{\bar{y} M^2}\,\varphi_{\rho_{\bot}}^{A4}(y)\right)\ \right] .
\nonumber
\end{eqnarray}
As expected, the twist-4 contribution  is suppressed by  one
 power of $1/M^2$.

Now, taking $q^2=0$ and introducing the integration variable
$s = yQ^2/\bar y$, we can write this term as
\begin{eqnarray}
A(Q^2) &=& - \frac{4\pi}{3Q^2} \frac{f_{\rho}^V}{m_{\rho}} \int_0^{\infty}
\frac{ds}{M^4} e^{-s/M^2}
 \left[\,a_{V1}\,f_{\rho}^V m_{\rho}
 \left(\varphi_{\rho_{\bot}}^{V1}\left(\frac{s}{s+Q^2}\right)
+
\frac{4 C_{[V1]4} (s+Q^2)}{ M^2Q^2}\,
\varphi_{\rho_{\bot}}^{[V1]4}\left(\frac{s}{s+Q^2}\right)\right) \right.
\nonumber \\ & & \left.
       + f_{\rho}^A
(1+2\frac{s+Q^2}{Q^2})\left(\varphi_{\rho_{\bot}}^{A}
\left(\frac{s}{s+Q^2}\right) +
\frac{4 C_{A4}(s+Q^2)}{ M^2
Q^2}\,\varphi_{\rho_{\bot}}^{A4}\left(\frac{s}{s+Q^2}\right)\right)\ \right] .
\end{eqnarray}

Note, that this representation has ``wrong'' powers
of the SVZ-Borel parameter $M^2$ compared to the canonical form
\begin{equation}
\Phi (M^2,Q^2)
=\frac{1}{\pi M^2} \int_0^\infty  e^{-s/M^2} {\rho}(s,Q^2) \, ds.
\label{eq:fph2}
\end{equation}
Using   the transformation
\begin{equation}
\int_0^{\infty} ds\ e^{-s/M^2}\,\frac{g(s)}{M^2} =
\int_0^{\infty} ds\ e^{-s/M^2} ( g(0)\,\delta(s) + g'(s) ),
\label{eq:specden1}
\end{equation}
\begin{equation}
\int_0^{\infty} ds\ e^{-s/M^2}\,\frac{g(s)}{M^4} =
\int_0^{\infty} ds\ e^{-s/M^2} ( g'(0)\,\delta(s) + g(0)\,\delta'(s) + g''(s) )
\label{eq:specden2}
\end{equation}
$etc.,$ we can always  cast  the terms with
the ``wrong'' powers  into the canonical form (\ref{eq:fph2}).

In particular, taking  the asymptotic forms (see Appendix C)
\begin{equation}
a_{V1}=\frac{1}{40},\, f_{\rho}^A=-\frac{f_{\rho}^V m_{\rho}}{4},\\
\varphi_{V1} = \varphi_{V1}^{as} = 60y\bar{y}\,(2y-1),\,
\varphi_A = \varphi_A^{as} = 6y\bar{y},
\label{eq:tw3asywf}
\end{equation}
as the simplest estimate
for the twist-2 contributions, we get
\begin{eqnarray}
g_{2,\rho}(s)& = &8\pi^2(f_{\rho}^V)^2 \frac{s\,Q^2}{(s+Q^2)^3}
\label{eq:g3rho}\\
\rho_{2}^{\rho}(s,Q^2)\, =\, g'_{2,\rho}(s)& = &
8\pi^2 (f_{\rho}^V)^2\, \frac{(Q^4-2sQ^2)}{(s+Q^2)^4}
\label{eq:rhotwist3}
\end{eqnarray}
Note, that the $s$-integral over the latter spectral density is zero.

The asymptotic  forms for the  twist-4 distribution amplitudes
can be directly extracted from the corresponding correlators:
\begin{equation}
\varphi_{[V1]4} = \varphi_{[V1]4}^{as} = 420(y\bar{y})^2\,(2y-1),\ \
\varphi_{A4} = \varphi_{A4}^{as} = 30y^2{\bar{y}}^2.
\label{eq:tw5asywf}
\end{equation}
From the equations of motion (see (C.4), Appendix C) it follows that
$C_{[V1]4}=5 C_{A4}/7$. Taking into account this expression and
eqs.(\ref{eq:tw5asywf}), we obtain:
\begin{equation}
A_4(Q^2)=\frac{1}{\pi M^2}\int_0^{\infty} ds\,e^{-s/M^2}
\frac{g_{4,\rho}(s)}{M^4} \  ,
\end{equation}
where
\begin{equation}
g_{4,\rho}(s)=-4\pi^2(f_{\rho}^V)^2\,40\,C_{A4} \frac{s^2\,Q^2}{(s+Q^2)^4} \ .
\label{eq:g5rho}
\end{equation}
Hence, the relevant spectral density is
\begin{equation}
\rho_{4}^{\rho}(s,Q^2)=
-8\pi^2(f_{\rho}^V)^2\,40\, C_{A4} Q^2\frac{(3s^2-6sQ^2+Q^4)}{(s+Q^2)^6}=
g^{\prime\prime}_{4,\rho}(s) .
\label{eq:rhotwist5}
\end{equation}

\subsection{Continuum contribution}

 As noted earlier,
 in the  basic  OPE for the small-$q^2$  kinematics,
one always deals with the difference  between  an ``exact''
bilocal correlator  $R$ and its perturbative analog $R^{PT}$
(see Fig.\ref{eq:fig8}). Since  the ultraviolet
behaviour of these two correlators is the same,
there is no need to explicitly specify a
subtraction  prescription for the correlators.

Now, incorporating
our model for the bilocal correlators, in which  the
contribution due to higher excited states
is approximated by the perturbative  spectral
density (see  (\ref{eq:SE2})), $i.e.,$ by the continuum  starting
at $s_{\rho}$, we can easily write down an expression for
the difference between the continuum contribution to $R$ and the perturbative
bilocal $R^{PT}$.  Then, substituting the result
into the original expansion (\ref{eq:sdII}) and performing some
straightforward calculations, we obtain:
\begin{eqnarray}
\lefteqn{\Phi
^{B(cont)} - \Phi^{B(PT)} =
\frac{1}{\pi}\, \int_0^1 dy
\frac{1}{ M^2}\ e^{{-Q^2 y}/{M^2 \bar{y}}}\,e^{{q^2 y}/{M^2}} } \nonumber\\
&\times & \left[\,\frac{2y}{M^2}\,
\left(q^2\ln{\frac{s_{\rho}+q^2}{q^2}} - s_{\rho}\right) +
\frac{y^2}{M^4}\,\left(q^4\ln{\frac{s_{\rho}+q^2}{q^2}} - q^2 s_{\rho}
+\frac{s_{\rho}^2}{2}\right)\ \right] \,  .
\label{eq:SE16}
\end{eqnarray}

The terms collected in  the ( \ ) brackets  correspond
 to contributions due to   operators with
twist 2 and 4.
 Note, that these  terms exactly  cancel the
logarithmic contributions $q^2\ln{q^2}, q^4\ln{q^2}$
present in the
coeficient function of  the unit operator for the  usual OPE
valid in  the large-$q^2$ kinematics.
As a result, the non-analytic terms are replaced  by  the
combinations $q^2\ln{(s_{\rho}+q^2)}$ and $ q^4\ln{(s_{\rho}+q^2)}$,
which are ``safe''  in the   $q^2\rightarrow 0$ limit.
On the other hand,  for  large $q^2$,  the  usual OPE
without additional terms must work,
$i.e.,$  the difference between ``exact''  bilocal term
and its perturbative analogue must vanish faster than any power of $1/q^2$
in the large-$q^2$ limit.
 It is easy to check that the terms  on the r.h.s. of eq.(\ref{eq:SE16})
behave like $1/q^2$ when $q^2 \to \infty$.
To get the total expression for the additional terms,
we should add the $\rho$-contribution  to eq.(\ref{eq:SE16}).
The  $\rho$-contribution also has the $1/q^2$-behaviour for large $q^2$.
However, to produce a perfect transition to the pure SD-case,
the  $1/q^2$-terms must  cancel.
Basically, this means that using  a  rough model for the correlator
one should not rely too heavily on the extrapolation of our
result (\ref{eq:SE16}) beyond the region $q^2 \lapprox m_{\rho}^2$.
However, we can require that the model, at least, should provide
the  cancellation of the $1/q^2$ terms.
If we choose the asymptotic form for the lowest-twist
distribution amplitudes, the
cancellation of the $1/q^2$-terms produces the estimate:
\begin{equation}
\widetilde{s}_{\rho}^2 = 8\pi^2 (f_{\rho}^V)^2 m_{\rho}^2.
\label{eq:duality}
\end{equation}

Our result (\ref{eq:SE16}) simplifies in the $q^2 \to 0$ limit:
\begin{equation}
\Phi^{B(cont)} - \Phi^{B(PT)}  =
\frac{1}{\pi M^2}\,
\int_0^{\infty} ds\ e^{-s/M^2}\frac{\bar{y}^2}{Q^2}
\left[-\frac{2y}{M^2} s_{\rho} + \frac{y^2}{M^4}\frac{s_{\rho}^2}{2}\right] ,
\label{eq:continq0}
\end{equation}
where  $s$ again is defined by $s=yQ^2/\bar y$.
The twist-2  contribution
can be represented in the canonical  form with the following spectral density:
\begin{equation}
\rho_{2}^{B}(s,Q^2)=
-2 s_{\rho} \frac{(Q^4-2sQ^2)}{(s+Q^2)^4}
\label{eq:twist3cont} .
\end{equation}

For the twist-4  contribution in eq.(\ref{eq:continq0}), we obtain:
\begin{equation}
\rho_4^{B}(s,Q^2)=
s_{\rho}^2 Q^2\frac{(3s^2-6sQ^2+Q^4)}{(s+Q^2)^6}=
g^{\prime\prime}_{4,\mbox{\scriptsize cont}}(s),
\label{eq:twist5cont}
\end{equation}
where
\begin{equation}
g_{4,\mbox{\scriptsize cont}}(s)=\frac{s_{\rho}^2}{2}\frac{s^2 Q^2}{(s+Q^2)^4}.
\label{eq:g5cont}
\end{equation}

Using (\ref{eq:duality}), the $\rho$-meson contribution from (\ref{eq:rhotwist3})
can be rewritten as
\begin{equation}
\rho_{2}^{\rho}(s,Q^2)\, =\, g'_{2,\rho}(s)=
\frac{\widetilde{s}_{\rho}^2}{m_{\rho}^2}\, \frac{(Q^4-2sQ^2)}{(s+Q^2)^4}.
\label{eq:rhotwist3dual}
\end{equation}

For the next-to-leading-twist  contributions, similar duality arguments
give  the following estimates for  the normalization
constants of  the relevant two-body distribution amplitudes:
\begin{equation}
\widetilde{C}_{A4}=
\frac{\widetilde{s}_{\rho}}{60}\simeq 2.28\,\cdot 10^{-2}\,\mbox{GeV}^2 \  ; \quad
\widetilde{C}_{[V1]4}=
\frac{\widetilde{s}_{\rho}}{84}\simeq 1.63\,\cdot 10^{-2}\,\mbox{GeV}^2
\label{eq:tw5estimation} .
\end{equation}
Hence, the expression (\ref{eq:rhotwist5}) can be written as
\begin{equation}
\rho_{4}^{\rho}(s,Q^2)=
-2\frac{{\widetilde{s}_{\rho}}^3}{3 m_{\rho}^2}
Q^2\frac{(3s^2-6sQ^2+Q^4)}{(s+Q^2)^6}.
\label{eq:rhotwist5dual}
\end{equation}
If the distribution amplitudes deviate from their asymptotic forms, then the duality
condition between the $\rho$-meson and the continuum should keep the form
of the expressions
(\ref{eq:rhotwist3dual}), (\ref{eq:rhotwist5dual}), but instead of
$\widetilde{s}_{\rho}$ we should get an effective duality interval.

Note that for $f_{\rho}^V=0.2\,\mbox{GeV}$, our estimates give
$\widetilde{s}_{\rho}\simeq 1.37\,\mbox{GeV}^2$ for the duality interval.
This value is
in good agreement with the  ``canonical'' one:
$s_{\rho}^{LD}=4\pi^2(f_{\rho}^V)^2\simeq 1.58\,\mbox{GeV}^2$. The latter
follows from the local duality considerations for the two-point
correlator of two vector currents.

\subsection{Twist-3 bilocals
for  two-propagator  coefficient functions}

Next in complexity is the contribution
related to the coefficient function
formed by a product of two propagators  $S(Y,Z)$ and $S(Z,0)$:
\begin{eqnarray}
\lefteqn{{\cal{F}}_{\alpha\mu\nu}^{B(2)}=\frac{2 \pi}{3} \int d^4Y\,e^{ipY}\
d^4Z\,\frac{(Y-Z)^\delta}{2\pi^2(Y-Z)^4}\ \frac{Z^\varepsilon}{2\pi^2Z^4}\
\sum_{n,m=0}^{\infty}\frac{1}{n!m!}\ Y^{\mu_1}\ldots Y^{\mu_n}\ Z^{\nu_1}\ldots
Z^{\nu_m} }
\label{eq:CF2sd2}\\
& &\times \int d^4X\,e^{-iq_1X}\ \langle 0 |T\{J_{\mu}(X)\,
\bar{u}(0)({\stackrel{\leftarrow}{\partial}}_{\mu_1}\ldots{\stackrel{\leftarrow}{\partial}}_{\mu_n})\gamma_{\nu}\gamma_{\delta}\
g_{\gamma}
g({\stackrel{\rightarrow}{\partial}}_{\nu_1}\ldots{\stackrel{\rightarrow}{\partial}}_{\nu_n}A_{\gamma}^b(0))
t^b\gamma_{\varepsilon}\gamma_5\gamma_{\alpha}u(0)\}| 0 \rangle  \ . \nonumber
\end{eqnarray}
Here we explicitly extracted the bilocal correlator
containing a composite operator composed of two  quark and one
gluonic  field.  Note that  the gluonic  field
$A_{\gamma}^b(Z)$ here may be treated as taken
in the Fock-Schwinger gauge, $i.e.,$ it can be substituted by
\begin{equation}
A_{\gamma}^b(Z) = Z_{\varphi} \int_0^1 \alpha\, G_{\varphi\gamma}^b(\alpha Z)d\alpha\, .
\label{eq:connect}
\end{equation}
As a result,  the  $\rho^o$-meson contribution   is determined by the
following  matrix elements:
\begin{eqnarray}
\langle 0|\bar{u}(Z_1)\gamma_{\beta}\gamma_5 g_s G_{\varphi\gamma}^b(Z_3) t^b u(Z_2) |
\rho_{\lambda}^o;\stackrel{\rightarrow}{p}\rangle\
&=& p_{\beta}\,\epsilon_{\varphi\gamma\theta\kappa}\,p_{\theta}\,\varepsilon_{\kappa}^{(\lambda)}\
  f_{3\rho}^A\,\phi_{3\rho}^A(Z_ip,\mu^2) \nonumber\\
& &+ \mbox{higher twist contributions}
\label{eq:SE18}
\end{eqnarray}
\begin{eqnarray}
\langle 0|\bar{u}(Z_1)\gamma_{\beta}\,i g_s G_{\varphi\gamma}^b(Z_3) t^b u(Z_2) |
\rho_{\lambda}^o;\stackrel{\rightarrow}{p}\rangle\
&=& p_{\beta}\,\left(p_{\varphi}\varepsilon_{\gamma}^{(\lambda)} -
  p_{\gamma}\varepsilon_{\varphi}^{(\lambda)}\right)\,
  f_{3\rho}^V\,\phi_{3\rho}^V(Z_ip,\mu^2) \nonumber\\
& &+ \mbox{higher twist contributions} \ .
\label{eq:SE19}
\end{eqnarray}
 In a standard way, we can introduce the momentum distribution
amplitudes\footnote{$[dy]_3\equiv dy_1 dy_2 dy_3\,\delta(1-\sum_{i} y_i)$}
$\varphi_{3\rho}^{V,A}(y_i)$:
\begin{equation}
\phi_{3\rho}^{V,A}(Z_ip,\mu^2) = \int_0^1 [dy]_3\
\varphi_{3\rho}^{V,A}(y_i)\,e^{-i\,\sum y_i\,(Z_i\, p)}
\label{eq:SE20}.
\end{equation}
They have the following  symmetry properties:
\begin{equation}
\varphi_{3\rho}^{A}(y_1,y_2;y_3) = \varphi_{3\rho}^{A}(y_2,y_1;y_3) \  ,  \
\varphi_{3\rho}^{V}(y_1,y_2;y_3) = -\varphi_{3\rho}^{V}(y_2,y_1;y_3) .
\end{equation}
In our definition, the normalization constants $f_{3\rho}^A=0.6\ \cdot
10^{-2}\,\mbox{GeV}^2,
f_{3\rho}^V=0.25\ \cdot 10^{-2}\,\mbox{GeV}^2$ \cite{CZ84} are factored out,
so that the  distribution amplitudes are normalized to unity:
\begin{equation}
\int_0^1 [dy]_3\ \varphi_{3\rho}^{A}(y_i) = 1 \ , \
\int_0^1 [dy]_3\ (y_1-y_2)\,\varphi_{3\rho}^{V}(y_i) = 1 \ .
\label{eq:SE21}
\end{equation}

Following the procedure described in  Sec.2.1,  we find the
$\rho^o$-meson contribution:
\begin{eqnarray}
\lefteqn{\Phi_{\rho}^{B(2)} =
\frac{8\pi}{3}\,\frac{f_{\rho}^V m_{\rho}}{m_{\rho}^2+q^2}\,
\int_0^1 d\alpha\,\alpha\,\int_0^1 d\beta\,\int_0^1 [dy]_3\
e^{\ b/{a M^2}}  } \label{eq:SE22}\\
&\times & \left\{f_{3\rho}^A\,\varphi_{3\rho}^{A}(y_1,y_2;y_3)\,
\left[\,\frac{c_1}{a^2 M^4} - \frac{d_1}{2 a^3 M^6}\,\right] -
  f_{3\rho}^V\,\varphi_{3\rho}^{V}(y_1,y_2;y_3)\,
\left[\,\frac{c_2}{a^2 M^4} - \frac{d_2}{2 a^3 M^6}\,\right]\right\},
\nonumber
\end{eqnarray}
where
\begin{eqnarray}
a&=&\alpha \beta y_3 + y_2  , \nonumber\\
b&=&-q^2\,(\alpha^2\beta y_3^2 + 2\alpha\beta y_2 y_3 - \alpha\beta y_3 + y_2^2 - y_2)
+ Q^2\,(\alpha\beta y_3 + y_2 - 1)
\label{eq:SE23}
\end{eqnarray}
and
\begin{eqnarray}
c_1&=&(\alpha \beta y_3)/(\alpha \beta y_3 + y_2) , \nonumber\\
d_1&=&c_1\,\left(\,
-q^2\,(\alpha^2\beta y_3^2 + 2\alpha y_2 y_3 + \alpha\beta y_3 + y_2^2 + y_2)
+ Q^2\,\right) , \nonumber\\
c_2&=&(\alpha \beta y_3 + 2\beta y_2)/(\alpha \beta y_3 + y_2) , \nonumber\\
d_2&=&c_1\,\left(\,
q^2\,(\alpha^2\beta y_3^2 + 2\alpha\beta y_2 y_3 + \alpha\beta y_3 +2\beta y_2^2 - y_2^2
+ y_2)
+ Q^2\,(1-2\beta)\,\right) .
\label{eq:SE231}
\end{eqnarray}

The perturbative spectral density for all of these correlators
is  suppressed by  $O(\alpha_s/\pi)$-factor, and for this reason
we neglect here the contribution due to  higher  states.

Taking  $q^2 = 0$, we get
\begin{eqnarray}
&&a=\alpha \beta y_3 + y_2  ,\quad b=-Q^2\,(1-a) \\
&\mbox{and}&\nonumber \\
&&c_1=(\alpha \beta y_3)/a , \quad d_1=c_1\,Q^2 , \nonumber\\
&&c_2=(\alpha \beta y_3 + 2\beta y_2)/a , \quad d_2=c_1\,Q^2\,(1-2\beta) .
\label{eq:SE230}
\end{eqnarray}
Introducing the variables:
\begin{equation}
s=Q^2(1-a)/a = Q^2 \frac{1-y_2-\alpha\beta y_3}{y_2+\alpha\beta y_3},\
v(s)=\frac{Q^2}{s+Q^2}
\end{equation}
and integrating over $\beta,\alpha$, we obtain  the representation
\begin{eqnarray}
\lefteqn{
\Phi_{\rho}^{B(2)}  =
\frac{8\pi}{3}\,\frac{f_{\rho}^V}{m_{\rho}}\,
\int_0^{\infty} ds\ e^{-s/M^2}\,\frac{(s+Q^2)}{Q^4 M^4}\,
\int_0^1 [dy]_3\ \theta(y_2\le v(s))\ \frac{(v(s)-y_2)}{y_3^2}\times
} \nonumber\\
&&\nonumber\\
&&\left\{f_{3\rho}^A\,\varphi_{3\rho}^{A}(y_1,y_2;y_3)\,
(y_2+y_3-v(s))\,\left[1-\frac{(s+Q^2)}{2 M^2}\right]  \right.
\label{eq:SE22q0}
\\
&& \nonumber\\
&&{}-f_{3\rho}^V\,\varphi_{3\rho}^{V}(y_1,y_2;y_3)\,
\left[\left(y_2+y_3-v(s)-2y_2 \ln\left(\frac{v(s)-y_2}{y_3}\right)\right)-{}
\right.\nonumber\\
&& \nonumber\\
&&{}-\left.\left.\frac{(s+Q^2)}{2 M^2}\left(y_2+y_3-v(s)+2(v(s)-y_2)
  \ln\left(\frac{v(s)-y_2}{y_3}\right)\right)\right]\right\}
\nonumber
\end{eqnarray}

To estimate  these contributions, we  use the asymptotic
forms of the corresponding three-body $\rho$-meson distribution amplitudes \cite{CZ84}:
\begin{eqnarray}
& &\varphi_{3A}(y_1 y_2 y_3) \to
 \varphi_{3A}^{as} (y_1 y_2 y_3) = 360 y_1 y_2 y_3^2,\nonumber  \\
& &\varphi_{3V} (y_1 y_2 y_3) \to \varphi_{3V}^{as}(y_1 y_2 y_3) =
7! (y_1-y_2)y_1 y_2 y_3^2 .
\label{eq:3particlas}
\end{eqnarray}

\subsection{Twist-2 bilocals for  three-propagator coefficient functions}

The bilocals associated with the coefficient functions given
by a product  of three propagators  can appear
in  the ${\langle\bar{\psi}\psi\rangle}^2$ quark condensate
diagrams of the unmodified  OPE   (see Figs.\ref{eq:fig4}$a-r$).
Furthermore,  for large and moderate
$q_1^2$, only the diagrams  \ref{eq:fig4}$a-d$ contribute to the invariant amplitude
$F$ we are interested in.
 So, let us  consider them  first.
In fact, among these diagrams, only \ref{eq:fig4}$b$ and \ref{eq:fig4}$c$
produce bilocals  with the three-propagator
coefficient function.
After some algebra, we  obtain
\begin{eqnarray}
\lefteqn{ {\cal{F}}_{\alpha\mu\nu}^{B(3)}=
\frac{32 \pi^2 \alpha_s \langle\bar{u} u\rangle}{27}\,
\int d^4Y e^{ip  Y} \left(\frac{p_{\alpha}}{p^2}\right)\,
\frac{Y_{\beta}}{8\pi^2 Y^2}
\sum_{n=0}^{\infty}\,\frac{1}{n!}\ Y^{\mu_1}\ldots Y^{\mu_n} }\nonumber\\
& &\times\int d^4X e^{-i q_1 X}\ \langle 0 |T\{J_{\mu}(X)\,
\bar{u}(0)({{\partial}}_{\mu_1}\ldots{{\partial}}_{\mu_n})
\gamma_{\nu}\gamma_{\beta}\gamma_5 u(0)\}| 0 \rangle .
\label{eq:SE24}
\end{eqnarray}
 Adding the charge conjugate contribution produces  an extra factor of 2.
The correlator which appeared in eq.(\ref{eq:SE24})
can also be treated as  a distribution amplitude $\varphi_{\gamma^*}(y, q^2)$
of a photon  with virtuality $q_1^2 = -q^2$.
The perturbative spectral density for this correlator is zero,
so the natural approximation  is
to model  $\varphi_{\gamma^*}(y, q^2)$
by explicit contributions from lowest resonances:
\begin{equation}
\varphi_{\gamma^*}(y, q^2) = \frac{m_{\rho}\,f_{\rho}^V f_{\rho}^{T}}{m_{\rho}^2 + q^2}
\varphi_{\rho}^{T}(y) + \frac{m_{\rho'}\,f_{\rho'}^V f_{\rho'}^{T}}{m_{\rho'}^2 + q^2}
\varphi_{\rho'}^{T}(y) + \ldots \  ,
\label{eq:rhotens}
\end{equation}
where the
$\rho$-contribution is determined by the  matrix element
\begin{equation}
\langle
0|\bar{\psi}(0)\sigma_{\nu\beta}\psi(Y) |
\rho_{\lambda}^o;\stackrel{\rightarrow}{p}\rangle\
= i\,(\,\varepsilon_{\nu}^{(\lambda)}\, p_{\beta} - \varepsilon_{\beta}^{(\lambda)}\, p_{\nu}\,)
\ f_{\rho}^{T}\,
\phi_{\rho}^{T}(Yp,\mu^2) + \mbox{higher twists} \  .
\label{eq:SE25}
\end{equation}
Here, as usual,  $\sigma_{\nu\beta}=\frac{i}{2}[\gamma_{\nu},\gamma_{\beta}],\ p=q_1+q_2$.
For the SVZ-transform $\Phi^{\gamma^*(3)}$ we thus get:
\begin{eqnarray}
 \Phi^{\gamma^*(3)}= 
-\frac{ 64\pi^2}{27}
\frac{\alpha_s \langle\bar{q} q\rangle}{M^6}\,
\int_0^1\!\!\int_0^1 dy d\beta\,e^{\,\beta(q^2 y\bar{y} - Q^2 y)/((1-y\beta)M^2)}
\frac{\beta\,\varphi_{\gamma^*}(y,q^2)}{(1-y\beta)^3},
\label{eq:SE26}
\end{eqnarray}
The distribution amplitude of the real photon
$\varphi_{\gamma}(y) \equiv \varphi_{\gamma^*}(y,q^2=0)$
was analyzed in ref.\cite{BBK}. It was argued there
that $\varphi_{\gamma}(y)$ has the ``asymptotic shape'':
\begin{equation}
\varphi_{\gamma}(y) = -\frac{2 k}{m_{\rho}^2} \langle\bar{q} q\rangle
\phi_{\gamma}(y)
\end{equation}
with $k \approx 1.3$ and $\phi_{\gamma}(y) = 6y(1-y)$
($\phi_{\gamma}(y)$ is the normalized photon  distribution amplitude
$i.e.,$ its zeroth moment equals 1).
 Taking $q^2=0$ and introducing  the variable
 $s=Q^2 y\beta/(1-y\beta)$, we then get:
\begin{equation}
\Phi^{\gamma (3)} =
\frac{128\pi^2}{27}
\frac{k \alpha_s {\langle\bar{q} q\rangle}^2}{M^6 Q^4 m_{\rho}^2}\,
\int_0^{\infty} \,e^{-s/M^2}\  g_{\gamma }(s) \, ds \,  ,
\label{eq:SE26q0}
\end{equation}
where
\begin{equation}
g_{\gamma }(s) = s \int \limits_{s/{s+Q^2}}^1\,dy\,
\frac{\phi_{\gamma}(y)}{y^2} .
\label{eq:gtensor}
\end{equation}
 For the  model
$\phi_{\gamma}(y) = 6y(1-y)$ advocated in \cite{BBK},  we obtain
\begin{equation}
g_{\gamma }(s) =  6s \left (\ln \frac{Q^2+s}{s} - \frac{Q^2}{Q^2+s} \right )  \,  .
\label{eq:gtensoras}
\end{equation}
Using the formulas (\ref{eq:specden1}),(\ref{eq:specden2}),
we can convert  the expression (\ref{eq:SE26q0}) into  the canonical form.
 Note, however, that  the second derivative of $g_{\gamma }(s)$
contains the $1/s$-singularity, and one  should be careful
when calculating the relevant spectral density  $\rho^{\gamma}(s, Q^2)$.
The simplest  procedure is to represent the
$\ln s$ term as $\lim_{\lambda^2 \to 0} \ln (s + \lambda^2)$.
Then application of eq.(\ref{eq:specden2}) is straightforward, and we get
\begin{equation}
\Phi^{\gamma} =
\frac{128\pi^2}{27}
\frac{\alpha_s {\langle\bar{q} q\rangle}^2}{M^2 Q^4 m_{\rho}^2}\,
\lim_{\lambda^2 \to 0} \left [ \ln \frac{Q^2}{\lambda^2}  -2
+ \int_0^{\infty} e^{-s/M^2} \left (
\frac{s^2+3sQ^2+4Q^4} {(s+Q^2)^3} - \frac1{s+\lambda ^2}
\right) ds \right ] .
\label{eq:fitensor}
\end{equation}
This trick amounts to using an alternative regularized form for the
$(1/s)_+$-distribution usually defined by
\begin{equation}
\int_0 f(s)  \left (\frac{1}{s_+} \right )  \, ds =
\int_0 \frac{f(s) - f(0)}{s} ds .
\label{eq:splus}
\end{equation}

\section{Bilocals and contact terms}

\setcounter{equation} 0

A special care must be taken about the correlators
containing the Dirac  operator $\gamma_{\mu}D^{\mu}$ acting on the
quark field $\psi$. Since the correlator is a $T$-product
of the electromagnetic current and a composite operator,
applying the equation of motion  one gets
the  $\delta^{(4)}(X)$-function, $i.e.,$  the external vertices
of the bilocal are  contracted into a single point and it
reduces to a  $q^2$-independent constant.

Let us sketch  a simple derivation for such terms
(see, $e.g.,$ \cite{BalDY}).
Using the functional representation for the  correlator
\begin{equation}
\langle 0|T\{\ldots\bar{\psi}(X)\hat{\nabla}\psi(0)\} |0\rangle
= \int\,{\cal{D}}[\bar{\psi}]\,{\cal{D}}[\psi]\,{\cal{D}}[A]\,
\{\ldots\bar{\psi}(X)\hat{\nabla}\psi(0)\}\,
\exp{\left(\,i\int\,{\cal{L}}(Z)\,d^4Z\,\right)} ,
\label{eq:ct1}
\end{equation}
where ${\cal{L}}(Z)=\bar{\psi}(Z)\,i\hat{\nabla}\psi(Z)+\cdots$ \,  ,
we can write
\begin{equation}
\hat{\nabla}\psi(0)\,\exp{\left(\,i\int\,{\cal{L}}(Z)\,d^4Z\,\right)}
=-\frac{\delta}{\delta\bar{\psi}(0)}\,
\exp{\left(\,i\int\,{\cal{L}}(Z)\,d^4Z\,\right)} .
\label{eq:ct2}
\end{equation}
Integrating by parts in (\ref{eq:ct1}) results in
\begin{equation}
\int\,{\cal{D}}[\bar{\psi}]\,{\cal{D}}[\psi]\,{\cal{D}}[A]\,
\left\{\cdots\frac{\delta\bar{\psi}(X)}{\delta\bar{\psi}(0)}\right\}\,
\exp{\left(\,i\int\,{\cal{L}}(Z)\,d^4Z\,\right)} .
\label{eq:ct3}
\end{equation}
It is the  derivative $\delta\bar{\psi}(X) / \delta\bar{\psi}(0)$
that produces the $\delta^{(4)}(X)$ term mentioned above.

The contact terms play an important role
in  all applications of the QCD sum rules to low-momentum behaviour
of hadronic form factors. In particular,  without them,
it is impossible to satisfy the Ward identities
fixing the pion form factor normalization
at zero momentum  transfer \cite{NeRa84b,BeiNeRa88}.

\subsection{Separating short-  and long-distance contributions}

Consider the  hard gluon exchange diagrams shown in
Figs.\ref{eq:fig4}$a-r$  which produce, in the $B$-regime,
the  bilocals associated with the
three-propagator coefficient function.
Take first  the diagrams \ref{eq:fig4}$e$,$f$.
Their  contributions are
\begin{equation}
5e\,\sim \frac{32i}{p^2q_1^2q_2^4}\ q_{1\mu}\,\epsilon_{\alpha\nu q_1q_2}\qquad , \quad
5f\,\sim \frac{32i}{p^4q_1^2q_2^2}\ q_{1\mu}\,\epsilon_{\alpha\nu q_1q_2} , \nonumber\\
\end{equation}
$i.e.$, they do not contribute to the
invariant form factor $F$. However, as we will see below,
in the  B-regime,  equations of motion ``extract''
the appropriate tensor structure, $i.e.,$
these diagrams cannot be ignored.
The relevant  term from the three-point correlation
function  can be written  as\footnote{To get the total
contribution, one should add also the
charge conjugate term.}
\begin{eqnarray}
{\cal{F}}_{\alpha\mu\nu}^{B,5ef}(q_1,q_2)&=&
 -\frac{32\pi^2 }{9}\frac{\alpha_s\langle\bar{u}u\rangle}{p^2q_2^2}\,
\epsilon_{\alpha\nu q_1q_2}\
\int d^4Z\,e^{ipZ}\,\frac{1}{4\pi^2\,Z^2}
\sum_{n=0}^{\infty}\,\frac{1}{n!}\ Z^{\mu_1}\ldots Z^{\mu_n} \nonumber\\
& &\times\int d^4X e^{-i q_1 X}\ \langle 0 |T\{J_{\mu}(X)\,
\bar{u}(0)({{\partial}}_{\mu_1}\ldots{{\partial}}_{\mu_n})
u(0)\}| 0 \rangle .
\label{eq:ct4}
\end{eqnarray}
Extracting the  bilocal term from (\ref{eq:ct4}),
one should pick out the traceless combination
of indices $\mu_1,\ldots\mu_n$, $i.e.,$ the
lowest-twist term which gives
the leading power contribution with respect to  $1/p^2,1/q_2^2$.
Introducing  the notation
\begin{equation}
\Pi_{\mu\{\mu_1\ldots\mu_n\}}(q_1)=\int d^4X e^{-i q_1 X}\
\langle 0 |T\{J_{\mu}(X)\,
\bar{u}(0)\{{{\partial}}_{\mu_1}\ldots{{\partial}}_{\mu_n}\}
u(0)\}| 0 \rangle ,
\label{eq:ct5}
\end{equation}
we can represent the correlator (\ref{eq:ct5})  in the following form:
\begin{eqnarray}
\Pi_{\mu\{\mu_1\ldots\mu_n\}}(q_1)&=&
A^{(n)}(q_1^2)\,q_{1\mu}\,{\{q_{1\mu_1}\ldots q_{1\mu_n}\}}+
B^{(n)}(q_1^2)\,{\{q_{1\mu},q_{1\mu_1}\ldots q_{1\mu_n}\}}+\nonumber\\
& &\qquad+C^{(n)}(q_1^2)\,g_{\mu}{\{{}_{\mu_1}q_{1\mu_2}\ldots q_{1\mu_n}\}}  ,
\label{eq:ct6}
\end{eqnarray}
where ${\{\ldots\}}$ denotes the traceless-symmetric part of a tensor.
Because of the electromagnetic current conservation,
we have the constraint
$q_{1\mu}\,\Pi_{\mu\{\mu_1\ldots\mu_n\}}(q_1)=0$ which
produces some relations between  the invariant
functions $A^{(n)},B^{(n)},C^{(n)}$. Using the formula from \cite{GutSop}
\begin{equation}
q_1^{\alpha}\,{\{q_{1\mu_1}\ldots q_{1\mu_{n-1}},q_{\alpha}\}} =
q_1^2\,\frac{n+1}{2n}\,{\{q_{1\mu_1}\ldots q_{1\mu_{n-1}}\}} \, ,
\end{equation}
we obtain
\begin{equation}
\left(A^{(n)}+\frac{(n+2)}{2(n+1)}\,B^{(n)}\right)\,q_1^2 + C^{(n)} = 0 .
\label{eq:ct7}
\end{equation}

Contracting  (\ref{eq:ct5}) with $g_{\mu\mu_1}$  gives
\begin{equation}
\Pi_{\mu\{\mu_1\ldots\mu_n\}}(q_1)\,g_{\mu\mu_1}=
\left(A^{(n)}\,q_1^2\,\frac{(n+1)}{2n} +
C^{(n)}\,\left(\frac{n+1}{n}\right)^2\right)\,
{\{q_{1\mu_2}\ldots q_{1\mu_n}\}}.
\label{eq:ct8}
\end{equation}
Furthermore, applying the technique  symbolized by  eqs. (\ref{eq:ct1}) -
(\ref{eq:ct3}) (see Appendix E) we obtain, in  the leading-twist approximation:
\begin{eqnarray}
\lefteqn{ \Pi_{\mu\{\mu_1\ldots\mu_n\}}(q_1)\,g_{\mu\mu_1}\simeq
(-i)^{n-1}\,{\{q_{1\mu_2}\ldots q_{1\mu_n}\}}\times }\nonumber\\
& &\times\left[-2\langle\bar{u}u\rangle
- \frac{1}{2}\,q_{1\varepsilon}\int d^4X e^{-i q_1 X}\
\langle 0 |T\{J_{\mu}(X)\,
\bar{u}(0)\{{{\partial}}_{\mu_1}\ldots{{\partial}}_{\mu_n}\}
\sigma_{\mu\varepsilon}u(0)\}| 0 \rangle\right] ,
\label{eq:cont9}
\end{eqnarray}
where the first contribution
inside the brackets is just the contact term,
while the second one contains the correlator
related to the photon distribution amplitude
$\varphi_{\gamma^*}(y,q^2)$ (\ref{eq:rhotens}):
\begin{eqnarray}
\Pi_{\mu\{\mu_1\ldots\mu_n\}}(q_1)\,g_{\mu\mu_1}=
(-i)^{n-1}\,{\{q_{1\mu_2}\ldots q_{1\mu_n}\}}
\left[-2\langle\bar{u}u\rangle
-\frac{3}{2}\,q^2\,
\int\, dy\, y^{n-1}\,\varphi_{\gamma^*}(y,q^2)\right].
\label{eq:cont10}
\end{eqnarray}
 Note that the term containing $\varphi_{\gamma^*}(y,q^2)$
vanishes for $q^2=0$.
Using  the formulas   from  Appendix D,  one can easily
perform the necessary contractions:
\begin{eqnarray}
\lefteqn{
Z_{\mu_1}\ldots Z_{\mu_n}\,
\Pi_{\mu\{\mu_1\ldots\mu_n\}}  =
A^{(n)}\,q_{1\mu}\,{\tau}^n\,C_n^{1}(\eta)
} \nonumber\\ & &
+ \frac{B^{(n)}}{n+1}\,
\left[-Z_{\mu}\frac{q_1^2}{2}\,{\tau}^{n-1}\,C_{n-1}^{2}(\eta) +
q_{1\mu}\,{\tau}^n\,C_n^{2}(\eta) \right] \nonumber\\
& & + \frac{C^{(n)}}{n}\,
\left[Z_{\mu}\,{\tau}^{n-1}\,C_{n-1}^{2}(\eta) -
q_{1\mu}\,\frac{Z^2}{2}\,{\tau}^{n-2}\,C_{n-2}^{2}(\eta) \right] ,
\label{eq:ct11}
\end{eqnarray}
where $C_n^{\lambda}(\eta)$ are the Gegenbauer polynomials and the
notation  $\eta=i\,(q_1Z)/ \sqrt{-Z^2q_1^2}$,
$\tau= -i \sqrt{-Z^2q_1^2}/2$ is introduced.

The tensor structure  ($\sim p_{\alpha}\,\epsilon_{\mu\nu q_1q_2}$)
we are interested in,
can be only produced in (\ref{eq:ct11}) by  the terms $\sim Z_{\mu}$.
Hence, other terms  can be ignored.
Combining now eqs. (\ref{eq:ct7}) -- (\ref{eq:cont10}), we get,
modulo the next-to-leading twist contributions:
\begin{eqnarray}
\lefteqn{ Z_{\mu}\,{\tau}^{n-1}\,C_{n-1}^{2}(\eta)\,
\left[ -\frac{q_1^2\,B^{(n)}}{2(n+1)} + \frac{C^{(n)}}{n} \right]\simeq }
\\
&\simeq& Z_{\mu}(-i\,q_1Z)^{n-1}\,\frac{2n^2}{(n+1)(n+2)}\,
\left[-2\langle\bar{u}u\rangle
- \frac{3}{2}\,q^2\,
\int\, dy\, y^{n-1}\,\varphi_{\gamma^*}(y,q^2)\right]. \nonumber
\label{eq:ct12}
\end{eqnarray}

Substituting (\ref{eq:ct12}) into (\ref{eq:ct4}),
integrating  over $d^4Z$ and summing
over $n$ by  using the generating function technique we get for
the  SVZ-transform of the contact terms:
\begin{equation}
\Phi^{5ef(C)} =
 -\frac{256\pi^2}{27}
\frac{\alpha_s{\langle\bar{q}q\rangle}^2}{Q^2 M^6}
\int\!\!\int_0^1\,d\beta dy\,
\frac{\beta(\bar{\beta}-\beta)y}{(1-y\beta)^3}\
e^{y\beta(q^2\bar{\beta}-Q^2)/((1-y\beta)M^2)}.
\label{eq:ct13}
\end{equation}
The non-contact terms give
\begin{equation}
\Phi^{5ef(\gamma^*)}=
 -\frac{256\pi^2}{27}
\frac{\alpha_s{\langle\bar{q}q\rangle}}{Q^2 M^6}\,
\frac{3}{4}\,q^2\,
\int\!\!\int\!\!\int_0^1\,d\alpha d\beta dy\,
\frac{y\alpha\beta(\overline{y\beta}-y\beta)}{(1-y \alpha \beta)^3}\
\varphi_{\gamma^*}(y,q^2)\
e^{y \alpha \beta(q^2\overline{y\beta}-Q^2)/
(1-y \alpha \beta)M^2}\label{eq:ct14}
\end{equation}
where  $\overline{y\beta} \equiv 1-y\beta,  \bar \beta \equiv 1- \beta$ .
Analyzing the remaining  bilocal contributions capable of
producing a coefficient function
of the three-propagator type (see diagrams of Fig.\ref{eq:fig4}
$b,c,i,l,o,p $), we found that they do not contain the
contact  terms  with   the   tensor structure
$p_{\alpha}\,\epsilon_{\mu\nu q_1q_2}$.
Furthermore, there are no contact terms in  the bilocals
corresponding to one- and
two-propagator coefficient functions.

For $q^2=0$,  the contribution of the non-contact terms
vanishes while the contact terms give:
\begin{equation}
\Phi^{C}(Q^2,M^2) = - \frac{256\pi^2
\alpha_s \langle \bar{q}q\rangle ^2}{27   Q^6  M^6}
\int_0^{\infty} e^{-s/M^2}
\left [ \ln{\frac{s+Q^2}{s}} - 2 \frac{Q^2}{s+Q^2} \right ]s ds  .
\label{eq:contactq0}
\end{equation}
 Representing again
$\ln s = \lim_{\lambda^2 \to 0} \ln (s + \lambda^2)$  and
using  eq.(\ref{eq:specden2}),  we obtain
\begin{equation}
\Phi^{C}(Q^2,M^2) = - \frac{256\pi^2
\alpha_s \langle \bar{q}q\rangle ^2}{27   Q^6  M^2}
\lim_{\lambda^2 \to 0} \left [ \ln \frac{Q^2}{\lambda^2}  -3
+ \int_0^{\infty} e^{-s/M^2} \left (
\frac{s^2+3sQ^2+6Q^4} {(s+Q^2)^3} - \frac1{s+\lambda ^2}
\right) ds \right ] .
\label{eq:ficont}
\end{equation}


\section{QCD sum rule in the
small-$q^2$ kinematics}

\setcounter{equation} 0

Collecting  now all the contributions, we obtain
the theoretical part (the modified OPE) of the  QCD sum rule
 for the form factor
\mbox{$F_{\gamma^*\gamma^*\rightarrow\pi^\circ}$}$(q^2,Q^2)$
(see Fig.\ref{eq:fig8}):
\begin{eqnarray}
\lefteqn{ \Phi(q^2,Q^2,M^2) = \Phi^{PT}(q^2,Q^2,M^2) +
\Phi_{3b}^{\langle GG \rangle} + \Phi_{3c}^{\langle GG \rangle} +
\Phi_{3f,g}^{\langle GG \rangle} + }\nonumber\\
& & + \Phi_{4g,h,i}^{\langle\bar{q}q\rangle} +
\Phi_{4d,e,f}^{\langle\bar{q}q\rangle} +
\Phi_{5a,b,c}^{\langle\bar{q}q\rangle} +
\Phi_{5d}^{\langle\bar{q}q\rangle} + \nonumber\\
& & + \Phi^{B(\rho)} +
\left(\,\Phi^{B(cont)}-
\Phi^{B(PT)} \right) -
\Phi_{3b}^{B} - \Phi_{3c}^{B} -
\Phi_{4d,f}^{B}+ \nonumber\\
& & + \Phi^
{\rho(2)} -
\Phi_{3f,g}^{B} - \Phi_{4e}^{B} -
\Phi_{5a}^{B} + \nonumber\\
& & + \Phi^{\gamma(3)} -
\Phi_{5b,c}^{B} +
\left[\,\Phi^{5ef(C)} +
\Phi^{5ef(\gamma^*)}\,\right],
\label{eq:mainsr}
\end{eqnarray}
where the first two rows correspond to the original OPE
valid for symmetric kinematics.
Each of the next rows represents the additional terms corresponding
to different types of the coefficient functions.
As explained in Section 4,  all the terms  of the
standard OPE, which are non-analytic in  the
$q^2 \to  0$ limit,  are cancelled  by  the
corresponding B-contributions. As a result,
 the coefficient functions of
the SD-regime are  analytic functions of $q^2$
(cf. \cite{tkachev}).
Substituting explicit expressions for all the terms
which appear in eq.(\ref{eq:mainsr}) gives the
following expression for the SVZ-Borel transformed
OPE for the three-point correlator valid in the region
of small $q^2$:
 \begin{eqnarray}
&\displaystyle \Phi(q^2,Q^2,M^2) = \frac{1}{\pi M^2} \left\{ \int_0^1 dy \ e^{{-Q^2 y}/{M^2 \bar{y}}}\,
\left\{(1+\frac{q^2y}{M^2}\,e^{{q^2 y}/{M^2}}) \right. \right. +&  \nonumber\\
& &  \nonumber\\
&\displaystyle + e^{{q^2 y}/{M^2}}\,\left[\,\frac{2y}{M^2}\,
\left(q^2\ln{\frac{(s_{\rho}+q^2)y}{M^2}} - s_{\rho}\right) +
\frac{y^2}{M^4}\,\left(q^4\ln{\frac{(s_{\rho}+q^2)y}{M^2}} - q^2 s_{\rho}
+\frac{s_{\rho}^2}{2}\right)\ \right] - &\nonumber\\
& &  \nonumber\\
&\displaystyle -\left.\sum_{n=1}^{\infty}\,\left(\frac{q^2y}{M^2}\right)^n
\frac{\psi(n)(n+1)}{(n-1)!} \right\} +& \nonumber\\
& &  \nonumber\\
&\displaystyle + \frac{\pi^2}{9}\,
\langle\frac{\alpha_s}{\pi} GG \rangle\,
\left[\,\frac{1}{2M^2Q^2} + \frac{1}{M^4}\,
\int_0^1 dy\,\frac{y}{{\bar{y}}^2}\,e^{-Q^2y/M^2\bar{y}}\,
\sum_{n=1}^{\infty}\,
\frac{1}{n!}\left(\frac{q^2 y}{M^2}\right)^{n-1}\,\right]
+& \nonumber\\
& &  \nonumber\\
&\displaystyle + \frac{64\pi^3}{243}\,\alpha_s
{\langle\bar{q}q\rangle}^2\,\frac{q^2}{Q^4M^2}\ + \
\frac{64\pi^3}{27}\,\alpha_s
{\langle\bar{q}q\rangle}^2\,\frac{1}{2Q^2M^4}\ +& \nonumber\\
& &  \nonumber\\
& &  \nonumber\\
&\displaystyle + \frac{4\pi^2}{3}
\frac{f_{\rho}^V m_{\rho}}{m_{\rho}^2+q^2}\, \int_0^1 dy
\frac{1}{\bar{y}^2 M^2}\, e^{{-Q^2 y}/{M^2 \bar{y}}}\,e^{{q^2 y}/{M^2}}&
\nonumber\\
&\displaystyle \times
\left[\,-a_{V1}\,f_{\rho}^V m_{\rho}\left(\varphi_{\rho_{\bot}}^{V1}(y) -
\frac{4 C_{[V1]4}}{\bar{y} M^2}\,\varphi_{\rho_{\bot}}^{[V1]4}(y)\right)
        - f_{\rho}^A (1+2\bar{y})\left(\varphi_{\rho_{\bot}}^{A}(y) -
\frac{4 C_{A4}}{\bar{y} M^2}\,\varphi_{\rho_{\bot}}^{A4}(y)\right)\ \right]&
\nonumber\\
& &  \nonumber\\
& &  \nonumber\\
&\displaystyle +\frac{8\pi^2}{3}\,\frac{f_{\rho}^V m_{\rho}}{m_{\rho}^2+q^2}\,
\int_0^1 d\alpha\,\alpha\,\int_0^1 d\beta\,\int_0^1 [dy]_3\
e^{\ b/{a M^2}} &  \nonumber\\
&\displaystyle \times
\left\{f_{3\rho}^A\,\varphi_{3\rho}^{A}(y_1,y_2;y_3)\,
\left[\,\frac{c_1}{a^2 M^2} - \frac{d_1}{2 a^3 M^4}\,\right] -
  f_{3\rho}^V\,\varphi_{3\rho}^{V}(y_1,y_2;y_3)\,
\left[\,\frac{c_2}{a^2 M^2} - \frac{d_2}{2 a^3 M^4}\,\right]\right\}&
\nonumber\\
& &  \nonumber\\
& &  \nonumber\\
&\displaystyle - \frac{64\pi^3}{27}
\frac{\alpha_s \langle\bar{q} q\rangle}{M^4}\,
\int\!\!\int_0^1 dy d\beta\,
\frac{\beta\,\varphi_{\gamma^*}(y,q^2)}{(1-y\beta)^3}\
e^{\beta(q^2 y\bar{y} - Q^2 y)/((1-y\beta)M^2)} &
\nonumber\\
& &  \nonumber\\
& &  \nonumber\\
&\displaystyle - \frac{256\pi^3}{27}
\frac{\alpha_s{\langle\bar{q}q\rangle}^2}{Q^2 M^4}\,
\int\!\!\int_0^1\,d\beta dy\,
\frac{\beta(\bar{\beta}-\beta)y}{(1-y\beta)^3}\
e^{y\beta(q^2\bar{\beta}-Q^2)/((1-y\beta)M^2)} &
\nonumber\\
& &  \nonumber\\
& &  \nonumber\\
&\displaystyle - \frac{64\pi^3}{9}
\frac{\alpha_s{\langle\bar{q}q\rangle}}{Q^2 M^4}\,
q^2\,
\left.\int\!\!\int\!\!\int_0^1\,d\alpha d\beta dy\,
\frac{y\alpha\beta(\overline{y\beta}- y\beta)}{(1-y \alpha\beta)^3}\
\varphi_{\gamma^*}(y,q^2)\
e^{y \alpha \beta(q^2\overline{y\beta}-Q^2)/(1-y\alpha \beta)M^2} \,\right\}  . &
\label{eq:modifope}
\end{eqnarray}

This expression can be used as a starting point
for constructing sum rules for the form factor
$F_{\gamma^*\gamma^*\pi^\circ}(q^2,Q^2)$ at $q^2=0$,
its derivative $\frac{\partial}{\partial q^2}
F_{\gamma^*\gamma^*\pi^\circ}(q^2,Q^2)|_{q^2=0}$
or even
for studying the $q^2$-dependence of
$F_{\gamma^*\gamma^*\pi^\circ}(q^2,Q^2)$ in the region of
small $q^2 \lapprox \, m_{\rho}^2$ for fixed values
of the large virtuality $Q^2 \gapprox \, 2\, GeV^2$.

For  the constants
present in eq.(\ref{eq:modifope}), we use the following  numerical values:
$f_{\rho}^V = 0.2\ \mbox{GeV}, m_{\rho} = 0.77\ \mbox{GeV}$ ; the constants
$f_{\rho}^A = -f_{\rho}^V\,m_{\rho}/4,\quad a_{V1} = 1/40$ are  obtained
from the equations of motion (see Appendix C), the values
$f_{3\rho}^A=0.6\ \cdot10^{-2}\,\mbox{GeV}^2,
f_{3\rho}^V=0.25\ \cdot 10^{-2}\,\mbox{GeV}^2$
are taken from the QCD sum rule estimates given in ref. \cite{CZ84},
and the value $k =1.3$ for the photon distribution amplitude
(\ref{eq:rhotens}) was taken from ref.\cite{BBK}.
The quark and gluon condensate values are standard:
$\langle (\alpha_s / \pi) GG \rangle = \, 0.012 \, GeV^4$,
$\alpha_s  \langle \bar q q \rangle^2  = \, 1.8\cdot 10^{-4} \, GeV^6$.
For the continuum threshold in the $\rho$-channel
we take the standard  value $s_{\rho}\simeq 1.5\, \mbox{GeV}^2$
obtained from
the QCD sum rule for the $\rho$-decay constant
$f_{\rho}^V$ \cite{svz}.
To estimate relative importance of various contributions, we  take
the asymptotic forms for $\varphi_{3V}(y_1,y_2,y_3)$,
$\varphi_{3A}(y_1,y_2,y_3)$ (eq.(\ref{eq:3particlas})) and
$\varphi_{V1}(y)$, $\varphi_A(y)$  (eqs.(\ref{eq:aswfVA}),(\ref{eq:aswfV1})).

Our numerical analysis shows that the  most important contributions in
(\ref{eq:modifope} come from:
$a)$ SD-regime (first five rows of (\ref{eq:modifope})),
$b)$ term with  the nondiagonal correlator (photon distribution
amplitude $\varphi_{\gamma}(y)$) and
$c)$ $\rho^o$-meson contribution with the leading twist wave functions
 (B-regime) in the diagonal correlator.
The terms associated with  three-particle twist-3 wave functions
are small: their contribution into the sum rule is
of the order  of a few percent.  We evaluated the
terms  corresponding to the
 next-to-leading two-particle wave functions
(of twist-4) and observed that they are also  suppressed.
The contact-type power corrections are  small as well.

Below,  we  consider only the simplest sum rule
for $F_{\gamma^*\gamma^*\pi^\circ}(q^2=0,Q^2)$,
keeping in it only the  most important terms listed above and
 contact terms.
All the necessary expressions
substituting  the terms from eq.(\ref{eq:modifope})
by their  $q^2=0$  limit were given in the preceding sections.
Combining them together, we  obtain the  QCD sum rule
for the $\gamma \gamma^* \to \pi^0$ form factor:
\begin{eqnarray}
& & \pi f_{\pi} F_{\gamma \gamma^* \pi^0}(Q^2) =
\int_0^{s_0}
\left \{ 
1 - 2 \frac{Q^2-2s}{(s+Q^2)^2}
\left (s_{\rho} - \frac{s_{\rho}^2}{2 m_{\rho}^2} \right )
\right.  \nonumber \\
& & + \left. 2\frac{Q^4-6s Q^2+3s^2}{(s+Q^2)^4} \left (\frac{s_{\rho}^2}{2}
 - \frac{s_{\rho}^3}{3  m_{\rho}^2} \right )
\right \} 
 e^{-s/M^2}
\frac{Q^2 ds }{(s+Q^2)^2}
 \nonumber \\
&+&\frac{\pi^2}{9}
{\langle \frac{\alpha_s}{\pi}GG \rangle}
\left \{ 
\frac{1}{2 Q^2 M^2} + \frac{1}{Q^4}
- 2 \int_0^{s_0} e^{-s/M^2} \frac{ds }{(s+Q^2)^3}
\right \} 
 \nonumber \\
& & +\frac{64}{27}\pi^3\alpha_s{\langle \bar{q}q\rangle}^2
\lim_{\lambda^2 \to 0}
\left \{ 
\frac1{2Q^2 M^4}
+ \frac{12 k}{Q^4 m_{\rho}^2 }
\left [ 
\log \frac{Q^2}{\lambda ^2} -2
\right.  \right. \nonumber \\
& & + \left.  \left. \int_0^{s_0} e^{-s/M^2}
\left ( 
\frac{s^2+3sQ^2+4Q^4} {(s+Q^2)^3} - \frac1{s+\lambda ^2}
\right) ds 
\right] 
\right. 
\nonumber \\
& & -
\left.  
\frac4{Q^6}
\left [ 
\log \frac{Q^2}{\lambda^2} -3+
\int_0^{s_0} e^{-s/M^2}
\left (  
\frac{s^2+3sQ^2+6Q^4} {(s+Q^2)^3} - \frac1{s+\lambda ^2}
\right) ds 
\right] 
\right \} .
\label{eq:finsr}
\end{eqnarray}
It is understood that the sum rule is  taken
in the limit $\lambda^2 \to 0$ for the parameter $\lambda ^2$
specifying the regularization which we used to calculate the integrals
with the $(1/s)_+$ distribution in eqs.(\ref{eq:fitensor}),(\ref{eq:ficont}).
Furthermore, in this sum rule, we model  the
continuum  by an effective spectral density
$ \rho^{eff}(s, Q^2)$ rather than
by $ \rho^{PT}(s, Q^2)$, including into  $ \rho^{eff}(s, Q^2)$
 all the spectral densities
which are nonzero for $s>0$, $i.e.,$ $ \rho_{2}^{B}(s, Q^2)$ (\ref{eq:twist3cont}),
$ \rho_{2}^{\rho}(s, Q^2)$ (\ref{eq:rhotwist3dual}), $ \rho_{4}^{B}(s, Q^2)$
(\ref{eq:twist5cont}),
$ \rho_{4}^{\rho}(s, Q^2)$ (\ref{eq:rhotwist5dual}),
$ \rho^{\gamma}(s, Q^2)$ (\ref{eq:fitensor}),
$ \rho^{C}(s, Q^2)$ (\ref{eq:ficont}) and
also an analogous contribution from the gluon condensate term.

We studied
the stability of our  sum rule
with respect to variations of the SVZ-Borel parameter $M^2$
in the region $M^2 > 0.6 \, GeV^2$.
A good stability was observed not only for the ``canonical'' value
$s_0^{\pi} \approx  0.7 \, GeV^2$,
but  also for smaller values
of $s_0$, even as small as  $0.4 \, GeV^2$.
Since our results are sensitive to  the  $s_0$-value,
we incorporated  a more detailed model for the spectral density,
treating the $A_1$-meson  as a separate resonance at
$s =1.6 \, GeV^2$,
with the continuum starting at some larger  value $s_A$.
The results obtained in this way
are very stable with respect to variations of  the SVZ-Borel parameter
 $M^2$ and, for $M^2 < 1.2 \, GeV^2$,
show no significant dependence on $s_A$. Numerically, they practically
coincide with the results obtained from the sum rule (\ref{eq:finsr})
for $s_0 = 0.7 \, GeV^2$.

\begin{figure}[thb]
\mbox{
   \epsfxsize=12cm
 \epsfysize=16cm
 \epsffile{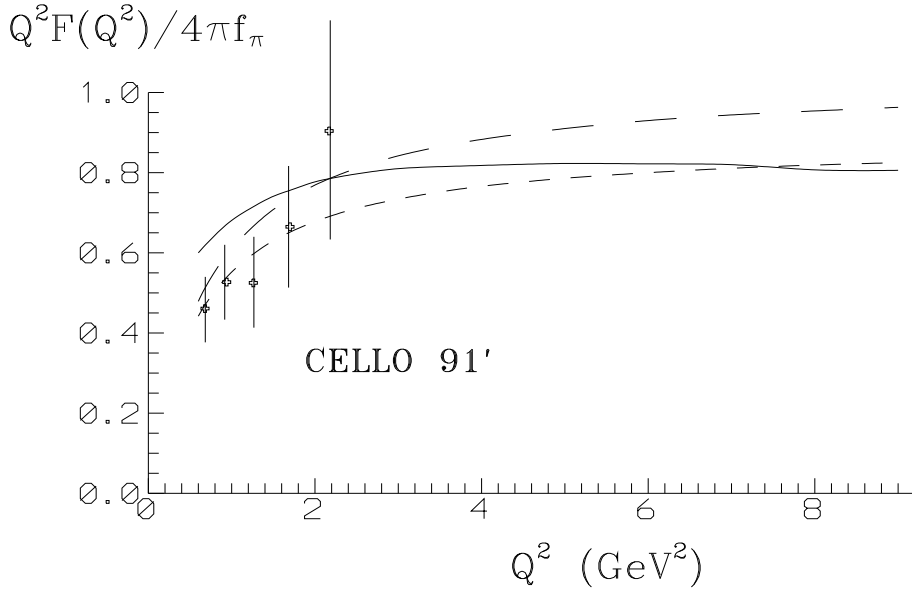}  }
  \vspace{-6.5cm}
{\caption{\label{figfin}
Combination  $Q^2 F_{\gamma \gamma^* \pi^0}(Q^2)/4\pi f_{\pi}$
as calculated from the QCD sum rule
(solid line), $\rho$-pole model (short-dashed line)
and Brodsky-Lepage interpolation (long-dashed line).
 }}
\end{figure}

In Fig.\ref{figfin}, we present a curve for
$Q^2F_{\gamma \gamma^* \pi^0}(Q^2)/4\pi f_{\pi}$
calculated from eq.(\ref{eq:finsr}) for $s_0 = 0.7 \, GeV^2$
and $M^2 = 0.8\, GeV^2$.
One can see that it is rather close to the curve corresponding to the
Brodsky-Lepage interpolation
formula  $\pi f_{\pi} F_{\gamma \gamma^* \pi^0}(Q^2) =
1/(1+Q^2/4\pi^2 f_{\pi}^2)$.
It is also close to the curve
based on the $\rho$-pole  approximation
$\pi f_{\pi} F(Q^2) = 1/(1+Q^2/m_{\rho}^2)$.
It should be noted, however,
that  the  closeness of our results to the
$\rho$-pole behaviour in the $Q^2$-channel  has nothing  to do
with the explicit use of the $\rho$-contributions
in our models for the correlators in the $q^2$-channel.
Recall, that we take $q^2=0$ there, hence, these correlators
simply specify some constants.
The basic reason for the  $Q^2$-dependence
of the $\rho$-pole type  is the fact that the pion
duality interval $s_0 \approx 0.7 \, GeV^2$
is numerically  close to $m_{\rho}^2\approx 0.6\,GeV^2$.

For $Q^2 < 3 \, GeV^2$, our curve
goes slightly  above those based on the  $\rho$-pole dominance
and BL-interpolation (which are close to the
data \cite{CELLO}). This  overshooting
is a consequence of our  assumption
that $Q^2$ can be treated as a large variable:
in some  terms $1/Q^2$ serves  as an expansion
parameter. Such an approximation for these terms
is invalid for small $Q^2$:  by analogy with
the $1/q^2$-terms, one may expect that,
in the low-$Q^2$ region, one should substitute
$1/Q^2$  by something like $1/(Q^2+m_{\rho}^2)$.
In fact, the change must be even more dramatic,
since, in the $Q^2 \to 0$ limit,
only the anomaly term ($i.e.,$ the uncorrected
contribution of the triangle diagram) should survive,
while  all  other terms should vanish.
Hence, using the  limiting form $1/Q^2$, one  appreciably
overestimates the corrections   for $Q^2 \sim 1 \, GeV^2$,
and this  produces  enlarged values for
$F_{\gamma \gamma^* \pi^0}(Q^2)$.

It is also worth noting that the
local duality approximation
\begin{equation}
 \pi f_{\pi} F_{\gamma \gamma^* \pi^0}^{LD}(Q^2) =
\int_0^{s_0} \rho^{PT}(s,Q^2,q^2=0) \,  ds
=\frac1{1+Q^2/s_0}
\end{equation}
(applied to  the original, uncorrected
perturbative spectral density (\ref{eq:rhoq20})
 exactly reproduces the Brodsky-Lepage
interpolation formula (\ref{eq:blin})
if the duality interval $s_0$
assumes the value $s_0=4 \pi^2 f_{\pi}^2 \approx 0.67 \, GeV^2$,
dictated by the
local duality for the two-point function
(see \cite{NeRa82}).

In the region  $Q^2 > 3 \, GeV^2$, our curve for
$Q^2F_{\gamma \gamma^* \pi^0}(Q^2)$
 is practically  constant, supporting
 the pQCD expectation (\ref{eq:ggpipqcd}).  Comparing the
absolute magnitude  of our result
with the  pQCD formula,  we conclude
that it corresponds to the estimate
  $I \approx 2.4$ for the $I$-integral.
Of course, this value has some uncertainty:
it will drift if we change our models
for the bilocals (photon distribution amplitudes).
The strongest sensitivity is to the choice of the
photon distribution amplitude
$\varphi_{\gamma}(y)$
in  the non-diagonal correlator
(\ref{eq:tensor}). However, it should be
emphasized that even switching to the  constant flat form
$\phi_{\gamma}(y)=1$
does  not increase  our result for $I$ by
more than 20\%.  The basic reason
for such a stability is that the
singular $1/q^2$  factor from  the relevant contribution
in  the original sum rule (\ref{eq:SR1})
is substituted  in (\ref{eq:tensor}) by a rather small
factor $k/m_{\rho}^2$.
In fact, having precise data, one can
obtain information on the shape of the
photon distribution amplitude  $\varphi_{\gamma}(y)$
(\ref{eq:rhotens}).

Recalling that
 $I^{as} = 3$ and $I^{CZ} = 5$, we
conclude  that our result $I \approx  2.4$  favours
a  pion  distribution amplitude
which is even narrower than the asymptotic form.
In particular, if we parameterize the width of $\varphi_{\pi}(x)$  by
a simple model $\varphi_{\pi}(x) \sim [x(1-x)]^n$,
we  find  that  $I=2.4$
corresponds to $n=2.5$.
The second moment $\langle \xi^2\rangle$ ($\xi$ is
the relative fraction
$\xi = x - \bar x$)
for such a function  is 0.125.
This  low value (recall that $\langle \xi^2\rangle^{as} =0.2$
while $\langle \xi^2\rangle^{CZ} = 0.43$) agrees, however,  with
the lattice calculation \cite{lattice} and
also with the recent result \cite{minn} obtained from
the analysis of a non-diagonal correlator.

 \section{Conclusions}

Our basic goal in the present
paper was to develop a regular  QCD sum rule
approach to the calculation
of the transition form factor $F_{\gamma \gamma^* \pi^0}(Q^2)$.
Our results  support the expectation that the $Q^2$-dependence
of the transition form factor $F_{\gamma \gamma^* \pi^0}(Q^2)$
is rather close to a simple interpolation
between its  $Q^2 =0$ value (determined  by the ABJ anomaly)
and  the large-$Q^2$ pQCD behaviour $F(Q^2) \sim Q^{-2}$.
Moreover, the QCD sum rule approach   enables  us to calculate
the absolute normalization of the
$Q^{-2}$ term.
The value produced by the QCD sum rule
 is close to that corresponding to
the asymptotic form $\varphi_{\pi}^{as}(x) = 6 f_{\pi} x (1-x)$ of
the pion distribution amplitude.
Though  a detailed comparison  with experimental data
is beyond the scope of this paper,
we would like to mention that our
 curve for $F_{\gamma \gamma^* \pi^0}(Q^2)$
 is  in satisfactory  agreement with the CELLO
data \cite{CELLO} and in good agreement with the
preliminary high-$Q^2$ results from CLEO \cite{CLEO}.
We interpret our findings as a  theoretical evidence
that  $\varphi_{\pi}(x)$ is a rather narrow function.

{\noindent} {\bf Acknowledgments.} We are
grateful to A.P. Bakulev, I. Balitsky, V.M. Braun,
W.W. Buck, H.G. Dosch, A.V. Efremov,
M. Frank, F.Gross, N.Isgur, X.Ji,
D.I. Kazakov, G. Marchesini
and S.V. Mikhailov  for useful  discussions and comments.
The work  of AR  was supported
by the US Department of Energy under contract DE-AC05-84ER40150,
the work of RR was  supported
by Russian Foundation for Fundamental
Research, Grant $N^o$ 96-02-17631 and
by International Science Foundation, Grant $N^o$ RFE300.
AR thanks the DOE's Institute for Nuclear Theory
at the University of Washington for its hospitality
and support  during the completion of  this paper.


\begin{appendix}
\appendix

\section{
Alpha-representation and asymptotic behaviour of the
three-point function}

\setcounter{equation} 0

To study the asymptotic behaviour of the perturbative amplitudes
in the limit when some momentum invariants are large,
one can  use different types of integral representations for the
relevant diagrams.
For the purposes of a  general analysis, one of
the most effective approaches is that using the
the ``alpha-representation'' for the relevant Feynman integral.
To get the alpha-representation, one should write
the denominator of each propagator of the Feynman diagram as
\begin{equation}
\frac1{m_{\sigma}^2 - k_{\sigma}^2 - i\epsilon} =
i \int_0^{\infty} \exp\{ i \alpha_{\sigma} ( k_{\sigma}^2 -
 m_{\sigma}^2 + i\epsilon) \}
d  \alpha_{\sigma} \,  ,
\end{equation}
where $\sigma$ numerates the lines of the diagram,  and then take the
resulting Gaussian integration over all the virtual momenta $k_{\sigma}$.
As a result, for each diagram contributing to $T(q_1,q_2)$
(see Fig.\ref{first}$a$),
one gets the
expression having the following structure:
\begin{eqnarray}
T(q_1,q_2) = \frac{P({\rm c.c.})}{(4\pi)^{zd/4}}
\int_0^{\infty} \prod_{\sigma} d\alpha_{\sigma} D^{-d/2}(\alpha)
\nonumber \\
G(\alpha, q_1,q_2;m_{\sigma})
\exp \left \{ i p^2 \frac{A_{0}( \alpha )}{D(\alpha) }
+i q_1^2 \frac{A_{1}(\alpha)}{D(\alpha) }
+i  q_2^2 \frac{A_{2}(\alpha)}{D(\alpha)} -
i \sum_{\sigma} \alpha_{\sigma} (m_{\sigma}^2- i\epsilon) \right \} \,  ,
\end{eqnarray}
where $d$ is the space-time dimension,  ${P({\rm c.c.})}$ is the relevant
 product of the coupling constants, $z$ is the number of loops of the diagram;
$D,Q,G$ are functions  of the $\alpha$-parameters uniquely determined by the
structure  of the diagram.
In particular,   $D(\alpha)$  is a sum of products of the
$\alpha$-parameters, with the number of the $\alpha$-factors
in each term of  the sum being equal to $z$. In our case,
all  the functions $A_{i}(\alpha)$ are also
 the sums of products of the
$\alpha$-parameters, with $z+1$ parameters $\alpha_{\sigma}$ in each product.
Hence, $D(\alpha)$ and all $A_{i}(\alpha)$
are positive for positive $\alpha$'s.
The preexponential factor $G(\alpha, q_1,q_2;m_{\sigma}) $ is a polynomial
in $\alpha$'s, $p^2$, $q_1^2$ and $q_2^2$.

In the region where one of the momentum variables $p_i^2$ is large,
all the contributions having a  power-type behaviour on that variable
can only come from the integration region  where the relevant
$A_i/D$ factor vanishes: if $A_i/D $ is larger than some constant $\rho$
in the region of integration, the resulting contribution
is $\sim \exp(ip_i^2 \rho)$, $i.e.,$ it is exponentially suppressed.

When all $A_i$'s are non-negative,  there are two basic possibilities to arrange
$A_i/D=0$. In the  first case, called the ``short-distance regime'',
$A_i$  vanishes faster than
$D$ when some of the $\alpha$-parameters tend to zero
(small $\alpha$ correspond to large virtualities $k^2$, $i.e.,$
to short distances).
The second possibility, called the
``infrared regime'', occurs if  $D$ goes to infinity faster than
$A_i$ when some of the $\alpha$-parameters tend  to infinity
(large $\alpha$ correspond to small  momenta $k$, $i.e.$
to the infrared limit).
One can also imagine a combined regime, when $A_i/D=0$
because some $\alpha$-parameters vanish and some are infinite.

In fact, there exists  a simple rule using which one can easily find the
lines $\sigma$ whose $\alpha$-parameters
 must be set to zero and those whose $\alpha$-parameters
  must be taken infinite
 to assure that  $A_i/D=0$.  First, one should realize that
$A_i/D=0$ means that the corresponding diagram
of a scalar theory (in which $G=1$)  has no dependence
on the momentum invariant $p_i^2$.
As the second step, one should incorporate the well-known analogy between
the Feynman diagrams and electrical circuits \cite{BjDrell}:
the $\alpha_{\sigma}$-parameters may  be interpreted  as the resistances
of the corresponding lines $\sigma$.
In other words,  $\alpha_{\sigma}=0$ corresponds to the
short-circuiting the line $\sigma$ while
$\alpha_{\sigma}= \infty$ corresponds to its removal from
the diagram.
Hence, the problem is to find the sets of lines $\{\sigma\}_{SD}$, $\{\sigma\}_{IR}$
whose contraction into point (for $\{\sigma\}_{SD}$)
or removal from the diagram (for $\{\sigma\}_{IR}$)  produces the diagram
which, in a scalar theory,  does not depend on $p_i^2$.

Thus, the rule determining possible
topological types of the short-distance factorizable
contributions is the following:
if the part of the diagram corresponding to
a  short-distance subprocess is contracted into point,
the resulting effective diagram should have no dependence
on the large momentum invariants.

The simplest situation is when the short-distance part
coincides with the whole diagram.
This configuration  is allowed for any
relation between $q_1^2,q_2^2$ and $p^2$.
In this case, all the  currents
$J_{\mu}(X)$, $J_{\nu}(0)$ and $j_{\alpha}^5(Y)$
(see Fig.\ref{eq:fig5}) are close to each other, $i.e.,$   all the
  intervals $X^2,Y^2,(X-Y)^2$\, are small.
However, if, say, the variable  $q_1^2$ is small,
the dependence on large variables $q_2^2$ and $p^2$
can be eliminated by contracting into point a subgraph
containing the vertices corresponding to momenta $q_2$ and $p$.
In this situation, the interval $Y^2$ is small while
$X^2$ and $(X-Y)^2$ are large. Such a configuration
is sensitive to the long-distance effects in the $q_1$-channel.
They can be described by introducing
distribution amplitudes for the $q_1$-photon.
Finally, another interesting situation is when
both the photon virtualities are much larger
than $p^2$. Then   there exists a
short-distance subprocess which   includes
only  the photon vertices:
the  interval $X^2$ is small, while $Y^2$ and $(X-Y)^2$  may be  large.
In this case, there are
long-distance effects in the axial current channel, which are usually
described/parameterized by the pion distribution amplitude.

\section{
 Calculation of some useful momentum integrals}

\setcounter{equation} 0

We calculate   integrals over the momentum
 using the dimensional regularization. The
basic integral is  well-known:
\begin{equation}
I(L,r)=\int d^D\hat{p} \frac{(p^2)^r}{[p^2 +S]^L}=
\frac{i(-1)^{r-L}}{(4\pi)^{D/2}} \frac{\mu^{4-D}}{\Gamma(L)}
\frac{\Gamma(r+D/2)}{\Gamma(D/2)} \frac{\Gamma(L-r-D/2)}{(-S)^{L-r-D/2}},
\end{equation}
where $D=4-2\varepsilon$ and $d^D\hat{p}\equiv{d^Dp}/{(2\pi)^D}$.
In fact, it is more
convenient to express the results in terms of  the integral
\begin{equation}
R(L,r)\equiv I(L,r) 2^r \frac{\Gamma(D/2)}{\Gamma(r+D/2)}.
\end{equation}
For our calculations, we need the following integrals:
\begin{equation}
\int d^D\hat{p} \frac{(pY)^n\{1,p_{\rho},p_{\rho}p_{\sigma},\ldots\}}
{(p^2)^{\alpha}(p-q)^{2\beta}}=\int_0^1 dx x^{\alpha-1} {\bar{x}}^{\beta-1}
\frac{\Gamma(\alpha+\beta)}{\Gamma(\alpha)\Gamma(\beta)}
\int d^D\hat{p} \frac{(pY)^n\{1,p_{\rho},p_{\rho}p_{\sigma},\ldots\}}
{[(p-\tilde{q})^2+S]^L},
\end{equation}
where $L=\alpha+\beta, \tilde{q}=q\bar{x}, S=q^2x\bar{x}, \bar{x}\equiv 1-x$.
Omitting, for a moment, the integration over $x$,
let us concentrate on
the integrals
\begin{equation}
\{J(L,n),J_{\rho}(L,n),J_{\rho\sigma}(L,n),\ldots\}=
\int d^D\hat{p} \frac{(pY)^n\{1,p_{\rho},p_{\rho}p_{\sigma},\ldots\}}
{[(p-\tilde{q})^2+S]^L} \, .
\end{equation}

Shifting   the integration variable,  we expand them in a standard way:
\begin{equation}
(p+\tilde{q}Y)^n=(\tilde{q}Y)^n + n(\tilde{q}Y)^{n-1}(pY) +
\frac{n(n-1)}{2}(\tilde{q}Y)^{n-2}(pY)^2 + \cdots  \  . \nonumber
\end{equation}
Now, the integration over $d^D\hat{p}$ is straightforward
 and, keeping the
terms  up to $Y^2$, we obtain
\begin{equation}
J(L,n)=(\tilde{q}.Y)^nR(L,0) + \frac{n(n-1)}{2}
(\tilde{q}Y)^{n-2}Y^2R(L,1) + \bar{O}(Y^4),
\end{equation}
\begin{eqnarray}
J_{\rho}(L,n)&=&{\tilde{q}}_{\rho}\left\{
(\tilde{q}Y)^nR(L,0) + \frac{n(n-1)}{2}(\tilde{q}Y)^{n-2}Y^2R(L,1) +
\bar{O}(Y^4)\right\} +\nonumber \\
&+&n Y_{\rho}\left\{
(\tilde{q}Y)^{n-1}R(L,1) + \frac{(n-1)(n-2)}{2}(\tilde{q}Y)^{n-3}Y^2R(L,2) +
\bar{O}(Y^4)\right\},
\end{eqnarray}
\begin{eqnarray}
J_{\rho\sigma}(L,n)&=&
{\tilde{q}}_{\rho}{\tilde{q}}_{\sigma}\left\{
(\tilde{q}Y)^nR(L,0) + \frac{n(n-1)}{2}(\tilde{q}Y)^{n-2}Y^2R(L,1) +
\bar{O}(Y^4)\right\} + \nonumber  \\
&+&n({\tilde{q}}_{\rho}Y_{\sigma}+{\tilde{q}}_{\sigma}Y_{\rho})\left\{
(\tilde{q}.Y)^{n-1}R(L,1) + \frac{(n-1)(n-2)}{2}(\tilde{q}Y)^{n-3}Y^2R(L,2) +
\bar{O}(Y^4)\right\} + \nonumber  \\
&+&g_{\rho\sigma}(\tilde{q}Y)^nR(L,1) + \frac{n(n-1)}{2}
(\tilde{q}Y)^{n-2}R(L,2)\left\{
2Y_{\rho}Y_{\sigma} + g_{\rho\sigma}Y^2\right\} + \nonumber  \\
&+&\frac{n(n-1)(n-2)(n-3)}{2}(\tilde{q}Y)^{n-4}R(L,3)
\left\{Y_{\rho}Y_{\sigma}Y^2 +
\bar{O}(Y^4)\right\}.
\end{eqnarray}

\section{
Equations of motion and  $\rho$-meson distribution amplitudes }

\setcounter{equation} 0

Here we demonstrate how one can use
equations of motion to obtain relations  between
the moments of the $\rho$-meson  distribution amplitudes.
 A similar analysis was
done in refs. \cite{Gor1} and
\cite{BraF90,brali}.

Consider the identity:
\begin{equation}
\langle 0 |{\bar{\psi}}_{\alpha}(0)\,(i\hat{\nabla} - m)_{\beta\rho}\,
\psi_{\rho}(z)|\rho \rangle = 0 ,
\label{eq:A1}
\end{equation}
where $i\hat{\nabla}_{\beta\alpha}=(i\hat{\partial}_z + g\hat{{\cal{A}}}(z))_{\beta\alpha}$.
Applying the Fiertz transformation we rewrite (\ref{eq:A1}) as
\begin{eqnarray}
\lefteqn{ (i\hat{\nabla} - m)_{\beta\alpha}\,S(z) +
\left[(i\hat{\nabla} - m)\,\gamma_5\right]_{\beta\alpha}\,P(z) +
\left[(i\hat{\nabla} - m)\,\gamma_{\mu}\right]_{\beta\alpha}\,V_{\mu}(z) - }\nonumber\\
& & \nonumber\\
&-&\left[(i\hat{\nabla} - m)\,\gamma_{\mu}\gamma_5\right]_{\beta\alpha}\,A_{\mu}(z) +
\frac{1}{2}\left[(i\hat{\nabla} - m)\,\sigma_{\mu\delta}\right]_{\beta\alpha}\,
T_{\mu\delta}(z) = 0 .
\label{eq:A2}
\end{eqnarray}

To obtain a relation  between the distribution amplitudes,  one should substitute
in (\ref{eq:A2}) the expressions for the bilocal matrix elements like
(\ref{eq:SE9}),(\ref{eq:SE10}),(\ref{eq:SE18}),(\ref{eq:SE19}),(\ref{eq:SE25}),
differentiate them with respect to $z$ and take $z^2=0$.
By contraction with $[\sigma_{\nu\rho}]_{\alpha\beta}$,  we pick out
a combination of
the V-, A- and T-projections. There are three independent tensor structures
$$z_{\nu}\varepsilon_{\rho}-z_{\rho}\varepsilon_{\nu},\
(\varepsilon z)\,(z_{\nu}p_{\rho}-z_{\rho}p_{\nu}),\ p_{\nu}\varepsilon_{\rho}-p_{\rho}\varepsilon_{\nu}, $$
and, as a result,  we get three systems of equations:
\begin{eqnarray}
f_{\rho}^A m_{\rho}^2\,{\langle x^{n+1} \rangle}_A& = &
 f_{3\rho}^A m_{\rho}^2\,\int_0^1\,d\beta\,\beta\,n\,
{\langle\ [x_3\beta + x_2]^{n-1}\rangle}_{3A}\,+ \nonumber\\
& & \nonumber\\
& &{} +2 f_{\rho}^A C_{A4}\,n\,{\langle x^{n-1} \rangle}_{A4}\,
-\, 2 C_{V4}\,f_{\rho}^V m_{\rho}\,{\langle x^n \rangle}_{V4}
+\,\frac{3}{2}\,f_{\rho}^V m_{\rho} C_{[V2]4}\,{\langle x^n \rangle}_{[V2]4},
\label{eq:bypreq}
\end{eqnarray}
\begin{equation}
f_{\rho}^A\,C_{A4}\,{\langle x^n \rangle}_{A4}\ = \
a_{V1}\,f_{\rho}^V m_{\rho}\,C_{[V1]4}\,{\langle x^n \rangle}_{[V1]4}
\,+\,\frac{1}{2}\,f_{\rho}^V m_{\rho} C_{[V2]4}\,{\langle x^{n+1} \rangle}_{[V2]4},
\label{eq:main1eq}
\end{equation}
\begin{eqnarray}
\lefteqn{(n+2)\,f_{\rho}^A\,{\langle x^n \rangle}_A  =
-\,f_{\rho}^V m_{\rho}\,{\langle x^{n+1} \rangle}_V \,
-\, a_{V1}\,f_{\rho}^V m_{\rho}\,{\langle x^n \rangle}_{V1}
}  \nonumber\\ & &
-\, f_{3\rho}^A\,\int_0^1\,d\beta\,\beta\,n\,{\langle\
[x_3\beta + x_2]^{n-1}\rangle}_{3A}
  \nonumber\\ & &
+\, f_{3\rho}^V\,\int_0^1\,d\beta\,\beta\,n\,
{\langle\ [x_3\beta + x_2]^{n-1}\rangle}_{3V}
\,+\, 4m_q\,f_{\rho}^T\,{\langle x^n \rangle}_T
\label{eq:maineq}
\end{eqnarray}
where we have used the notation:
${\langle x^n \rangle}_A  = \int_0^1 dx x^n \varphi_{\rho}^A(x,\mu^2)$,
etc.
It  should be noted that eqs.(\ref{eq:bypreq}),(\ref{eq:main1eq}) are new.
Eq.(\ref{eq:maineq}) was
derived also in \cite{Gor1}, but the constant $a_{V1}$ was missed there.
In the chiral limit, we can neglect the last term in  (\ref{eq:maineq}).
Taking the infinite limit for the
renormalization parameter,   $\mu^2 \to \infty$ in (\ref{eq:maineq}), we
obtain the equation
relating  the moments of the asymptotic  leading-twist
distribution amplitudes:
\begin{equation}
(n+2)\,f_{\rho}^A\,{\langle x^n \rangle}_A\ = \
-\,f_{\rho}^V m_{\rho}\,{\langle x^{n+1} \rangle}_V\,
-\, a_{V1}\,f_{\rho}^V m_{\rho}\,{\langle x^n \rangle}_{V1} .
\label{eq:asypteq}
\end{equation}

The asymptotic distribution amplitudes $\varphi_V^{as}(x)$,
 $\varphi_A^{as}(x)$
were originally obtained in \cite{CZ84}:
\begin{equation}
\varphi_V^{as}=\frac{3}{2}(1-2x\bar{x}),\
\varphi_A^{as}=6x\bar{x},
\label{eq:aswfVA}
\end{equation}
Taking into account the normalization conditions (\ref{eq:norm}),
 we conclude that there exists only one
solution:
\begin{equation}
\varphi_{V1}^{as}=60x\bar{x}\,(2x-1),\
\label{eq:aswfV1}
\end{equation}
with
\begin{equation}
(f_{\rho}^A)^{as} = -\frac{f_{\rho}^V\,m_{\rho}}{4},\quad
(a_{V1})^{as} = \frac{1}{40} .
\label{eq:consts}
\end{equation}
Note, that $\varphi_V^{as},\varphi_{V1}^{as}$,  given by eqs. (\ref{eq:aswfVA})
and (\ref{eq:aswfV1}), obey the
condition that, for a longitudinally polarized $\rho^o$-meson
($i.e.,$ when $\varepsilon_{\sigma}^{\lambda=0}\simeq i\,{p_{\sigma}}/{m_{\rho}}+
{\cal{O}}\left({m_{\rho}}/{p_z}\right), \mbox{as}\ p_z\rightarrow\infty$),
the leading-twist part in eq.(\ref{eq:SE9}) provides the well known
asymptotic twist-2 vector wave function  (cf. eq.(\ref{eq:SE4})).
The value of $f_{\rho}^A$ was also calculated within the SR method \cite{CZ84},
and the result is in a good agreement with that dictated by
 equations of motion.

Substituting eq.(\ref{eq:consts}) into   eq.(\ref{eq:asypteq}), we get
\begin{equation}
\frac{(n+2)}{4}\,{\langle x^n \rangle}_A\ = \
{\langle x^{n+1} \rangle}_V\,
+\, \frac{1}{40}\,{\langle x^n \rangle}_{V1} .
\label{eq:aseqfin}
\end{equation}
This formula is an analog of the well known relations \cite{WaWi}
(see also \cite{brali,brball}) between the moments of the
spin-dependent structure functions $g_1(x)$,
$g_2(x)$.

The asymptotic form for the next-to-leading two-body distribution amplitudes
can be directly extracted from the corresponding correlators (\ref{eq:defbilocr}):
\begin{equation}
\varphi_{A4}^{as} = 30y^2{\bar{y}}^2,\
\varphi_{[V1]4}^{as} = 420(y\bar{y})^2\,(2y-1),\ \
\varphi_{[V2]4}^{as} = 30y^2{\bar{y}}^2\  .
\label{eq:tw5aswfall}
\end{equation}
Using (\ref{eq:main1eq}) for $n=0$ and $n=1$ gives  the
following relations:
\begin{equation}
C_{[V2]4} = - C_{A4}\quad  \mbox{and}\quad  C_{[V1]4} = \frac{5}{7}\, C_{A4}.
\end{equation}
In the main text of the paper we gave the estimates (\ref{eq:tw5estimation})
for $C_{A4}$  and $ C_{[V1]4}$ based on the local duality
arguments. These estimates satisfy the  second relation above.
The first one can be used to estimate $C_{[V2]4}$.

Combining the SR method for the correlators (\ref{eq:defbilocr}) and  expansion
of the relevant  composite operators in the conformal basis\footnote{
To one loop accuracy, only the operators with the same conformal spin mix
under the renormalization.} one can calculate the
corrections to (\ref{eq:aswfVA})--(\ref{eq:consts}).
It should be noted  that this  expansion should
be consistent with the equations of motion
(\ref{eq:bypreq}) -- (\ref{eq:maineq}). Such a program for the case
of the nonleading
$\pi$-meson distribution amplitudes  was performed in
\cite{Gor1,BraF90} and it was shown there that
the deviation from the asymptotic form is  small.

\section{
Some properties of  traceless combinations}

\setcounter{equation} 0

To construct
the orthogonal projection operators $P_{(n)}$ onto the subspace of traceless
symmetric Lorentz tensors of rank $n$,
we use the techniques similar to those in \cite{GutSop}.
Here  we list some useful formulas concerning these
projectors as well as some contractions that appear in the paper.

By definition, for an arbitrary Lorentz tensor $T$ we have \cite{GutSop}:
\begin{equation}
\left[P_{(n)} T\right]^{\mu_1\ldots\mu_n} =
{P_{(n)}}_{\nu_1\ldots\nu_n}^{\mu_1\ldots\mu_n}\,T^{\,\nu_1\ldots\nu_n} .
\end{equation}
It is straightforward to derive the formula
\begin{equation}
\left(P_{(n)} T\right)^{\mu_1\ldots\mu_n} =
\frac{1}{n}\,\sum_{i=1}^{n}\,T^{\,\mu_1\ldots[\mu_i]\ldots\mu_n\mu_i} -
 \frac{1}{n^2}\,\sum_{i<j}^{n}\,g^{\mu_i\mu_j}\,
T^{\,\mu_1\ldots[\mu_i]\ldots[\mu_j]\ldots\mu_n\alpha\,\alpha} ,
\end{equation}
where $T^{\,\nu_1\ldots\nu_{n-1}\,\alpha}$ is now traceless and symmetric in
its first $n-1$ indices and $[\mu_i]$ means that the corresponding
index is absent. Choosing $T^{\,\{\nu_1\ldots\nu_{n-1}\}\,\alpha}
\equiv s^{\alpha}\,\{q_1^{\nu_1}\ldots q_1^{\nu_{n-1}}\}$ we have:
\begin{eqnarray}
\{s^{\mu_1}q_1^{\mu_2}\ldots q_1^{\mu_n}\}& = &
\frac{1}{n}\,\sum_{i=1}^n\,
s^{\mu_i}\,\{q_1^{\mu_1}\ldots[q_1^{\mu_i}]\ldots q_1^{\mu_n}\} -
\nonumber\\
&-& \frac{1}{n^2}\,\sum_{i<j}^n\,g^{\mu_i\mu_j}\,
s^{\alpha}\,
\{q_1^{\alpha}q_1^{\mu_1}\ldots[q_1^{\mu_i}]\ldots[q_1^{\mu_j}]
\ldots q_1^{\mu_n}\}  .
\label{eq:skill1}
\end{eqnarray}

Making use of the Nachtmann's \cite{Nacht} contraction
\begin{equation}
z^{\mu_1}\ldots z^{\mu_n}\,\{q_1^{\mu_1}\ldots q_1^{\mu_n}\}_n =
\left( \frac{ q_1^2 z^2}{4}\right)^{n/2}\,C_n^{1}(\eta)
\label{eq:skill2}
\end{equation}
and some recursion
relations for the Gegenbauer polynomials $C_n^{\lambda}(\eta)$ \cite{Prudn},
one can  derive the formula:
\begin{eqnarray}
z^{\mu_1}\ldots z^{\mu_{n-1}}\,
\{q_1^{\alpha}q_1^{\mu_1}\ldots q_1^{\mu_{n-1}}\} &=&
\frac{1}{n}\,\frac{\partial}{\partial z^{\alpha}}\,
z^{\mu_1}\ldots z^{\mu_n}\,\{q_1^{\mu_1}\ldots q_1^{\mu_n}\} = \nonumber\\
&=&\frac{1}{n}\left[
\frac{z^{\alpha}}{Z^2}\,{\tau}^n \,(-2\,C_{n-2}^{2}(\eta)\,) +
q_1^{\alpha}\,{\tau}^{n-1}\,C_{n-1}^{2}(\eta)\right] ,
\label{eq:skill3}
\end{eqnarray}
where $\eta=i\,(q_1z)/\sqrt{-z^2 q_1^2}
\tau= -i \sqrt{-z^2 q_1^2}/2$.

Using (\ref{eq:skill1}) -- (\ref{eq:skill3}),  one gets for an arbitrary
4-vector $s$:
\begin{equation}
z^{\mu_1}\ldots z^{\mu_n}\,\{s^{\mu_1}q_1^{\mu_2}\ldots q_1^{\mu_n}\} =
\frac{(z s)}{n}\,{\tau}^{n-1}\,C_{n-1}^{2}(\eta) -
\frac{(q_1 s)}{n}\,\frac{z^2}{2}\,{\tau}^{n-2}\,C_{n-2}^{2}(\eta) .
\label{eq:skill4}
\end{equation}

\section{
Contact terms  }

\setcounter{equation} 0

Here we derive eq.(\ref{eq:cont9}). Before considering the relevant
contraction, let us note that, incorporating   the relation
\begin{equation}
\{z \partial\}^n = (z \partial)\{z \partial\}^{n-1} -
\frac{\partial^2 z^2}{4} \frac{(n-2)}{n}\, \{z \partial\}^{n-2},
\end{equation}
and neglecting  higher twist contributions,  one can
substitute the original correlator (\ref{eq:ct5})  by
\begin{equation}
\Pi_{\mu\{\mu_1\ldots\mu_n\}}(q_1) =
\Pi_{\mu\mu_1\{\mu_2\ldots\mu_n\}}(q_1) + \cdots \  .
\end{equation}

As a  result,
\begin{eqnarray}
\lefteqn{ \Pi_{\mu\{\mu_1\ldots\mu_n\}}(q_1)\,g_{\mu\mu_1} \simeq
\Pi_{\mu\mu\{\mu_2\ldots\mu_n\}}(q_1) }\nonumber\\
& & = \int dx e^{-i q_1 x}\,
\langle 0|T\{ J_{\mu}(x)\,
\bar{u}(0){{\partial}}_{\mu}\{{{\partial}}_{\mu_2}\ldots{{\partial}}_{\mu_n}\}
u(0)\}| 0 \rangle \nonumber\\
& & = -\frac{1}{2} \int dx e^{-i q_1 x}\,
\langle 0|T\{ J_{\mu}(x)\,\bar{u}(0)
\widehat{\stackrel{\leftarrow}{\partial}}\gamma_{\mu}\{{{\partial}}_{\mu_2}\ldots{{\partial}}_{\mu_n}\}
u(0)\}| 0 \rangle
\label{eq:contact}\\
& & +\frac{1}{2} \int dx e^{-i q_1 x}\,
\langle 0|T\{ J_{\mu}(x)\,\bar{u}(0)
\gamma_{\mu}\{{{\partial}}_{\mu_2}\ldots{{\partial}}_{\mu_n}\}\widehat{\stackrel{\rightarrow}{\partial}}
u(0)\}| 0 \rangle
\label{eq:zero}\\
& & +\frac{i}{2}\,q_{1\varepsilon}\,\int dx e^{-i q_1 x}\,
\langle 0|T\{ J_{\mu}(x)\,\bar{u}(0)
\gamma_{\mu}\gamma_{\varepsilon}\{{{\partial}}_{\mu_2}\ldots{{\partial}}_{\mu_n}\}
u(0)\}| 0 \rangle ,
\label{eq:wftens}
\end{eqnarray}
where we have made use of the identity:
\begin{eqnarray}
\int dx e^{-i q_1 x}\,\left[
\langle 0|T\{ J_{\mu}(x)\,\bar{u}(0)
{\stackrel{\leftarrow}{\partial}}_{\varepsilon}\hat{\Gamma} u(0)\}| 0 \rangle +
\langle 0|T\{ J_{\mu}(x)\,\bar{u}(0)
\hat{\Gamma} {\stackrel{\rightarrow}{\partial}}_{\varepsilon} u(0)\}| 0 \rangle \right] &=& \nonumber\\
 = i\,q_{1\varepsilon}\, \int dx e^{-i q_1 x}\,
\langle 0|T\{ J_{\mu}(x)\,\bar{u}(0)
\hat{\Gamma} u(0)\}| 0 \rangle. \qquad& &
\label{eq:idnt}
\end{eqnarray}

Applying (\ref{eq:ct3}) to (\ref{eq:contact}) and  (\ref{eq:zero}), and
integrating  by parts in (\ref{eq:contact}) we get
\begin{equation}
({\rm E}.3) = 2\,(-i)^{n-1}\,\{q_{1\mu_2}\ldots q_{1\mu_n}\}\,
\langle\bar{u} u \rangle .
\end{equation}
Now, taking into account that
$\langle 0|\bar{u}(0)
\{{{\partial}}_{\mu_2}\ldots{{\partial}}_{\mu_n}\}
u(0)\}| 0 \rangle = 0 $ for all $n$, we obtain:
\begin{equation}
({\rm E}.4) = 0 \ .
\end{equation}

\end{appendix}

\end{document}